\documentclass[11pt]{article}
\usepackage{amsmath,amssymb}
\usepackage[T1]{fontenc}
\usepackage{lmodern}
\usepackage{verbatim}
\font\dm=cmr9

\numberwithin{equation}{section}
\let\dsp\displaystyle
\let\ov\overline
\let\os\overrightarrow
\let\pa\partial
\let\q\quad
\def\qh#1{\quad\hbox{#1}\quad}
\let\ul\underline
\let\wt\widetilde
\let\td\tilde
\let\a\alpha
\let\b\beta
\let\d\delta
\let\D\varDelta
\let\ve\varepsilon
\let\g\gamma
\let\G\varGamma
\let\la\lambda
\let\n\nabla
\let\o\omega
\let\O\varOmega
\let\vf\varphi
\let\si\sigma
\let\t\theta
\def\U{\text{U}}
\let\z\zeta
\def\(#1){{(#1)}}
\def\[#1]{{[#1]}}
\def\m{{\mu\nu}}
\def\cL{\mathcal L}
\def\toG{\wt{\ov \G}{}}
\def\ton{\wt{\ov \nabla}{}}
\def\oG{\overline\varGamma}
\def\dg #1,#2,#3,{{#1{}_{#2}}^{\!#3}}
\def\gd #1,#2,#3,{{#1{}^{#2}}_{\!#3}}
\def\gdg #1,#2,#3,#4,{{{#1{}^{#2}}_{\!#3}}{}^{\!#4}}
\def\dgd #1,#2,#3,#4,{{{#1{}_{#2}}^{\!#3}}{}_{\!#4}}
\def\nad#1#2{\overset{\rm #1}{#2}}
\def\falg{{\mathrel{\lower5pt\hbox{${\scriptstyle\sim}$}\hskip-5pt g}}}
\def\falj{{\mathrel{\lower5pt\hbox{${\scriptstyle\sim}$}\hskip-4pt j}}}
\def\fal#1{{\mathrel{\lower4pt\hbox{${\scriptstyle\sim}$}\hskip-7.5pt #1}}}
\def\Tem_#1{\nad{em}T_{\!#1}}
\def\tl{\mathopen{\hbox{\dm[}}}
\def\tp{\mathclose{\hbox{\dm]}}}
\def\ddf#1#2{\frac{d#1}{d#2}}
\def\pap#1#2{\frac{\pa #1}{\pa #2}}
\def\pae#1#2#3{\X2(\frac{\pa #1}{\pa #2}\Y2)^#3}
\def\sgn{\mathop{\rm sgn}}
\catcode`@=11
\def\X#1#2{\mathopen{\hbox{$\left#2\vbox to\ifcase#1\or9.5\p@\or11.5\p@\or14.5\p@\or17.5\p@\fi{}\right.\n@space$}}}
\def\Y#1#2{\mathclose{\hbox{$\left.\vbox to\ifcase#1\or9.5\p@\or11.5\p@\or14.5\p@\or17.5\p@\fi{}\right#2\n@space$}}}
\catcode`@=12

\let\ti\textit
\def\beq#1 #2\e{\begin{equation}\label{#1}#2\end{equation}}
\def\beqg#1\e{\begin{equation*}#1\end{equation*}}
\def\bea#1 #2\e{\begin{align}\label{#1}#2\end{align}}
\def\bml#1 #2\e{\begin{multline}\label{#1}#2\end{multline}}
\def\bg#1 #2\e{\begin{gather}\label{#1}#2\end{gather}}
\def\bgg#1\e{\begin{gather*}#1\end{gather*}}
\let\bal\aligned \let\eal\endaligned
\let\bga\gathered \let\ega\endgathered
\def\ben{\begin{enumerate}}
\def\een{\end{enumerate}}
\def\labelenumi{{\rm\arabic{enumi})}}
\let\nn\nonumber
\def\lb#1 {\label{#1}}
\let\er\eqref

\def\up#1{\uppercase{#1}}
\def\E{\expandafter\up}
\def\as{axially-symmetric}
\def\cf{coefficient}
\def\cfn{confinement}
\def\cn{connection}
\def\cd{coordinate}
\def\co{cosmolog}
\def\ci{covariant}
\def\cvt{curvature}
\def\cy{cylindrical}
\def\df{definition}
\def\dv{derivative}
\def\di{di\-men\-sion}
\def\elm{electromagneti}
\def\e{equation}
\def\ev{equivalen}
\def\fb{fibre bundle}
\def\fw{following}
\def\f{function}
\def\fn{fundamental}
\def\gz{generalized}
\def\gm{geometrization}
\def\gr{grav\-ita\-tion}
\def\gelm{gravito-electromagneti}
\def\Hm{Higgs' mechanism}
\def\ia{interaction}
\def\iv{invarian}
\def\JT{(Jordan--Thiry) }
\def\KK{Kaluza--Klein }
\def\KW{Kerner--Wong--Kopczy\'nski }
\def\Lg{lagrangian}
\def\LC{Levi-Civita }
\def\MR{Moffat--Ricci}
\def\nA{non-Abelian}
\def\nh{nonholonomic}
\def\nos{nonsymmetric}
\def\NK#1{Nonsymmetric Kaluza--Klein #1Theory}
\def\pc{particle}
\def\pt{potential}
\def\pp{principle}
\def\qe{quintessen}
\def\sf{satisf}
\def\so{solution}
\def\spt{space-time}
\def\ssb{spontaneous symmetry breaking}
\def\sy{stationary}
\def\st{such that }
\def\s{symmetr}
\def\tf{transformation}
\def\un{unification}
\def\v{variation}
\def\wrt{with respect to }
\def\YM{Yang--Mills}
\let\TM\texttrademark

\author{M. W. Kalinowski}
\title{Axially-symmetric, stationary fields,
gravito-electromagnetic waves\\
and a possibility to travel in higher dimensions\\
in the Nonsymmetric Kaluza--Klein Theory}
\begin{document}
\maketitle

\vskip20pt
\begin{abstract}
We consider in the paper axially symmetric and stationary fields and cylindrically symmetric gravito-electromagnetic waves in the Nonsymmetric
Kaluza--Klein Theory. Using symbolic manipulations we write down all important quantities in the theory. We write field equations for both cases and
partially integrate them. We consider also a generalized Kerner--Wong--Kopczy\'nski equation in GSW model.
\end{abstract}

\section{Introduction}
In the paper we consider \sy\ and \as\ fields and field \e s in the \NK{\JT}. We consider only 5-\di al, \elm c case.
A~difference between \NK{} and \E\nos\ Jordan--Thiry Theory consists in introducing a scalar field in \E\nos\ Jordan--Thiry Theory. This is
the same as in the classical \KK \JT Theory. The \E\nos\ \KK \JT Theory \sf ies a Bohr correspondence principle to the classical one.

The \NK{\JT} has been created and developed (see Refs \cite1, \cite2, \cite3, \cite{3a}, also \cite4, \cite{4a}, \cite m). The theory uses a
\nos\ metrization of a fibre bundle. On a \spt\ theory uses a formalism of Einstein Unified Field Theory with Moffat interpretation as an extended
theory of \gr\ (see Ref.~\cite5). The theory has been considered in 5-\di al case (Refs \cite1,~\cite3) and also in $(n+4)$-\di al case (see Refs
\cite1,~\cite2), including also \ssb\ and \Hm\ (see Ref.~\cite1). In 5-\di al case the theory is bi-invariant \wrt $\U(1)$-group action on a fibre
bundle. In a general \nA\ case we have only right-invariance of an action of a group~$G$ on a~bundle. The theory has been applied to a problem of
an anomalous acceleration of Pioneer 10 and~11 spacecraft (see Ref.~\cite6). We develop a theory finding spherically \s ic \so s (5-\di al \elm c
case) with \pc-like properties and \gz\ plane gravito-\elm c waves (see Refs \cite3,~\cite7). A~dielectric model of a \cfn\ of charge (see
Ref.~\cite7) has been posed and developed. Afterwards such a model has been extended to a \nA\ dielectric model of color \cfn\ (see Ref.~\cite8).
The \NK{} with \ssb\ and \Hm\ has been developed. In particular, a~bosonic part of GSW (Glashow--Salam--Weinberg) has been derived with correct
Weinberg angle and masses of $W^\pm,Z^0$ and Higgs' \pc\ agreed with experiment (see Ref.~\cite8). A problem of hierarchy of \s y breaking in the
\NK{} has been derived (see Ref.~\cite9). \E\co ical models in the \NK{\JT} with \qe ce and \qe tial inflation have been obtained (\cite{10}).
All of the above investigations are in line of the so-called programme of \gm\ and \un\ of \fn\ physical \ia s (see
Refs \cite{11},~\cite{12}).

The paper has been organized as follows. In the second section we write down all important formulas in the \NK{} (5-\di al case), which we should use
in Sections~3 and~4. Section~3 is devoted to \as\ \sy\ fields in (5-\di al) \NK{}. In Section~4 we consider \cy\ gravito-\elm c waves in the \NK{}. We
use results calculated by our programmes written in \ti{Mathematica}. We write down field \e s for the mentioned cases. In the fifth section we
consider \gz\ \KW \e\ in GSW model as a possibility to travel in higher \di s.

In \ti{Conclusions and Prospects} we give some prospects for further investigations. The programmes have been quoted in Appendix~A and Appendix~B.
In Appendix~A for \as\ and \sy\ case. In Appendix~B for \cy\ gravito-\elm c waves.

Let us remind to the reader that someone
considers \cy\ waves in Einstein's unified field theory (see Refs \cite{l2},~\cite{l3}) and in Bonnor's and Schr\"odinger's unified field theory
(see Ref.~\cite{l1}).
In~NGT (\E\nos\ \E\gr\ Theory) the problem of \gr al radiation from axi\s ic sources has been taken into account (see Refs \cite{l4},
\cite{l5},~\cite{l6}).

We use such system of units that $G_N=c=1$.

\section{The \NK{}}
Let us give some details of the \NK{}. The all full theory has been described in Refs \cite1, \cite2, \cite3, \cite{3a}, \cite4,~\cite{4a}.
In this paper we do not repeat our considerations from previous papers. The \NK{} unifies gravity and \elm sm. Gravity is described by NGT
(\E\nos\ \E\gr al Theory, see Ref.~\cite5). In this way in the theory we have \nos\ metric
\beq2.1
g_{\a\b}=g_\(\a\b)+g_\[\a\b]
\e
and two \cn s $\gd\ov W,\a,\b,$ and $\gd \ov\o,\a,\b,$ \st
\beq2.2
\gd\ov W,\a,\b,=\gd\ov\o,\a,\b,-\frac23\,\ov W\gd\d,\a,\b,
\e
where $\ov W$ is a one form on $E$ (a \spt),
\beq2.3
\ov W=\ov W_\mu \ov\t^\mu,
\e
where $\ov\t{}^\mu$ is a frame on $E$
\beq2.4
\ov W{}_\mu=\frac12 (\gd\ov W,\si,\mu\si,-\gd\ov W,\si,\si\mu,).
\e
For \cn\ $\gd\ov\o,\a,\b,$ we suppose the \fw\ conditions
\beq2.5
\bga
\ov Dg_{\a\b} - g_{\a\d}\gd\ov Q,\d,\b\g,(\ov\G)\ov\t{}^\g=0\\
\gd\ov Q,\a,\b\a,(\ov\G)=0
\ega
\e
where $\ov D$ is an exterior \ci\ \dv\ \wrt a \cn\ $\gd\ov\o,\a,\b,$ and $\gd\ov Q,\a,\b\g,(\ov\G)$ its torsion.
\beq2.6
\gd\ov\o,\a,\b, = \gd\ov\G,\a,\b\g,\ov\t{}^\g.
\e
According to our general scheme we metrize \nos ally an \elm c fibre bundle $\ul P$ over a \spt~$E$. The metrization is of course bi\iv t \wrt an
action of a~group $\U(1)$.

We introduce two types of affine \cn s on~$\ul P$ (a~metrized \elm c \fb) and calculate a scalar \cvt\ for those affine \cn s. One gets
\beq2.7
\sqrt{-g}\,R(W) = \sqrt{-g}\,\X1( \ov R(\ov W)+(2(g^\[\m]F_\m)^2 - H^{\a\mu}F_{\a\mu})\Y1)
\e
where $H^{\a\mu}$ is a tensor on $E$ \st
\bg2.8
H^{\a\mu} = g^{\a\b}g^\m H_{\b\nu}\\
H_{\b\nu}=-H_{\nu\b} \lb2.9 \\
g_{\d\b}g^{\g\d} H_{\g\si} + g_{\a\d}g^{\d\g}H_{\b\g} = 2g_{\a\d}g^{\d\g}F_{\b\g} \lb2.10
\e

We have as usual
\beq2.11
\ov R(\ov W)= \ov R(\ov\G)+\frac23\,g^\[\m]W_\[\mu,\nu].
\e
From Palatini \v al \pp\ \wrt $\gd\ov W,\a,\b,$, $g_{\a\b}$ and $A_\mu$ one gets
\bg2.12
\ov R_{\a\b}(\ov W) - \frac12\,g_{\a\b}\ov R(\ov W) = 8\pi\nad{em}T_{\a\b},\\
\gd\falg,\[\m],{,\nu}=0, \lb2.13 \\
g_{\m,\si} - g_{\xi\nu}\gd\ov\G,\xi,\mu\si, - g_{\mu\xi}\gd\ov\G,\xi,\si\nu, = 0, \lb2.14 \\
\pa_\nu(\fal H^{\a\mu}) = 2\falg^\[\si\b] \pa_\b(g^\[\m]F_\m), \lb2.15
\e
or
\beq2.15a
\ov\n_\mu (H^{\a\mu}) = 2g^{\a\b} \ov\n_\b(g^\[\m]F_\m).
\e

We have
\bg2.16
\Tem_{\a\b} = \frac1{4\pi}\,\X3(g_{\g\b}g^{\tau\mu}g^{\ve\g}H_{\mu\a}H_{\tau\ve} - 2g^\[\m]F_\m F_{\a\b}
- \frac14\,g_{\si\b}\X2(H^\m F_\m-2(g^\[\m]F_\m)^2\Y2)\Y3), \\
\falg^\[\m] = \sqrt{-g}\,g^\[\m], \lb2.17 \\
\fal H^{\mu\a} = \sqrt{-g}\,g^{\b\mu}g^{\g\a}H_{\b\g}. \lb2.18
\e
\E\e s \er{2.12}--\er{2.15} can be rewritten in the \fw\ form:
\bg2.19
\ov R_\(\a\b)(\ov\G) = 8\pi\Tem_{(\a\b)}, \\
\ov R_\[[\a\b],\g](\ov\G) - 8\pi\Tem_{[[\a\b],\g]}=0, \lb2.20 \\
\ov\G_\mu = 0, \lb2.21 \\
g_{\m,\si} - g_{\xi\nu}\gd\ov\G,\xi,\mu\si, - g_{\mu\xi}\gd\ov\G,\xi,\si\nu, = 0, \lb2.22 \\
\pa_\mu\X2(\fal H^{\si\mu} - 2g^\[\si\mu](g^\[\nu\b]F_{\nu\b})\Y2) = 0, \lb2.23
\e
where $\ov R_{\a\b}(\ov\G)$ is a \MR\ tensor for the \cn
\beq2.24
\bga
\gd\ov\o,\a,\b, = \gd\ov\G,\a,\b\g,\ov \t{}^\g \\
\oG_{\mu} = \gd\oG,\a,[\mu\a],.
\ega
\e
For an energy-momentum tensor of an \elm c field we have three \ev t forms.
\bg2.25
\nad{em(1)}T_{\!\!\a\b} = \Tem_{\a\b} \q \text{(from Eq.~\er{2.16})} \\
\nad{em(2)}T_{\!\!\m} = \frac1{4\pi}\X2\{ g^{\a\b}H_{\b\nu}H_{\si\mu} - 2g^\[\a\b]F_{\a\b}F_\m
-\frac14\,g_\m \X1[H^{\a\b}H_{\a\b} - 2(g^\[\a\b]F_{\a\b}^2\Y1]\Y2\} - \frac1{8\pi}\,J_\m, \lb2.26 \\
J_\m = 4H_{\a\mu}H_{\b\nu}g^\[\a\b] - 4H_{\a\mu}H_{\tau\ve}g^{\tau\si}g_{\b\nu}g^\[\ve\b], \lb2.27 \\
\nad{em(3)}T_{\!\!\a\b} = \frac1{4\pi}\,\X2\{g_{\g\b}H^{\mu\si}F_{\mu\si} - 2g^\m F_\m F_{\a\b}
-\frac14\,g_{\a\b}\X1[H^\m F_\m - 2(g^\[\m]F_\m)^2\Y1]\Y2\}. \lb2.28
\e
It is easy to see that
\beq2.29
g^{\a\b}\nad{em(1)}T_{\!\!\a\b} = g^{\a\b}\nad{em(2)}T_{\!\!\a\b} = g^{\a\b}\nad{em(3)}T_{\!\!\a\b} =0.
\e

All forms of an energy-momentum tensor for an \elm c field are \ev t modulo \e~\er{2.10} and also modulo the conditions:
\bgg
g^\[\m]H_\m = g^\[\m]F_\m ,\\
g^{\a\o}g^{\b\mu}H_{\a\b}H_{\o\mu} = g^{\a\o}g^{\b\mu}H_{\a\b}F_{\o\mu}, \\
g^{\si\nu}g^{\a\mu}H_{\si\a}F_\mu + g^{\mu\si}g^{\nu\b} H_{\b\si}F_\m = 2g^{\mu\si}g^{\nu\b}F_\m F_{\b\si}
\e
derived from the condition \er{2.10}. The condition \er{2.10} can be solved (see Ref.~\cite7). Moreover, from a theoretical point of view it is more
convenient to work with the tensor $H_\m$.

\E\e\ \er{2.10} can be solved (as we mentioned before) and one gets
\beq2.30
H_{\nu\mu} = F_{\nu\mu} - \wt g{}^\(\tau\a)F_{\a\nu}g_\[\mu\tau] - \wt g{}^{\tau\a}F_{\a\mu}g_\[\nu\tau].
\e
In this way a \Lg\ for an \elm c field in the \NK{}
\beq2.31
\cL_{\rm em} = -\frac1{8\pi}\X1(H^\m F_\m -2(g^\[\m]F_\m)^2\Y1)
\e
can be rewritten in the form
\beq2.32
\cL_{\rm em} = -\frac1{8\pi}\X2((g^{\mu\a}g^{\nu\b} - g^{\nu\b}\wt g{}^\(\mu\a) + g^{\nu\b}g^{\mu\o}\wt g{}^\(\mu\a)g_{\o\tau})
F_{\a\b}F_\m - 2(g^\[\m]F_\m)^2\Y2)
\e
or
\beq2.33
\cL_{\rm em} = -\frac1{8\pi}\X2(F^\m F_\m - 2(g^\[\m]F_\m)^2 + \X1(g^{\nu\b}g^{\mu\o}\wt g{}^\(\tau\a)g_{\o\tau} - g^{\nu\b}\wt g{}^\(\mu\a)\Y1)
F_{\a\b}F_\m\Y2)
\e
where
\beq2.34
F^\m = g^{\mu\a}g^{\nu\b}F_{\a\b}.
\e

Using \er{2.30} we can reconsider an energy-momentum tensor for an \elm c field. One gets
\beq2.35
\Tem_{\a\b} = \nad oT_{\a\b} + \frac1{4\pi}\,t_{\a\b},
\e
where
\beq2.36
\nad oT_{\a\b}=\frac1{4\pi}\X2( \gd F,\tau,\a, F_{\tau\b} - \frac14\,g_{\a\b}F^\m F_\m\Y2)
\e
is an energy-momentum tensor of the \elm c field in NGT,
\beq2.37
\gd F,\tau,\a, = g^{\tau\g}F_{\g\a} = -\dg F,\a,\tau,
\e
and
\bml2.38
t_{\a\b} = g_{\g\b}\dg F,\nu,\tau, F_{\o\tau}g^{\ve\g}\wt g{}^\(\rho\nu)\wt g{}^\(\d\o) g_\[\a\rho]g_\[\ve\d]
-g_{\g\b}\wt g^\(\rho\nu)\X2(F^{\mu\g}F_{\nu\mu}g_\[\a\rho] + F_{\mu\ve}\dg F,\nu,\mu, g^{\ve\g}g_\[\a\rho]\Y2) \\
{}-2g^\[\m]F_\m F_{\a\b} + \frac14\,g_{\a\b}\X1(2(g^\[\m]F_\m)^2
-\X1(g^{\nu\d}g^{\mu\o}\wt g{}^\(\tau\ve) g_{\o\tau} - g^{\nu\d}\wt g^\(\mu\ve)\Y1)F_{\ve\d}F_\m.
\e

The second part of Maxwell \e s in the \NK{} (i.e.\ Eq.~\er{2.15} or Eq.~\er{2.15a}) can be rewritten in the \fw\ form:
\beq2.39
\ov\n_\mu F^{\a\mu} = \gd J,\a,p, + \gd J,\a,t,
\e
where $\gd J,\a,t,$ is a topological current and $\gd J,\a,p,$ is a polarization current
\bml2.40
\gd J,\a,p, = 4\pi \ov\n_\mu M^{\a\mu} = \frac{4\pi}{\sqrt{-g}}\,\pa_\mu(\sqrt{-g}\,M^{\a\mu})
=\ov\n_\mu \X1(g^{\a\b}g^{\mu\g}\wt g{}^\(\tau\rho)(F_{\rho\g}g_\[\b\tau]-F_{\rho\b}g_\[\g\tau])\Y1)\\
{}=\frac1{\sqrt{-g}}\,\pa_\mu\X1(\falg^{\a\b}g^{\mu\g}\wt g{}^\(\tau\rho)(F_{\rho\g}g_\[\b\tau] - F_{\rho\b}g_\[\g\tau])\Y1).
\e
$M^{\a\mu}$ is a polarization tensor induced by a \nos\ part of the metric
\beq2.41
H_\m =F_\m -4\pi M_\m.
\e
From Eq.~\er{2.30} one gets
\beq2.42
M_\m = \frac1{4\pi}\,\wt g{}^\(\tau\a)\X1(F_{\a\mu}g_\[\nu\tau] - F_{\a\nu} g_\[\mu\tau]\Y1).
\e
This tensor has an interpretation as a tensor of torsion in the fifth \di
\beq2.43
\gd Q,5,\m, = 8\pi M_\mu.
\e
The topological current
\beq2.44
\gd J,\a,t, = 2\ov\n_\mu\X1(g^\[\a\mu](g^\[\nu\b]F_{\nu\b})\Y1) = 2g^\[\nu\b]\ov\n_\mu(g^\[\nu\b]F_{\nu\b})
\e
is conserved by its \df. Its current density is equal to
\beq2.45
\gd\fal J,\a,t, = 2\pa_\mu\X2(\sqrt{-g}\,g^\[\a\mu](g^\[\nu\b]F_{\nu\b})\Y2) = 2\falg^\[\a\b] \pa_\b(g^\[\m]F_\m)
\e
and one gets
\beq2.46
\pa_\a \gd\fal J,\a,t,=0.
\e

The first part of Maxwell \e\ is of course a Bianchi identity (due to this a four\pt\ exists)
\beq2.47
d\O=0
\e
where
\bg2.48
\O=d\a = \frac12\,\pi^\ast (F_\m\ov\t{}^\mu \land \ov \t{}^\nu)\\
F_\m = \pa_\mu A_\nu - \pa_\nu A_\mu, \q e^\ast=A_\mu\ov\t{}^\mu, \lb2.49
\e
$\a$ is an \elm c \cn\ defined on an \elm c bundle $\ul P$, $e$~is a local section of a bundle~$\ul P$ ($e: E\supset U\to P$).
We have as usual
\beq2.50
g_{\a\b}g^{\g\b} = g_{\b\a}g^{\b\g} = \dg \d,\a,\g,
\e
and
\bg2.51
g = \det g_{\a\b} \ne 0\\
\wt g = \det g_\(\a\b)\ne0 \lb2.52 \\
\wt g^\(\a\b)g_\(\a\mu) = \gd \d,\b,\mu,, \nn
\e
$\wt g^\(\a\b)$ is an inverse tensor of $g_\(\a\b)$.

According to Refs \cite{13}, \cite{14} a condition \er{2.21}--\er{2.22} or \er{2.5} can be exactly solved. One gets
\beq2.53
\gd \ov\G,\la,\m, = \gd\toG,\la,\m, + \gd\ov Q,\la,\m, + \gd\D,\la,\m,
\e
where $\gd\toG,\la,\m,$ is a \LC \cn\ induced by $g_\(\a\b)$ on~$E$ and
\beq2.53a
\gd\ov Q,\la,\m, = \frac12\,\X1(\dg K,\g\mu,\nu, - 2\dgd g,\tl[\mu,\a,],K_{\g\tp\a\b} \cdot g_\[\nu\b]\Y1)
\e
is a torsion of an affine \cn\ $\gd\oG,\la,\m,$.
\bg2.54
\gd\D,\nu,\g\mu, = \wt g{}^\(\nu\d) \X2\{\dg K,\d(\g,\a, g_\[\mu)\a] +
\dgd g,[\rho,\b,], \X2[\dgd g,([\mu,\rho,], K_{\g)\a\b}\dgd g,[\d,\a,], - K_{\d\a\b}\dgd g,([\g,\a,], \dgd g,[\mu),\rho,],\Y2]\Y2\}\\
K_{\a\b\g} = -\ton_\a g_\[\b\g] - \ton_\a g_\[\b\g] + \ton_\g g_\[\a\b] \lb2.55
\e
where $\ton$ is a \ci\ \dv\ \wrt a \LC \cn\ $\gd\toG,\a,\b\g,$ defined on~$E$ and induced by a tensor $g_\(\a\b)$. In particular $\gd \toG,\a,\b\g,$
can be a Christoffel symbol built for $g_\(\a\b)$.

One can rewrite the remaining Einstein \e s and the second pair of Maxwell \e s in the \fw\ way.
\bg2.56
\wt{\ov R}_{\b\g} = 8\pi \Tem_{(\b\g)} + \frac34\,\ton_\d \gd\D,\d,\b\g, - \frac14\,\ton_{(\g} \gd\D,\a,\b)\a, \\
\frac14\,\ton _{\tl[\g}\gd\D,\a,\b]{|\a|,}\mu\tp, - \frac12\,\ton_\d \gd\ov Q,\d,{[\b\g,\mu]}, =
8\pi \Tem_{\tl[\b\g],\mu\tp} \lb2.57 \\
\ton_\mu F^{\a\mu} = \gd\D,\mu,\d\mu, F^{\d\a} - \gd\ov Q,\a,\d\mu, F^{\d\mu} \hskip100pt \nn \\
\hskip40pt {}+\ov\n_\mu \X1(g^{\a\b}g^{\m}\wt g{}^\(\tau\rho)(F_{\rho\g}g_\[\b\tau] - F_{\rho\b} g_\[\g\tau])\Y1)
+2g^\[\a\b]\ov\n_\mu (g^\[\nu\b]F_{\nu\b}) \lb2.58
\e
In this way we get field \e\ in a GR (General Relativity) shape.

$\wt{\ov R}_{\b\g}$ is a Ricci tensor for a \LC \cn\ defined on $E$, induced by $g_\(\a\b)$.

Moreover, in future applications we use full field \e\ \er{2.19}--\er{2.23} and with a known \so\ for $H_{\nu\mu}$ (Eq.~\er{2.30}) and
a~\so\ \er{2.53} for a \cn\ $\gd\oG,\la,\m,$ (\so\ of conditions \er{2.21}--\er{2.22} or \er{2.5}). We will work in holonomical or in
unholonomical frames.

In the case of a holonomic frame we have the following points to introduce or calculate symbolically:
\ben
\item $g_\m$
\item $g^\m$, $g^\[\m]$, $\falg^\[\m]=\sqrt{-g}\,g^\[\m]$, $\sqrt{-g}\,g^\m=\falg^\m$
\item $g=\det g_\m$
\item $g_\(\m)$, $g_\[\m]$
\item $\wt g=\det g_\(\m)$, $\wt g{}^\(\a\b)$
\item $F_\m$
\item $H_\m=F_\m - \wt g{}^\(\tau\a)F_{\a\nu}g_\[\mu\tau] + \wt g{}^\(\tau\a)F_{\a\mu}g_\[\nu\tau]$
\item $H^\m = g^{\b\nu}g^{\g\nu}H_{\b\g}$, $\fal H^\m = \sqrt{-g}H^\m$
\item $8\pi\Tem_{\a\b} = 2\X2[g_{\g\b}g^{\tau\mu}g^{\ve\g}H_{\mu\a}H_{\tau\ve} - 2g^\[\m]F_\m F_{\a\b} - \frac14\,g_{\a\b}\X1(H^\m F_\m
-2(g^\[\m]F_\m)^2\Y1)\Y2]$
\item $\gd \toG,\a,\b\g, = \frac12\,\wt g{}^\(\a\d)\X2(\pap{g_\(\d\b)}{x^\g} + \pap{g_\(\d\g)}{x^\b} - \pap{g_\(\b\g)}{x^\d}\Y2)$
\item $\ton_\a g_\[\b\g] = \pa_\a g_\[\b\g] - \toG{}^\d_{\b\a}g_\[\d\g] - \toG{}^\d_{\b\a} g_\[\b\d]$
\item $K_{\a\b\g} = -\ton_\a g_\[\b\g] - \ton_\b g_\[\g\a] + \ton_\g g_\[\a\b]$
\item $\dg K,\g\mu,\nu, = K_{\g\mu\a}\wt g{}^\(\a\nu)$
\item $\dgd g,[\rho,\b,], = \wt g{}^\(\a\b) g_\[\rho\a]$
\item $\gd \ov Q,\nu,\g\mu, = \frac12 \, \X1(\dg K,\g\mu,\nu, - 2\dgd g,\tl[\mu,\a,], K_{\g\tp\a\b}\wt g{}^\(\nu\rho) \wt g^\(\b\d)g_\[\rho\d]\Y1)$
\item $\gd\D,\nu,\g\mu, = \wt g{}^\(\nu\d) \X2\{\dg K,\d(\g,\a, g_{[\mu)\a]} + \dgd g,[\rho,\b,], \X1[ \dgd g,([\mu,\rho,], K_{\g)\a\b}
\dgd g,[\d,\a,],- K_{\d\a\b}\dgd g,([\g,\a,], \dgd g,[\mu),\rho,],\Y1]\Y2\}$
\item $\gd \ov\G,\nu,\g\mu, = \gd\wt{\ov \G},\nu,\g\mu, +\gd\ov Q,\nu,\g\mu, + \gd\D,\nu,\g\mu,$
\item $\gd\ov R,\a,\b\g\d, = 2\gd\ov\G,\a,\b{[\d,\g]}, +2\gd\ov\G,\a,\rho[\g, \gd\ov\G,\rho,|\b|\d],$
\item $\ov R_\m = \gd\ov R,\a,\m\a, + \frac12\,\gd\ov R,\a,\a\m,$ (a \MR\ tensor)
\item $\ov R_\(\m)$
\item $\ov R_\[\m]$
\item $8\pi \Tem_{(\a\b)}$
\item $8\pi \Tem_{[\a\b]}$
\item $\ov R_\(\a\b) = 8\pi \Tem_{(\a\b)}$
\item $\ov R_{\tl[\m],\g\tp} = 8\pi \Tem_{\tl[\m],\g\tp}$
\item $2\falg^\[\a\b] \pa_\b(g^\[\m] F_\m)$
\item $\pa_\mu(\fal H^{\a\mu})$
\item $\pa_\mu(\fal H^{\a\mu}) = 2\falg^\[\a\b] \pa_\b(g^\[\m]F_\m)$
\item $\gd\falg,[\m],{,\nu},=0$.
\een

In the case of unholonomic frames we have some differences concerning points 10) and 18). In this case $_{,\nu}$ means an action of a vector field
dual to the unholonomic frame. Section~3 gives us an example of this situation.

Let us introduce external (material) sources in the \NK{} as in Refs \cite1, \cite3. Let us notice that one introduces material (external) sources
in the \KK Theory (\KK Theory with torsion) in Ref.~\cite m (see also Refs \cite{aa}, \cite{ab}, \cite{ac}).

In order to do this we write down a material sources \Lg\ in the \fw\ way:
\beq2.61
L_m = -8\pi g^{\m}\fal T^\m + \falj^\mu A_\mu.
\e
One gets
\bg2.62
\fal T^\m=-\frac1{8\pi}\,\frac{\d L_m}{\d g^\m} \\
\falj^\mu = \frac{\d L_m}{\d A_\mu} \lb2.63
\e
where
\beq2.64
\fal T^\m = \sqrt{-g}\,T^\m, \q \falj^\mu = \sqrt{-g}\,j^\mu.
\e
$T^\m$ is an energy momentum tensor of material sources, $j^\mu$~an electric current and $A_\mu$ a four-\pt.

Usually we suppose a conservation of an electric charge
\beq2.65
\pa_\mu \falj^\mu=0.
\e
One can consider also $T^\m$ as a sum of a hydrodynamical tensor and an \ia\ term
\bg2.66
T^\m = u^\mu h^\nu - pg^\m + \nad{\rm int}T{}^\m \\
\nad{\rm int}T{}^\m = g^\m \falj^\a A_\a \lb2.67
\e
where $u^\mu$ is the four-velocity of a fluid, $p$~its pressure, $h^\nu$~a~four-vector of an enthalpy. In this way in Eqs \er{2.12}, \er{2.19},
\er{2.20} we should pass from $\Tem_\m$ to $\nad{\rm eff}T_{\!\m}$, where
\beq2.68
\nad{\rm eff}T_{\!\m}=\Tem_\m + T_\m .
\e
Eq.\ \er{2.23} should be rewritten as
\beq2.69
\pa_\mu(\fal H^{\si\mu}-2g^\[\si\mu](g^{\nu\b}F_{\nu\b})) = \falj^\si.
\e
In some cases we can consider only
\beq2.70
\nad{\rm eff}T_{\!\m}=\Tem_\m + \nad{\rm int}T_\m .
\e
neglecting the hydrodynamic part. An electric current can be written also as
\beq2.71
j^\a = q\rho u^\a
\e
where $q$ is an electric charge and $\rho$ the density of a fluid.

External sources in \KK Theory (any type---\nos, \s ic, with torsion etc.) are in some sense against a unification approach, where everything has been
geometrized. Moreover, those external sources can be obtained from more unified theory (also \KK or Jordan--Thiry type) where 5-\di al theory (an
\elm c theory unified with gravity) is only a~part of a full theory which is \nA\ \KK theory with Higgs' fields and \ssb. This theory geometrized
all physical \ia s (a~bosonic part of the physical world) and only as external sources remain fermion which are coupled to the multi\di al geometry
via a~minimal coupling scheme. It means they are coupled by new type of gauge \dv s. All fermions are unified in one multi\di al spinor.
All physical \ia s are described by only one affine \cn\ defined on the \KK \JT manifold. In this setting a \co ical
constant which we consider in the 5-\di al case is also a~part of external sources which can be derived from more extended theory.

Let us remind to the reader what we mean by a \gm\ in the Einstein programme. In this programme all physical quantities concerning \fn\ physical
\ia\ should get geometrical interpretation. They should be geometrical quantities as \cn s, \cvt s, metric tensors, torsions, nonmetricity tensors
etc. The best way to achieve it
is to construct a~\Lg\ of the theory as a scalar \cvt\ of some affine \cn\ defined on a many\di al manifold. In this way classical \e s of fields
obtained from \v al \pp\ are vacuum \e s. On the right-hand side we have only zero, on the left-hand side geometrical quantities only. If this \Lg\
contains all physical \ia s unified in this one affine \cn, we get simultaneously a~\un\ of those \ia s. Moreover, 5-\di al theory which describes
only a \un\ of a gravity and an electrodynamics, even it is given by a \Lg\ which is a scalar \cvt\ of an affine \cn\ is not  a full story. Thus
external sources are not breaking the idea of a full \gm, even they are not of geometrical nature and in the field \e s they are on the right hand
side of field \e s.

They can be derived from more unified theory (in higher \di s, more than five \di s). 5-\di al theory in our \gm\ and \un\ is pure geometrical
because $\Tem_\m$ (an energy-momentum tensor) is derived from geometry and the \Lg\ of the theory is a scalar \cvt. We have the same in more
advanced theory i.e.\ a~bosonic part of GSW-model (see Ref.~\cite8).

\section{\E\as\ and stationary fields\\ in \NK{}}
Let us consider \as\ and \sy\ fields in the \NK{}. First of all we write down a \s ic part of the metric (see Refs \cite{15}--\cite{27})
\beq3.1
ds^2 = f^{-1}\X1(e^{2\g}(d\rho^2+dz^2) + \rho^2\,d\vf^2 \Y1) - f(dt -\o\,d\vf)^2
\e
where $\rho$ and $z$, $\vf$ are usual cylindrical \cd s. A~signature is here ${}+{+}+{-}$. We introduce a new nonholonomical frame
\beq3.2
\bal
\ov\t{}^4 &= dt - \o\,d\vf\\
\ov\t{}^2 &= \frac1{\sqrt f}\,e^\g\,d\rho \\
\ov\t{}^1 &= \frac1{\sqrt f}\,e^\g\,dz\\
\ov\t{}^3 &= d\vf.
\eal
\e
(\f s are \f s of $\rho$ and $z$ only). In this new frame a metric \er{3.1} has a form
\beq3.3
ds^2 = \ov\t{}^1 \otimes \ov\t{}^1 + \ov\t{}^2 \otimes \ov\t{}^2 + f^{-1}\rho^2 \ov\t{}^3 \otimes \ov\t{}^3 -f \ov\t{}^4 \otimes \ov\t{}^4.
\e
Let us introduce new \f s $r_1$ and $r_2$ \st
\beq3.4
\bal
r_1^2 &=f^{-1}\rho^2\\
r_2^2 &=f
\eal
\e
and $r_1^2r_2^2=\rho^2$. Simultaneously we change for a convenience the signature to ${}-{-}-{+}$. One gets
\beq3.5
ds^2 = -\ov\t{}^1 \otimes \ov\t{}^1 -\ov\t{}^2 \otimes \ov\t{}^2 -r_1^2\ov\t{}^3 \otimes \ov\t{}^3 + r_2^2\ov\t{}^4 \otimes \ov\t{}^4.
\e

In this way a \s ic part of a metric has the \fw\ shape
\bg3.6
g_\(\m) = \left(\begin{array}{c|c|c|c}
\ \ -1 & 0 & 0 & 0 \\\hline 0 &\ \ -1\ \ & 0 & 0 \\\hline 0 & 0 &\ -r_1^2\ & 0 \\\hline 0 & 0 & 0 &\ r_2^2 \
\end{array}\right);
\\
g^\(\m) = \left(\begin{array}{c|c|c|c}
\ \ -1 \ \ & 0 & 0 & 0 \\\hline 0 & \ \ -1 \ \  & 0 & 0 \\\hline 0 & 0 &\ -\frac1{r_1^2}\ & 0 \\\hline 0 & 0 & 0 & \ \frac1{r_2^2}\
\end{array}\right)
\e
is an inverse tensor for $g_\(\m)$.

Now we rename variables \st $1-x$, $2-y$ and of course $r_1^2r_2^2=y^2$. One gets
\bg3.7
\ov\t{}^1 = f^{-1/2}e^\g\,dx = e^n\,dx, \quad \ov\t{}^2 = f^{-1/2}e^\g\,dy = e^n\,dy,\\
e^{2n} = f^{-1}e^{2\g}. \lb3.8
\e
The frame considered by us is nonholonomical and we get
\bea3.9
d\ov\t{}^1 &= -e^{-n} \,\pap ny\, \ov\t{}^1 \land \ov\t{}^2 \\
d\ov\t{}^2 &= -e^{-n} \,\pap nx\, \ov\t{}^1 \land \ov\t{}^2 \lb3.10 \\
d\ov\t{}^3 &= 0 \nn \\
d\ov\t{}^4 &= -e^n \,\pap \o x\, \ov\t{}^1 \land \ov\t{}^3 - e^n \,\pap\o y\, \ov\t{}^2\land \ov\t{}^3.
\e

We can recapitulate it in  a one formula
\bea3.12
d\ov\t{}^\mu &= \frac12\, \gd C,\mu,\a\b, \ov\t{}^\a \land \ov\t{}^\b \\
\gd C,1,12, &= -\gd C,1,21, = -2e^{-n}\,\pap ny \lb3.13 \\
\gd C,2,12, &= -\gd C,2,21, = 2e^{n} \, \pap nx \lb3.14 \\
\gd C,4,13, &= -\gd C,4,31, = -2\, \pap\o x\,e^{-n} \lb3.15 \\
\gd C,4,23, &= -\gd C,4,32, = -2\,\pap \o y\,e^{-n}. \lb3.16
\e
The remaining \cf s are zero.

\allowdisplaybreaks
Now we calculate a \LC \cn\ \cf s induced by \er{3.6} in our \nh\ frames from the formulas
\bg3.17
0 = \wt{\ov D} g_\m = dg_\m - \gd\toG,\a,\mu\b, g_{\a\nu}\ov\t{}^\b - \gd\toG,\a,\nu\b, g_{\mu\a}\ov\t{}^\b \\
0= \wt{\ov D}\ov\t{}^\mu. \lb3.17a
\e
One gets
\bea3.18
\gd\toG, 3, 31, &=\gd\toG,3,13, = \frac{r_{1,1}}{r_1} = e^{-n} \,\pap{r_1}x\,\frac1{r_1} \nn \\
\gd\toG, 3, 32, &=\gd\toG,3,23, = \frac{r_{1,2}}{r_1} = e^{-n}\,\pap{r_1}y \,\frac1{r_1} \nn \\
\gd\toG, 1, 34, &=\gd\toG,1,43, = \frac12\, r_2^2\gd C,4,13, = -\frac12\,e^{-n}r_2^2\,\pap \o x \nn \\
\gd\toG, 2, 34, &=\gd\toG,2,43, = \frac12\, r_2^2\gd C,4,23, = -\frac12\,e^{-n}r_2^2\,\pap \o y \nn \\
\gd\toG, 4, 13, &=-\gd\toG,4,31, = \frac12\,\gd C,4,13, = -\frac12\,e^{-n}\,\pap \o x \nn \\
\gd\toG, 4, 23, &=-\gd\toG,4,32, = \frac12\,\gd C,4,23, = -\frac12\,e^{-n}\,\pap\o y \nn \\
\gd\toG, 1, 33, &=-r_1r_{1,1} = -e^{-n}r_1\,\pap{r_1}x  \nn \\
\gd\toG, 2, 33, &=-r_1r_{1,2} = -e^{-n}r_1\,\pap {r_1}y \nn \\
\gd\toG, 1, 44, &=r_2r_{2,1} = e^{-n}r_2\,\pap{r_2}x \nn \\
\gd\toG, 2, 44, &=r_2r_{2,2} = e^{-n}r_2\,\pap{r_2}y \nn \\
\gd\toG, 3, 41, &=\gd\toG,3,14, = -\frac{r_2^2}{2r_1^2}\,\gd C,4,13, = \frac{r_2^2}{2r_1^2}\,e^{-n}\,\pap \o x \nn \\
\gd\toG, 3, 42, &=\gd\toG,3,24, = -\frac{r_2}{2r_1^2}\,\gd C,4,23,= \frac{r_2^2}{2r_1^2}\,e^{-n}\,\pap\o y \nn \\
\gd\toG, 4, 41, &=\gd\toG,4,14, = \frac{r_{2,1}}{r_2} = \frac1{r_2}\,e^{-n}\,\pap{r_2}x \nn \\
\gd\toG, 4, 42, &=\gd\toG,4,24, = \frac{r_{2,2}}{r_2} = \frac1{r_2}\,e^{-n}\,\pap{r_2}y \nn \\
\gd\toG, 1, 21, &=e^{-n}\,\pap ny \nn \\
\gd\toG, 2, 11, &=-e^{-n}\,\pap ny \nn \\
\gd\toG, 2, 12, &=e^n\,\pap nx \nn \\
\gd\toG, 1, 22, &=-e^{-n}\,\pap nx. \nn \\
\e

Let us introduce a skew-\s ic part of the metric in the mentioned \nh\ frame
\beq3.19
g_\[\m] = \left(\begin{array}{c|c|c|c}
0 & 0 & 0 & \ b\sqrt f\ \\\hline 0 & 0 & 0 &\ 0 \\\hline 0 & 0 & 0 & 0 \\\hline \ -b\sqrt f\ &\ 0\ &\ 0\ &0
\end{array}\right)
\e
In this way our \nos\ metric (\as\ and \sy) takes a form
\beq3.20
g_\m= \left(\begin{array}{c|c|c|c}
-1 & 0 & 0 &\ b\sqrt f\ \\\hline 0 & -1 & 0 & 0 \\\hline 0 & 0 &\ -r_1^2\ & 0 \\\hline \ -b\sqrt f\ & \ 0\ & 0 & r_2^2
\end{array}\right).
\e
$b$ is a \f\ of $z$ and $\rho$ (or~$r$). In our contemporary notation it is a \f\ of $x$ and~$y$ only.

If we define $g_\[\m]=K_\m$ we can define a very important object
\beq3.21
K_{\o\m} = \ton_\o K_{\nu\mu} + \ton_\mu K_{\o\nu} + \ton_\nu K_{\o\mu}
\e
and we get
\beq3.22
\bal
K_{214}&=-\ton_2K_{14}\\
K_{141}&=2\ton_1 K_{14}\\
K_{241}&=-K_{214} = \ton K_{14}.
\eal
\e
The remaining $K_{\o\m}$ equal zero.

$K_{\o\m}$ are important in order to construct a \cn\ $\gd\ov\G,\mu,\a\b,$ compatible with a full \nos\ metric $g_\m$.

In order to calculate a \cvt\ tensor for the \cn~$\gd\toG,\mu,\a\b,$ (we should remember that we are working in an unholonomic frame) we write down
\bg3.23
\gd\ov\O{},\a,\b, = \gd d\ov \o{},\a,\b, + \gd\o{},\a,\g,\land \gd\o,\g,b, \\
\gd\ov\o,\a,\b, = \gd\ov\G,\a,\b\mu, \ov\t{}^\mu. \lb3.24
\e
One gets
\beq3.25
\gd\ov\O{},\a,\b, = -\gd\ov\G,\a,{\b[\rho,\si]},\ov\t{}^\rho \land \ov\t + \frac12\,\gd\ov\G,\a,\b\g,\gd C,\g,\rho\si,\ov\t{}^\rho\land \ov\t{}^\si
+ \gd\ov\G,\a,\g[\rho, \gd\ov\G,\g,|\b|\si], \ov\t{}^\rho \land \ov\t{}\si.
\e
For
\beq3.26
\gd\ov\O,\a,\b, = \frac12\,\gd \ov R,\a,\b\rho\si, \ov\t{}^\rho \land \ov\t{}^\si
\e
one gets
\bml3.27
\gd\ov R,\a,\b\rho\si, = -2\ov\G_{\b[\rho\si]} + \gd\ov\G,\a,\b\g,\gd C,\g,\rho\si, + 2\gd\ov\G,\a,\g[\rho, \gd\ov\G,\g,|\b|\si],\\
{}= -\gd\ov\G,\a,{\b\rho,\si}, + \gd\ov\G,\a,{\b\si,\rho}, + \gd\ov\G,\a,\g\rho,\gd\ov\G,\g,\b\si, - \gd\ov\G,\a,\g\si,\gd\ov\G,\g,\b\rho,
+ \gd\ov\G,\a,\b\g, \gd C,\g,\b\si, ,
\e
$_{,\g}$ means an action of a vector field dual to $\ov\t{}^\g$. We suppose the \fw\ shape of $F_\m$ tensor
\beq3.28
F_\m = \left(\begin{array}{c|c|c|c}
0 & 0 & 0 & \ e^{-m}f \pap{a_3}x\ \\\hline 0 & 0 & 0 & \ e^{-m}f \pap{a_3}y\ \\\hline 0 & 0 &\ \ 0\ \ & 0 \\\hline
\ -e^{-m}f\pap{a_3}x \ & \ -e^{-m}f\pap{a_3}y \ & 0 & 0
\end{array}\right)
\e
We have the \fw\ relation:
$$
n=m-\tfrac12\log f.
$$

In Appendix A we calculate using symbolic manipulations in \ti{Mathematica} all important quantities in our theory.

Let us write down field \e s for \as, \sy\ field configurations in the \NK{}. All independent \e s are written down below. We use of course our
programme to calculate $\ov R_\(\a\b)$, $\ov R_\[\a\b]$, $\Tem_{(\a\b)}$, $\Tem_{[\a\b]}$, $W^\m$. We get some zeros $\Tem_{(13)}=\Tem_{(34)}
=\Tem_{[13]}=\Tem_{[23]}=\Tem_{[34]}=W^{13}=W^{23}=W^{34}=0$.
$$
\ov R_{11}=8\pi\Tem_{11}
$$
or
\bml A.1
\frac1{8 \rho^2 f^2}\, \X4(4 \rho^2 b f^5
\X2(2 \pap bz \pap mz + b^3 \X2(\pap mz\Y2)^2 + b \pap{^2m}{z^2}\Y2) +
e^{2 m} \rho^2 b^2 \X3((1 + b^2) \X2(\pap fz\Y2)^2 + 2 b^2 \X2(\pap f\rho\Y2) ^2\Y3)\\
{}+e^m \rho^2 b f\X3(b (1 + 2 b^2) (4 + 2 b^2 + b^4) \X2(\pap fz\Y2)^2 +
4 b^5 \X2(\pap f\rho\Y2)^2 \\
{}+ 2 e^m \X3(2 (1 + b^2) \pap bz \pap fz +
b^2 \pap f\rho  \X2(4 \pap b\rho  + b \pap f\rho \Y2)\Y3)\Y3) \\
{}+2 f^6 \X3(-2 b^2 \X2(\pap \o z\Y2)^2 + (-2 - b^2 + 6 b^4) \X2(\pap\o \rho\Y2)^2\Y3) +
\rho^2 f^2 \X3(4 e^{2 m} (1 + b^2) \X2(\pap bz\Y2)^2 \\
{}+2 e^m b^3 (-2 + 5 b^2 + 2 b^4) \pap bz \pap fz +
2 \X1(4 + b^2 (2 + b^2)^2\Y1) \X2(\pap fz\Y2)^2 \\
{}+2 (6 + 5 b^2 - 2 b^4 + b^8) \X2(\pap f\rho\Y2)^ 2 +
8 e^{2 m} b^2 \pap b\rho  \X2(\pap b\rho  + b \pap f\rho \Y2) \\
{}+ e^m b \X3(-16 \pap b\rho  \pap f\rho  + b \X2((2 + 2 b^2 + b^4)
\X2(\pap fz\Y2)^2 - 2 (1 + 4 b^2 + 2 b^4) \pap fz \pap mz -
2 (1 + 2 b^2) \pap fz  \\
{}+ 8 b^3 \pap b\rho  \pap f\rho  +
2 (2 + 2 b^2 + b^4) \X2(\pap f\rho\Y2)^2 - 4 \pap {^2f}{\rho^2}\Y2)\Y3)\Y3) \\
{}+ 2 \rho f^3 \X3(4 e^{2 m} \rho b^2 \X2(\pap b\rho\Y2)^ 2 - e^m
\X3(2 (\rho + 6 \rho b^2) \X2(\pap bz\Y2)^2 + \rho b \pap bz
\X2(-(-2 + 2 b^2 + b^4) \pap fz \\
{}+ 2 (1 + 4 b^2 + 2 b^4)
\pap mz\Y2) + 4 \rho \X2(\pap b\rho\Y2)^2 - 2 \rho b^5 \pap b\rho  \pap f\rho  +
4 \rho b^3 \X2(\pap {^2b}{z^2} - \pap b\rho  \pap f\rho \Y2) \\
{}+ b^4 \X2(\rho \pap fz \pap mz + 2 \pap f\rho  \X2(1 + \rho \pap m\rho \Y2)\Y2) +
b^2 \X2(\rho \pap fz \pap mz +
\rho \pap {^2f}{z^2} + 2 \pap f\rho  \X2(1 + \rho \pap m\rho \Y2)\Y2) \\
{}+ 2 \rho b \X2(\pap {^2b}{z^2} - 2 \pap b\rho  \pap f\rho  + 2 \pap{^2b}{z^2}\Y2)\Y3) +
\rho \X2(b \pap fz \X2(-4 \pap bz - b (4 + 6 b^2 + 5 b^4 + 2 b^6) \pap mz\Y2)\\
{} - 2 (2 + b^2) \pap {^2f}{z^2} -
4 b (3 + 2 b^2) \pap b\rho  \pap f\rho  -
b^2 (2 + 3 b^2 + b^4) \X2(\pap f\rho\Y2)^2 + 2 b^2 (1 + b^2)
\pap f\rho  \pap m\rho  - 2 (2 + 3 b^2 + b^4) \pap {^2f}{\rho^2}\Y2)\Y3)\\
{}+ 2 \rho f^4 \X3(-2 e^m \rho \X2(\pap bz\Y2)^2 - 2 \rho b
\X1(-2 + e^m + (-8 + e^m) b^2\Y1) \pap bz \pap mz +
\rho b^6 \pap mz \X2(-\pap fz + 4 \pap mz\Y2) \\
{}- 4 \pap m\rho  - 4 e^m b^3 \pap b\rho  \X2(1 + \rho \pap m\rho \Y2) + 2 b
\X2(-e^m \rho \pap{^2b}{z^2} - 2 (-2 + e^m) \pap b\rho  \X2(1 + \rho \pap m\rho \Y2)\Y2) \\
{}+2 b^4 \X3(\pap f\rho  + \rho \X2(-\pap fz \pap mz +
\X2(\pap mz\Y2)^2 + 2 \pap{^2m}{z^2} + \pap f\rho  \pap m\rho \Y2)\Y3) +
4 \rho \X2(\pap{^2m}{z^2} + \pap{^2m}{\rho^2}\Y2) \\
{}+ 2 b^2 \X2(\pap f\rho  - 2 \pap m\rho  + \rho \X2(\pap mz \X2(\pap fz + \pap mz\Y2) +
\pap{^2m}{z^2} + \pap f\rho  \pap m\rho  + 2 \pap{^2m}{\rho^2}\Y2)\Y2)\Y3)\Y4) \\
{}=\frac1{(-1+b^2)^2}\,f \X3((1+2b^2-5b^4+2b^6)\X2(\pap{a_3}z\Y2)^2 + (-1+3b^2)\X2(\pap{a_3}\rho\Y2)^2\Y3).
\e

\goodbreak
$$
\ov R_\(12)=8\pi \Tem_{(12)}
$$
or
\bml{A.2}
\frac1{16 \rho^2 f^2}\,
\X4(4 \rho^2 b^8 f^2 \pap{f}{z} \pap{f}{\rho} + 2 e^m \rho b^5
f^2 \X3(4 \rho \pap{f}{z} \pap{b}{\rho} + \pap{b}{z} \X2(-9 \rho \pap{f}{\rho} +
f \X2(4 + \rho \pap{f}{\rho} + 6 \rho \pap{m}{\rho}\Y2)\Y2)\Y3) \\
{}+ 2 \rho b^3 f \X3(e^{2 m} \rho (-1 + f) \X2(\pap{f}{z} \pap{b}{\rho} + \pap{b}{z} \pap{f}{\rho}\Y2) -
8 \rho f^2 \X2(\pap{f}{z} \pap{b}{\rho} + \pap{b}{z} \pap{f}{\rho}\Y2) \\
{}+ e^m f \X2(4 \rho \pap{f}{z} \pap{b}{\rho} - 7 \rho \pap{b}{z} \pap{f}{\rho} -
2 f^2 \X2(-\rho \pap{m}{z} \pap{b}{\rho} + \pap{b}{z} \X2(1 + 2 \rho\, \pap{m}{\rho}\Y2)\Y2) +
f \X2(2 \rho \X2(-\pap{f}{z} + \pap{m}{z}\Y2) \pap{b}{\rho} \\
{}+ \pap{b}{z} \X2(6 + 5 \rho \pap{f}{\rho} + 2 \rho \pap{m}{\rho}\Y2)\Y2)\Y2)\Y3) +
8 f^2 \X2(\rho f \pap{f}{z} - 2 \rho f^2 \pap{m}{z} -
e^{2 m} \rho^2 \pap{b}{z} \pap{b}{\rho} +
e^m \rho^2 (-1 + f) f \pap{b}{z} \pap{b}{\rho} \\
{}+ \rho^2 \pap{f}{z} \pap{f}{\rho} - f^4 \pap{\o}{z} \pap{\o}{\rho}\Y2) +
b^6 f \X2(2 \rho f^2 \X2(-\rho \pap{f}{z} \pap{f}{\rho} +
\pap{m}{z} \X2(-4 f + \rho (9 + f) \pap{f}{\rho}\Y2) +
2 \rho \X2(\pap{f}{z} - 3 f \pap{m}{z}\Y2) \pap{m}{\rho}\Y2) \\
{}+ e^m \rho \pap{f}{z} \X2(-5 \rho \pap{f}{\rho} + f \X2(4 + \rho \pap{f}{\rho} + 6 \rho \pap{m}{\rho}\Y2)\Y2) -
2 f^5 \pap{\o}{z} \pap{\o}{\rho}\Y2) + 4 \rho b f
\X3(-e^{2 m} \rho \X2(\pap{f}{z} \pap{b}{\rho} + \pap{b}{z} \pap{f}{\rho}\Y2) \\
{}+ 2 f^2 \X2(2 f \pap{b}{z} - 2 \rho \X2(\pap{f}{z} - f \pap{m}{z}\Y2)
\pap{b}{\rho} - \rho (1 + f) \pap{b}{z} \pap{f}{\rho}\Y2) -
e^m f \X2(2 \rho \pap{f}{z} \pap{b}{\rho} + 3 \rho \pap{b}{z} \pap{f}{\rho} +
f \X2(-\pap{b}{z} \X2(1 + 6 \rho \pap{f}{\rho}\Y2) \\
{}+ 2 \rho \pap{^2b}{\rho \pa z}\Y2) + f^2 \X2(\pap{b}{z} \X2(1 + 4 \rho \pap{m}{\rho}\Y2) -
2 \rho \X2(\pap{m}{z} \pap{b}{\rho} + \pap{^2b}{\rho \pa z}\Y2)\Y2)\Y2)\Y3) +
b^4 \X3(-e^{2 m} \rho^2 \pap{f}{z} \pap{f}{\rho} \\
{}+ e^m (-3 + e^m) \rho^2 f \pap{f}{z} \pap{f}{\rho} - 4 \rho f^5\, \pap{m}{z} \X2(1 + 2 \rho \pap{m}{\rho}\Y2) +
2 \rho f^4 \X2(\pap{m}{z} \X2(-2 + 5 \rho \pap{f}{\rho} - 2 \rho \pap{m}{\rho}\Y2) \\
{}+ \pap{f}{z} \X2(2 + 4 \rho \pap{m}{\rho}\Y2)\Y2) +\rho f^2 \X2(2 e^m \rho \pap{m}{z} \pap{f}{\rho}
+ \pap{f}{z} \X2((-14 + 3 e^m) \rho \pap{f}{\rho} + 2 e^m \X2(3 + \rho \pap{m}{\rho}\Y2)\Y2)\Y2) \\
{}+ 12 f^6 \pap{\o}{z} \pap{\o}{\rho} - 2 \rho f^3
\X2(\pap{f}{z} \X2(2 + e^m + 5 \rho \pap{f}{\rho} + 2 (-3 + e^m) \rho \pap{m}{\rho}\Y2) -
\rho \X2((5 + e^m) \pap{m}{z} \pap{f}{\rho} - 4 \pap{^2f}{\rho \pa z}\Y2)\Y2)\Y3) \\
{}+ 2 b^2 \X3(e^{2 m} \rho^2 \X2(2 (-1 + f) f^2 \pap{b}{z} \pap{b}{\rho} -
\pap{f}{z} \pap{f}{\rho}\Y2) - e^m \rho f \X2(\rho \pap{f}{z} \pap{f}{\rho} +
f \X2(-\pap{f}{z} \X2(1 + 2 \rho \pap{f}{\rho}\Y2) + 2 \rho \pap{^2f}{\rho \pa z}\Y2) \\
{}+ f^2 \X2(\pap{f}{z} \X2(1 + 4 \rho \pap{m}{\rho}\Y2) -
2 \rho \X2(\pap{m}{z} \pap{f}{\rho} + \pap{^2f}{\rho \pa z}\Y2)\Y2)\Y2) +
2 f^3 \X2(-\rho f^2 \pap{m}{z} + \rho \pap{f}{z} \X2(2 + f -
2 \rho \pap{f}{\rho} + 2 \rho (1 + f) \pap{m}{\rho}\Y2)\\
{} + f^3 \pap{\o}{z} \pap{\o}{\rho} + 3 \rho^2 \X2(\pap{m}{z} \pap{f}{\rho} - \pap{^2f}{\rho \pa z}\Y2) - \rho f
\X2(\pap{m}{z} \X2(3 + 2 \rho \pap{m}{\rho}\Y2) + \rho \X2(\pap{^2f}{\rho \pa z} - 2 \pap{^2m}{\rho \pa z}\Y2)\Y2)\Y2)\Y3)\Y4)\\
{}=\frac{2f\pap{a_3}{z}\pap{a_3}{\rho}}{-1+b^2}.
\e

\goodbreak
$$
\ov R_\(13)=0
$$
or
\bml A.3
\frac1{16 \rho^2}\, \sqrt f \X3(8 \rho^2 f^2 \pap{b}{z} \pap{\o}{z} -
4 \rho^2 b^7 f \pap{f}{\rho} \pap{\o}{\rho} - 2 e^m b^4 f
\X2((2 \rho^2 + f) \pap{b}{z} \pap{\o}{z} + 4 \rho^2 \pap{b}{\rho} \pap{\o}{\rho}\Y2) \\
{}+2 b^2 f \X2(e^m (-\rho^2 + (-1 + \rho^2) f) \pap{b}{z} \pap{\o}{z} +
4 \rho^2 (e^m - 3 f) \pap{b}{\rho} \pap{\o}{\rho}\Y2) +
4 \rho b f \X2(\rho \X2(5 \pap{f}{z} - 2 f \pap{m}{z}\Y2) \pap{\o}{z} \\
{}+ 2 \rho f \pap{^2\o}{z^2} + 2 f \X2(-1 + \rho \pap{m}{\rho}\Y2) \pap{\o}{\rho}\Y2) +
b^5 \X2(-\X1(4 \rho^2 f + e^m (2 \rho^2 + f)\Y1) \pap{f}{z} \pap{\o}{z} -
4 e^m \rho^2 \pap{f}{\rho} \pap{\o}{\rho} \\
{}+ 2 f \X2(f (2 \rho^2 + f)
\pap{m}{z} \pap{\o}{z} - 3 \rho \X2(-2 f + \rho \pap{f}{\rho}\Y2) \pap{\o}{\rho}\Y2)\Y2) +
b^3 \X3(\X2(e^m (-\rho^2 + (-1 + \rho^2) f) \\
{}- 2 f (3 \rho^2 + (1 + \rho^2) f)\Y2)
\pap{f}{z} \pap{\o}{z} + 4 e^m \rho^2 \pap{f}{\rho} \pap{\o}{\rho} +
2 f \X2((1 + \rho^2) f^2 \pap{m}{z} \pap{\o}{z} -
16 \rho^2 \pap{f}{\rho} \pap{\o}{\rho} \\
{}+ \rho f \X2(3 \rho \pap{m}{z} \pap{\o}{z} +
2 \X2(7 + 4 \rho \pap{m}{\rho}\Y2) \pap{\o}{\rho} - 4 \rho \pap{^2\o}{\rho^2}\Y2)\Y2)\Y3)\Y3)=0
\e

\goodbreak
$$
\ov R_\(14)=8\pi\Tem_{(14)}
$$
or
\bml A.4
\frac1{32 \rho^4 f^{3/2}}\,
\X4(8 \rho^4 b^7 f^2 \X2(\pap{f}{\rho}\Y2)^2 + 16 e^m \rho^4 b^4 f^2
\X2(\pap{b}{z} \pap{f}{z} + 2 \pap{b}{\rho} \pap{f}{\rho}\Y2) \\
{}+ 4 \rho^4 b^2 f \X2(e^m \pap{b}{z} \X2(\X2(2 + f - 2 f^2 + e^m (1 + f)\Y2) \pap{f}{z} +
2 f (-1 + 2 f^2) \pap{m}{z}\Y2) \\
{}+ 2 \X2(4 e^{2 m} + 2 e^m f + 3 f^2\Y2) \pap{b}{\rho} \pap{f}{\rho}\Y2) +
8 \rho^2 b^5 f \X3(\rho^2 (e^m + 2 f) \X2(\pap{f}{z}\Y2)^2 -
4 \rho^2 f^2 \pap{f}{z} \pap{m}{z}\\
{} + \rho^2 (2 e^m + f) \X2(\pap{f}{\rho}\Y2)^2 +
2 f^5 \X2(\pap{\o}{\rho}\Y2)^2\Y3) + 16 \rho^4 f^3
\X2(\pap{b}{z} \X2(-\pap{f}{z} + f \pap{m}{z}\Y2) + e^m \pap{^2b}{\rho^2}\Y2) \\
{}+ 4 \rho^2 b f \X3(e^{2 m} \rho^2 f \X3((1 + f) \X2(\pap{b}{z}\Y2)^2 + 8 \X2(\pap{b}{\rho}\Y2)^2\Y3) -
2 f \X3(-\rho^2 \X2(\pap{f}{z}\Y2)^2 - 3 \rho^2 f \pap{f}{z} \pap{m}{z}\\
{} +2 \X3(\rho^2 f^2 \X2(\X2(\pap{m}{z}\Y2)^2 - \pap{^2m}{z^2}\Y2) - \rho^2 \X2(\pap{f}{\rho}\Y2)2 +
\rho f \X2(\rho \pap{^2f}{z^2} + \pap{f}{\rho} \X2(1 + \rho \pap{m}{\rho}\Y2)\Y2) +
f^4 \X2(\X2(\pap{\o}{z}\Y2)^2 + \X2(\pap{\o}{\rho}\Y2)^2\Y2)\Y3)\Y3)\\
{} + e^m \rho^2 \X2(-3 \X2(\pap{f}{\rho}\Y2)^2 + 2 f \pap{^2f}{\rho^2}\Y2)\Y3) +
b^3 \X3(\rho^4 \X2(e^m (4 + e^m) + f
\X2(8 + e^m (2 + e^m) + 4 f (2 - e^m + f)\Y2)\Y2) \X2(\pap{f}{z}\Y2)^2 \\
{}- 4 \rho^4 f \X1(e^m + f (4 + f (5 - 2 e^m + 2 f))\Y1)
\pap{f}{z} \pap{m}{z} + 4 \X3(3 \rho^4 f^4 \X2(\pap{m}{z}\Y2)^2 - \rho^4 f^5 \X2(\pap{m}{z}\Y2)^2 \\
{}+ 2 e^{2 m} \rho^4 \X2(\pap{f}{\rho}\Y2)^2 + 2 e^m \rho^4 f \X2(\pap{f}{\rho}\Y2)^2 +
5 \rho^4 f^2 \X2(\pap{f}{\rho}\Y2)^2 + \rho^2 f6 \X2(\X2(\pap{\o}{z}\Y2)^2 + 2 \X2(\pap{\o}{\rho}\Y2)^2\Y2) \\
{}+ f^7 \X2(\X2(\pap{\o}{z}\Y2)^2 + 4 \X2(\pap{\o}{\rho}\Y2)^2\Y2) + 2 \rho3 f^3
\X2(\rho \X2(\pap{m}{z}\Y2)^2 - \pap{f}{\rho} \X2(2 + 3 \rho \pap{m}{\rho}\Y2) + \rho \pap{^2f}{\rho^2}\Y2)\Y3)\Y3)\Y4)\\
{}=-2bf^{3/2}\X2(\pap{a_3}{z}\Y2)^2.
\e

\goodbreak
$$
\ov R_{22}=8\pi\Tem_{22}
$$
or
\bml A.5
\frac1{16 \rho^2 f^2}\, \X4(e^{2 m} \rho^2 b^2 (-2 - 3 b^2 + b^4 + 2 b^6) \X2(\pap{f}{z}\Y2)^2 +
e^m \rho^2 b f \pap{f}{z} \X2(4 e^m (-2 - 3 b^2 + b^4 + 2 b^6) \pap{b}{z} \\
{}+ b \X1(4 + b^2 (10 + 4 b^4 - e^m (1 + b^2))\Y1) \pap{f}{z}\Y2) +
4 (-2 + b^2) f^6 \X2(\pap{\o}{z}\Y2)^2 + 4 \rho^2 b f^5
\X2(4 \pap{b}{z} \pap{m}{z} + b^3 (-1 + b^2) \X2(\pap{m}{z}\Y2)^2 \\
{}+ 2 b \pap{^2m}{z^2}\Y2) + 2 \rho^2 f^2
\X3(2 e^{2 m} (-2 - 3 b^2 + b^4 + 2 b^6) \X2(\pap{b}{z}\Y2)^2 - 2 e^m b
\X1(2 + b^2 \X1(7 - 2 b^4 + e^m (1 + b^2)\Y1)\Y1) \pap{b}{z} \pap{f}{z} \\
{}+ 2 (6 - 3 b^2 - 2 b^4 + b^6 + b^8) \X2(\pap{f}{z}\Y2)^2 -
4 (-2 + b^2) \X2(\pap{f}{\rho}\Y2)^2 - e^m b^2
\X2(-(-2 + 2 b^2 + b^4) \X2(\pap{f}{z}\Y2)^2 \\
{}+ 2 (1 - b^4 + 2 b^6) \pap{f}{z} \pap{m}{z} + (2 + 4 b^2) \pap{^2f}{z^2} + 4 \X2(\pap{f}{\rho}\Y2)^2\Y2)\Y3) -
4 \rho f^3 \X3(e^m \rho \X1(2 + b^2 (12 + e^m (1 + b^2))\Y1) \X2(\pap{b}{z}\Y2)^2 \\
{}- 2 \rho b^6 \pap{f}{z} \pap{m}{z} + 2 \rho b^8 \pap{f}{z} \pap{m}{z} +
\rho b \pap{b}{z} \X2(\X1(4 + 8 b^2 - e^m (2 + 2 b^2 + b^4)\Y1)
\pap{f}{z} + 2 e^m (1 - b^4 + 2 b^6) \pap{m}{z}\Y2) \\
{}+ 4 e^m \rho b^3 \pap{^2b}{z^2} + \rho b^4
\X2(\pap{f}{z}2 + (-5 + e^m) \pap{f}{z} \pap{m}{z} + 2 \pap{^2f}{z^2}\Y2) +
2 e^m \rho b \X2(\pap{^2b}{z^2} + 2 \pap{b}{\rho} \pap{f}{\rho}\Y2) \\
{}- b^2 \X2((6 + e^m) \rho \pap{f}{z} \pap{m}{z} + (-2 + e^m) \rho \pap{^2f}{z^2} +
2 \pap{f}{\rho} \X2(-2 + \rho \X2(\pap{f}{\rho} + (1 + e^m) \pap{m}{\rho}\Y2)\Y2)\Y2) \\
{}+4 \X2(\pap{f}{\rho} + \rho \X2(\pap{^2f}{z^2} + \pap{^2f}{\rho^2}\Y2)\Y2)\Y3) +
4 \rho f^4 \X3(2 e^m \rho \X2(\pap{b}{z}\Y2)^2 + 2 \rho b^8 \X2(\pap{m}{z}\Y2)^2 -
\rho b^6 \pap{m}{z} \X2(\pap{f}{z} + 3 \pap{m}{z}\Y2) \\
{}- 2 \rho b \pap{b}{z} \X2(2 \pap{f}{z} + \X1(-2 - e^m + (-8 + e^m) b^2\Y1) \pap{m}{z}\Y2) +
\rho b^4 \X2(2 \pap{f}{z} \pap{m}{z} - \X2(\pap{m}{z}\Y2)^2 + 4 \pap{^2m}{z^2}\Y2) \\
{}+ 2 e^m \rho b \X2(\pap{^2b}{z^2} + 2 \pap{b}{\rho} \pap{m}{\rho}\Y2) + 2 b^2
\X2(\rho \X2(\X2(\pap{f}{z} - \pap{m}{z}\Y2) \pap{m}{z} - \pap{^2f}{z^2} + \pap{^2m}{z^2}\Y2) +
\X2(2 - \rho \pap{f}{\rho}\Y2) \pap{m}{\rho}\Y2)\\
{} + 4 \X2(\pap{m}{\rho} + \rho \X2(\pap{^2m}{z^2} + \pap{^2m}{\rho^2}\Y2)\Y2)\Y3)\Y4)
=\frac{f\X1((1-4b^2+3b^4)(\pap{a_3}{z})^2 - (1+b^2)(\pap{a_3}{\rho})^2\Y1)}{(-1+b^2)^2}\,.
\e

\goodbreak
$$
\ov R_\(23)=0
$$
or
\bml A.6
\X4(-8 \rho^2 f^2 \pap{b}{z} \pap{\o}{\rho} + 2 e^m b^4 f
(7 \rho^2 + (2 + \rho^2) f) \pap{b}{z} \pap{\o}{\rho} \\
{}- 2 b^2 f \X1(12 \rho^2 f + e^m (-5 \rho^2 + (-2 + \rho^2) f)\Y1) \pap{b}{z} \pap{\o}{\rho} +
b^5 \X2((-2 \rho^2 f (1 + f) + e^m (7 \rho^2 + (2 + \rho^2) f)\Y1) \pap{f}{z} \\
{}+ 2 f^2 (\rho^2 + (-2 + \rho^2) f) \pap{m}{z}\Y2) \pap{\o}{\rho} +
4 \rho b f \X3(-\rho \X2(2 \pap{\o}{z} \pap{f}{\rho} + 3 \pap{f}{z} \pap{\o}{\rho}\Y2) +
2 f \X2(2 \pap{\o}{z} \X2(-1 + \rho \pap{m}{\rho}\Y2) - \rho \pap{^2\o}{\rho\pa z}\Y2)\Y3) \\
{}+ b^3 \X3(-e^m (-5 \rho^2 + (-2 + \rho^2) f) \pap{f}{z} \pap{\o}{\rho} -
2 f \X3(\rho^2 \X2(\pap{\o}{z} \pap{f}{\rho} + 9 \pap{f}{z} \pap{\o}{\rho}\Y2) +
f^2 \X2(2 \pap{\o}{z} \pap{m}{\rho} + (-2 + \rho^2) \pap{m}{z} \pap{\o}{\rho}\Y2) \\
{}- f \X2(2 \pap{\o}{z} \X2(\rho + \pap{f}{\rho}\Y2) + \X2((-2 + \rho^2) \pap{f}{z} +
3 \rho^2 \pap{m}{z}\Y2) \pap{\o}{\rho} - 4 \rho^2 \pap{^2\o}{\rho\pa z}\Y2)\Y3)\Y3)\Y4)=0
\e

\goodbreak
$$
\ov R_\(24) = 8\pi\Tem_{(24)}
$$
or
\bml A.7
\frac1{16 \rho^4 f^{3/2}}\, \X4(4 e^m \rho^4 b^6 f^2 \pap{b}{z} \pap{f}{\rho} + 2 \rho^4 b^7
f \X2((e^m + 2 f) \pap{f}{z} - 2 f^2 \pap{m}{z}\Y2) \pap{f}{\rho} \\
{} + 2 e^m \rho^4 b^4 f \X3(2 (e^m + 2 f) \pap{f}{z} \pap{b}{\rho} +
2 e^m \pap{b}{z} \pap{f}{\rho} - f \X3(9 \pap{b}{z} \pap{f}{\rho} + f \X2(4 \pap{m}{z} \pap{b}{\rho} +
\pap{b}{z} \X2(\pap{f}{\rho} - 10 \pap{m}{\rho}\Y2)\Y2)\Y3)\Y3)\\
{}+ 2 \rho^4 b^2 f \X3(6 f^2 \pap{b}{z} \pap{f}{\rho} -
e^{2 m} (-1 + f) \X2(\pap{f}{z} \pap{b}{\rho} + \pap{b}{z} \pap{f}{\rho}\Y2) +
e^m \X2(-4 \pap{b}{z} \pap{f}{\rho} + f\X2(2 (4 + f) \pap{f}{z} \pap{b}{\rho} \\
{}- 5 \pap{b}{z} \pap{f}{\rho} +
f \X2(-6 (1 + f) \pap{m}{z} \pap{b}{\rho} + \pap{b}{z}
\pap{f}{\rho}\Y2) + 2 (2 + 3 f) \pap{b}{z} \pap{m}{\rho}\Y2)\Y2)\Y3) \\
{}+ \rho^2 b^5 \X3(\rho^2 \pap{f}{z} \X2(\X2(2 e^{2 m} + 2 (-2 + f) f^2 -
e^m f (5 + f)\Y2) \pap{f}{\rho} + 10 f^2 (e^m + 2 f) \pap{m}{\rho}\Y2) \\
{}+ 2 f^2 \X2(-\rho^2 \pap{m}{z} \X2((2 e^m + (-5 + f) f) \pap{f}{\rho} + 10 f^2
\pap{m}{\rho}\Y2) + f^4 \pap{\o}{z} \pap{\o}{\rho}\Y2)\Y3) \\
{} + 4 \rho^4 f^2 \X3(2 f \X2(-f \pap{m}{z} \pap{b}{\rho} + \pap{b}{z} \pap{f}{\rho}\Y2) +
e^m \X2(-\pap{f}{z} \pap{b}{\rho} + \pap{b}{z} \pap{f}{\rho} + 2 f \pap{^2b}{\rho \pa z}\Y2)\Y3) \\
{} + b^3 \X3(e^{2 m} \rho^4 \X2(8 f^2 \pap{b}{z} \pap{b}{\rho} -
(-1 + f) \pap{f}{z} \pap{f}{\rho}\Y2) + e^m \rho^4 \X2(-6 f^2 (1 + f) \pap{m}{z} \pap{f}{\rho} \\
{}+ \pap{f}{z} \X2((-4 + 3 f (1 + f)) \pap{f}{\rho} + 2 f (2 + 3 f) \pap{m}{\rho}\Y2)\Y2) +
2 f \X2(-4 \rho^4 \pap{f}{z} \pap{f}{\rho} + \rho^3 f^3 \pap{m}{z}
\X2(-4 + 3 \rho \pap{f}{\rho} - 6 \rho \pap{m}{\rho}\Y2) \\
{}+ \rho^4 f \X2(-\X2(\pap{f}{z} - 4 \pap{m}{z}\Y2) \pap{f}{\rho} + 4 \pap{f}{z} \pap{m}{\rho}\Y2) +
2 \rho^2 f^5 \pap{\o}{z} \pap{\o}{\rho} + 2 f^6 \pap{\o}{z} \pap{\o}{\rho} +
\rho^4 f^2 \X2(-\X2(\pap{f}{z} - 7 \pap{m}{z}\Y2) \pap{f}{\rho} \\
{} -4 \pap{m}{z} \pap{m}{\rho} + 2 \pap{^2f}{\rho \pa z}\Y2)\Y2)\Y3) +
2 \rho^3 b f \X3(-2 e^{2 m} \rho (-1 + f) f \pap{b}{z} \pap{b}{\rho} +
e^m \rho \X2(-3 \pap{f}{z} \pap{f}{\rho} + 2 f \pap{^2f}{\rho \pa z}\Y2)\\
{}+ 2 f \X2(\rho \pap{f}{z} \pap{f}{\rho} - \rho f
\X2(\pap{m}{z} \pap{f}{\rho} + 4 \pap{f}{z} \pap{m}{\rho} - 2 \pap{^2f}{\rho \pa z}\Y2) +
2 f^2 \X2(\pap{m}{z} \X2(-1 + \rho \pap{m}{\rho}\Y2) - \rho \pap{^2m}{\rho \pa z}\Y2)\Y2)\Y3)\Y4)\\
{}=-\frac{2bf^{3/2}\pap{a_3}{z} \pap{a_3}{\rho}}{-1+b^2}\,.
\e

\goodbreak
$$
\ov R_{33} = 8\pi\Tem_{33}
$$
or
\bml A.8
\frac1{8 \rho^2 f^2}\, \X4(4 e^m \rho^4 b^3 f \pap{b}{z} \pap{f}{z} - 2 e^m \rho^3 b
f \X2(\rho (-1 + f) \pap{b}{z} \pap{f}{z} + 2 f \pap{b}{\rho}
\X2(-2 f + \rho \pap{f}{\rho}\Y2)\Y2) \\
{}+ 4 \rho^2 b^8 f^5 \X2(\pap{\o}{\rho}\Y2)^2 +
4 b^6 f^5 (-2 \rho^2 + f) \X2(\pap{\o}{\rho}\Y2)^2 +
2 b^4 \X2(\rho^4 (e^m + f) \X2(\pap{f}{z}\Y2)^2 -
2 \rho^4 f^2 \pap{f}{z} \pap{m}{z} - 2 f^6 \X2(\pap{\o}{\rho}\Y2)^2\Y2) \\
{}+ b^2 \X3(-\rho^4 (e^m (-1 + f) - 2 f (1 + f)) \X2(\pap{f}{z}\Y2)^2 -
2 \rho^4 f^2 (1 + f) \pap{f}{z} \pap{m}{z} +
2 f \X3(-e^m \rho^4 \X2(\pap{f}{\rho}\Y2)^2 \\
{}- 2 \rho^3 f^2 \X2(\pap{f}{\rho} - 2 \pap{m}{\rho}\Y2) +
\rho^3 f \pap{f}{\rho} \X2(2 e^m + \rho \pap{f}{\rho} - 2 \rho \pap{m}{\rho}\Y2) +
2 \rho^2 f^4 \X2(\pap{\o}{\rho}\Y2)^2 + f^5 \X2(\X2(\pap{\o}{z}\Y2)^2 + 2 \X2(\pap{\o}{\rho}\Y2)^2\Y2)\Y3)\Y3) \\
{}- 4 \rho^2 f \X3(-\rho^2 \X2(\pap{f}{z}\Y2)^2 - \rho^2 \X2(\pap{f}{\rho}\Y2)^2 + f^4 \X2(\X2(\pap{\o}{z}\Y2)^2 +
\X2(\pap{\o}{\rho}\Y2)^2\Y2) + \rho f \X2(\pap{f}{\rho} + \rho \X2(\pap{^2f}{z^2} + \pap{^2f}{\rho^2}\Y2)\Y2)\Y3)\Y4)\\
{}=\frac{\rho^2(1+b^2)\X1((-1+b^2)(\pap{a_3}{z})^2 -(\pap{a_3}{\rho})^2\Y1)^2}{(-1+b^2)^2}
\e

\goodbreak
$$
\ov R_\(34)=0
$$
or
\bml A.9
\X4(-8 \rho^3 b^8 f^2 \pap{f}{\rho} \pap{\o}{\rho} + 4 \rho b f
\X2(\X1(-2 f^3 + e^m \rho^2 (-1 + (-1 + f) f)\Y1) \pap{b}{z} \pap{\o}{z} \\
{}+2 ((1 + e^m) \rho^2 - 2 f) f^2 \pap{b}{\rho} \pap{\o}{\rho}\Y2) -
2 e^m \rho b^5 f^2 \X2(-f \pap{b}{z} \pap{\o}{z} +
4 \rho^2 (1 + f) \pap{b}{\rho} \pap{\o}{\rho}\Y2)\\
{} + 2 \rho b^3 f^2 \X2(-e^m (6 \rho^2 + f^2) \pap{b}{z} \pap{\o}{z} + 2 \X1(-8 \rho^2 f +
e^m (3 \rho^2 + (\rho^2 - 2 f) f)\Y1) \pap{b}{\rho} \pap{\o}{\rho}\Y2) \\
{}+ \rho b^6 f \X3(-2 f^3 \pap{m}{z} \pap{\o}{z} +
2 \rho f \X2(11 \rho \pap{f}{\rho} + 2 f \X2(-1 + \rho \pap{f}{\rho} - 2 \rho \pap{m}{\rho}\Y2)\Y2)
\pap{\o}{\rho} + e^m \X2(f \pap{f}{z} \pap{\o}{z} -
4 \rho^2 (1 + f) \pap{f}{\rho} \pap{\o}{\rho}\Y2)\Y3) \\
{}+ 8 \rho^2 f^2 \X3(2 \rho \pap{f}{z} \pap{\o}{z} + 2 \rho \pap{f}{\rho} \pap{\o}{\rho} +
f \X2(-\pap{\o}{\rho} + \rho \X2(\pap{^2\o}{z^2} + \pap{^2\o}{\rho^2}\Y2)\Y2)\Y3) \\
{}+ b^4 f \X3(-\rho (8 \rho^2 f - 2 f^3 + e^m (6 \rho^2 + f^2)) \pap{f}{z} \pap{\o}{z} +
2 e^m \rho (3 \rho^2 + (\rho^2 - 2 f) f) \pap{f}{\rho} \pap{\o}{\rho} \\
{}- 2 f \X3(\rho f^3 \pap{m}{z} \pap{\o}{z} + 7 \rho^3 \pap{f}{\rho} \pap{\o}{\rho} +
2 f^2 \X2(2 - \rho \pap{f}{\rho} + 2 \rho \pap{m}{\rho}\Y2) \pap{\o}{\rho} \\
{}+ f \X2(-6 \rho^3 \pap{m}{z} \pap{\o}{z} + \rho \X2((-6 + \rho^2) \pap{f}{\rho} -
2 \rho^2 \pap{m}{\rho}\Y2) \pap{\o}{\rho} + 4 \rho^3 \pap{^2\o}{\rho^2}\Y2)\Y3)\Y3)\\
{} +2 b^2 \X4(e^m \rho^3 \X2((-1 + (-1 + f) f) \pap{f}{z} \pap{\o}{z} +
2 f^2 \pap{f}{\rho} \pap{\o}{\rho}\Y2) -
2 f \X3(\rho \X1(\rho^2 + f (\rho^2 + (2 + \rho^2) f)\Y1) \pap{f}{z} \pap{\o}{z}\\
{} +f \X3(-\rho^3 (1 + f + f^2) \pap{m}{z} \pap{\o}{z} +
\rho^3 \pap{f}{\rho} \pap{\o}{\rho} + f^2 \X2(\rho \pap{^2\o}{z^2} - 4 \pap{\o}{\rho} +
2 \rho \pap{^2\o}{\rho^2}\Y2) \\
{}+ \rho f \X2(\X2((4 + \rho^2) \pap{f}{\rho} -
\rho \X2(1 + 2 \rho \pap{m}{\rho}\Y2)\Y2) \pap{\o}{\rho} - \rho^2 \pap{^2\o}{\rho^2}\Y2)\Y3)\Y3)\Y4)\Y4)=0
\e

\goodbreak
$$
\ov R_{44} = 8\pi\Tem_{44}
$$
or
\bml A.10
\frac1{16 \rho^4 f^2}\, \X4(8 e^m \rho^4 b^7 f^3 \pap{b}{z} \pap{f}{z} + 4 e^m \rho^4 b^5
f^2 \X2((1 + 3 f + 2 e^m (1 + f)) \pap{b}{z} \pap{f}{z}\\
{}+ 2 f (3 + 2 f) \pap{b}{\rho} \pap{f}{\rho}\Y2) +
4 \rho^4 b^8 f^2 \X2((e^m + 2 f) \X2(\pap{f}{z}\Y2)^2 -
2 f^2 \pap{f}{z} \pap{m}{z} + 2 f \X2(\pap{f}{\rho}\Y2)^2\Y2) \\
{}+ 2 \rho^3 b^6 f \X3(\rho \X1(e^m (1 + e^m) + (1 + e^m) (2 + e^m) f - 4 f^3\Y1)
\X2(\pap{f}{z}\Y2)^2 + 2 \rho f^2 (-1 + f + 4 f^2) \pap{f}{z} \pap{m}{z} \\
{}+ 2 f \X3(e^m \rho (3 + 2 f) \X2(\pap{f}{\rho}\Y2)^2 - 2 f
\X2(\rho f^2 \X2(\pap{m}{z}\Y2)^2 + \rho f^3 \X2(\pap{m}{z}\Y2)^2 + 4 \rho \X2(\pap{f}{\rho}\Y2)^2 +
f \pap{f}{\rho} \X2(-2 + \rho \pap{f}{\rho} - 2 \rho \pap{m}{\rho}\Y2)\Y2)\Y3)\Y3) \\
{}+ 4 \rho^3 b^3 f \X3(\rho \pap{b}{z} \X2(\X2(16 f^3 - e^m f
(-1 + (-2 + f) f) + e^{2 m} (1 + f^2)\Y2) \pap{f}{z} + 2 (3 e^m - 8 f) f^3 \pap{m}{z}\Y2) \\
{}+ 4 f \X3(4 \rho f^2 \pap{b}{\rho} \pap{f}{\rho} + e^{2 m} \rho (1 + f) \pap{b}{\rho} \pap{f}{\rho} +
e^m f \X3(-3 \rho \pap{b}{\rho} \pap{f}{\rho} + f \X2(\rho \pap{^2b}{z^2} +
\pap{b}{\rho} \X2(1 - \rho \pap{f}{\rho} + \rho \pap{m}{\rho}\Y2)\Y2)\Y3)\Y3)\Y3) \\
{}+ 4 \rho^3 b f^2 \X3(\rho \pap{b}{z} \X2(-(2 e^{2 m} - 4 f + e^m (2 + f + f^2))
\pap{f}{z} + 2 (e^m - 2 f) f (1 + f) \pap{m}{z}\Y2) - 4 e^{2 m} \rho \pap{b}{\rho} \pap{f}{\rho} \\
{}- 4 \rho f^2 \pap{b}{\rho} \pap{f}{\rho} + 2 e^m f \X3(\rho (1 + f) \pap{^2b}{z^2} + \pap{b}{\rho}
\X2(-2 \rho \pap{f}{\rho} + f \X2(2 - \rho \pap{f}{\rho} + 2 \rho \pap{m}{\rho}\Y2)\Y2) +
2 \rho f \pap{^2b}{\rho^2}\Y3)\Y3)\\
{} + 8 \rho^2 f^3 \X2(\rho^2 \X2(\pap{f}{z}\Y2)^2 - f^4 \X2(\pap{\o}{z}\Y2)^2 - \rho^2 f \pap{^2f}{z^2} -
e^{2 m} \rho^2 \X2(\X2(\pap{b}{z}\Y2)^2 + 2 \X2(\pap{b}{\rho}\Y2)^2\Y2)\\
{} + e^m \rho^2 \X2((1 + f) \X2(\pap{b}{z}\Y2)^2 + 2 f \X2(\pap{b}{\rho}\Y2)^2\Y2) -
\rho f \pap{f}{\rho} + \rho^2 \X2(\pap{f}{\rho}\Y2)^2 -
f^4 \X2(\pap{\o}{\rho}\Y2)^2 - \rho^2 f \pap{^2f}{\rho^2}\Y2) \\
{}+ \rho^2 b^4 \X4(e^{2 m} \rho^2 \X3(\X2(\pap{f}{z}\Y2)^2 + f \X3(f \X2(8 f (1 + f)
\X2(\pap{b}{z}\Y2)^2 + \X2(\pap{f}{z}\Y2)^2\Y2) + 4 (1 + f) \pae{f}{\rho}2\Y3)\Y3) \\
{}- 2 e^m \rho f \X3(\rho (-1 + f (10 + f)) \pae{f}{z}2 -
6 \rho f^2 \pap{f}{z} \pap{m}{z} - 4 f \X3(-3 \rho \pae{f}{\rho}2 \\
{}+ f \X2(\rho \pap{^2f}{z^2} + \pap{f}{\rho} \X2(1 - \rho \pap{f}{\rho} + \rho \pap{m}{\rho}\Y2)\Y2)\Y3)\Y3) \\
{}+ 4 f^3 \X3(-6 \rho^2 \pae{f}{z}2 + \rho^2 (1 + f)^2 \pap{f}{z} \pap{m}{z} -
\rho^2 f^3 \pae{m}{z}2 - 2 \rho^2 f^2 \X2(\pae{m}{z}2 + 2 \pap{^2m}{z^2}\Y2) \\
{}- 3 \rho^2 \pae{f}{\rho}2 + 2 f^4 \pae{\o}{\rho}2 +
\rho f \X2(-\rho \pae{m}{z}2 + 4 \rho \pap{^2f}{z^2} +
\pap{f}{\rho} \X2(2 + \rho \pap{f}{\rho} - 2 \rho \pap{m}{\rho}\Y2) + 4 \rho \pap{^2f}{\rho^2}\Y2)\Y3)\Y4) \\
{}+ 2 b^2 f \X3(e^{2 m} \rho^4 \X2(2 (f + f^3) \pae{b}{z}2 - \pae{f}{z}2 + 8 f^2 (1 + f) \pae{b}{\rho}2 - 2 \pae{f}{\rho}2\Y2) \\
{} + e^m \rho^3 \X3(24 \rho f^3 \pae{b}{z}2 - 6 \rho \pae{f}{z}2 + \rho f \X2(-5 \pae{f}{z}2 +
2 \pap{f}{z} \pap{m}{z} + 2 \X2(\pap{^2f}{z^2} - 6 \pae{f}{\rho}2\Y2)\Y2) \\
{}+ f^2 \X2(-\rho \pae{f}{z}2 + 2 \rho \pap{f}{z} \pap{m}{z} + 2 \rho \pap{^2f}{z^2} +
2 \pap{f}{\rho} \X2(2 - \rho \pap{f}{\rho} + 2 \rho \pap{m}{\rho}\Y2) + 4 \rho \pap{^2f}{\rho^2}\Y2)\Y3) \\
{}+ 2 f \X3(\rho^4 (-4 + f^2) \pae{f}{z}2 + \rho^4 f (2 + f - f^2) \pap{f}{z} \pap{m}{z} +
f \X2(-2 \rho^4 f^2 \X2(\pae{m}{z}2 + \pap{^2m}{z^2}\Y2)\\
{}+ \rho^4 \X2(2 \pap{^2f}{z^2} + 5 \pae{f}{\rho}2\Y2) + \rho^2 f^4 \X2(\pae{\o}{z}2 - \pae{\o}{\rho}2\Y2) + f^5
\X2(\pae{\o}{z}2 + 2 \pae{\o}{\rho}2\Y2) \\
{}- \rho^3 f \X2(2 \rho \pap{^2m}{z^2} + \pap f\rho
\X2(4 - \rho \pap{f}{\rho} + 2 \rho \pap{m}{\rho}\Y2) + 2 \rho \pap{^2f}{\rho^2}\Y2)\Y2)\Y3)\Y3)\Y4)\\
{}=-\frac{(-1+3b^2)f^2\X1((-1+b^2)(\pap{a_3}{z})^2 - (\pap{a_3}{\rho})^2 \Y1)}{(-1+b^2)^2}\,.
\e

\goodbreak
$$
\ov R_\[34] = {\rm const.}
$$
or
\bml A.11
\frac1{16 \rho^3} \, e^{-2 m} b f
\X4(-2 \rho b^3 f^4 \pap{m}{z} \pap{\o}{z} + 2 e^m \rho^3 b
\X2((1 + b^2) \pap{f}{z} \pap{\o}{z} + (2 + b^2) \pap{f}{\rho} \pap{\o}{\rho}\Y2) \\
{} + \rho f \X3(2 \rho^2 \X2(2 b^3 \pap{f}{z} \pap{\o}{z} +
b (-2 + 5 b^2 + b^4) \pap{f}{\rho} \pap{\o}{\rho}\Y2) +
e^m \X2(\X2(4 \rho^2 (1 + b^2) \pap{b}{z} + b^5 \pap{f}{z}\Y2) \pap{\o}{z} \\
{}+ 2 \rho^2 \X2(2 (2 + b^2) \pap{b}{\rho} - b^3 \pap{f}{\rho}\Y2) \pap{\o}{\rho}\Y2)\Y3) +
2 f^3 \X3(-\rho (4 + e^m b^2) \pap{b}{z} \pap{\o}{z} -
\rho b^5 \pap{m}{z} \pap{\o}{z} - 8 \rho \pap{b}{\rho} \pap{\o}{\rho} \\
{}-4 e^m \rho b^2 \pap{b}{\rho} \pap{\o}{\rho} + b^3 \X2(\rho \pap{f}{z}
\pap{\o}{z} + 2 \X2(-2 + \rho \pap{f}{\rho} - 2 \rho \pap{m}{\rho}\Y2) \pap{\o}{\rho}\Y2) -
2 b \X2(\rho \pap{^2\o}{z^2} - 4 \pap{\o}{\rho} + 2 \rho \pap{^2\o}{\rho^2}\Y2)\Y3) \\
{}+ \rho f^2 \X3(-e^m b^2 \X2(b \X2(-2 b \pap{b}{z} + \pap{f}{z}\Y2)
\pap{\o}{z} + 4 \X2(\rho^2 \pap{b}{\rho} + b \pap{f}{\rho}\Y2) \pap{\o}{\rho}\Y2) -
2 \X3(2 \rho b^5 \pap{\o}{\rho} + 4 \rho^2 \pap{b}{\rho} \pap{\o}{\rho} \\
{}+ b^3 \X2(2 \rho^2 \pap{m}{z} \pap{\o}{z} - \X2((6 + \rho^2) \pap{f}{\rho} - 2 \rho^2 \pap{m}{\rho}\Y2)\pap{\o}{\rho}\Y2) \\
{}+ 2 b \X2(2 \pap{f}{z} \pap{\o}{z} + \rho^2 \pap{m}{z}
\pap{\o}{z} - \X2(\rho - 4 \pap{f}{\rho}\Y2) \pap{\o}{\rho} + \rho^2 \pap{^2\o}{\rho^2}\Y2)\Y3)\Y3)\Y4)
=c={\rm const.}
\e

Eq.\ \er{A.11} $(\ov R_\[34]={\rm const})$ is coming from the \e s $\pap{}\rho (\ov R_\[34]-8\pi\Tem_{[34]})=\pap{}z(\ov R_\[43]-8\pi\Tem_{[34]})=0$
and $\Tem_{[34]}=0$.

\goodbreak
$$
\pap{}\rho \ov R_\[23] - \pap{}z \ov R_\[13]=0
$$
or
\bml A.12
\pap {}\rho\,\X4(\frac1{16 \rho^2}\,e^{-2 m} \sqrt f \X4(16 e^m \rho^2 f \pap{\o}{z} \pap{b}{\rho} + 8 e^m \rho^2 b
\pap{\o}{z} \pap{f}{\rho} - 16 e^m \rho^2 b^6 f \pap{b}{z} \pap{\o}{\rho} \\
{}+ 2 e^m b^4 f (-5 \rho^2 + (-2 + \rho^2) f) \pap{b}{z} \pap{\o}{\rho} -
2 b^2 f \X1(-24 \rho^2 f + e^m (-\rho^2 + (2 + \rho^2) f)\Y1) \pap{b}{z} \pap{\o}{\rho} \\
{}- 8 \rho^2 b^7 \X2((e^m + f) \pap{f}{z} - 2 f^2 \pap{m}{z}\Y2) \pap{\o}{\rho} +
b^5 \X2(\X1(-2 \rho^2 (-1 + f) f + e^m (-5 \rho^2 + (-2 + \rho^2) f)\Y1) \pap{f}{z} \\
{}+ 2 f^2 (-3 \rho^2 + (2 + \rho^2) f) \pap{m}{z}\Y2) \pap{\o}{\rho} +
b^3 \X3(-e^m (-\rho^2 + (2 + \rho^2) f) \pap{f}{z} \pap{\o}{\rho} +
2 f \X3(\rho^2 \X2(3 \pap{\o}{z} \pap{f}{\rho} + 11 \pap{f}{z} \pap{\o}{\rho}\Y2) \\
{}+ f^2 \X2(2 \pap{\o}{z} \pap{m}{\rho} - (2 + \rho^2) \pap{m}{z} \pap{\o}{\rho}\Y2) +
f \X2(2 \pap{\o}{z} \X2(\rho - \pap{f}{\rho}\Y2) + \X2((2 + \rho^2) \pap{f}{z} -
9 \rho^2 \pap{m}{z}\Y2) \pap{\o}{\rho} + 8 \rho^2 \pap{^2\o}{\rho z}\Y2)\Y3)\Y3)\Y4)\Y4)\\
{}-\pap{}z \X4(\frac1{16 \rho^2}\,e^{-2m} \sqrt f \X4(-2 b^3 (1 - \rho^2 + b^2) f^3 \pap{m}{z} \pap{\o}{z} -
e^m \rho^2 b \X2((4 + b^2) \pap{f}{z} \pap{\o}{z} +
4 (-1 - b^2 + b^4) \pap{f}{\rho} \pap{\o}{\rho}\Y2) \\
{}+ f \X3(e^m \X2(\X2(-2 \rho^2 (4 + b^2) \pap{b}{z} + b^3 (1 + \rho^2 + b^2) \pap{f}{z}\Y2)
\pap{\o}{z} + 8 \rho^2 (1 + b^2 - b^4) \pap{b}{\rho} \pap{\o}{\rho}\Y2) \\
{}+ 2 \rho^2 b^3 \X2(-\pap{f}{z} \pap{\o}{z} + (12 + 5 b^2 - 2 b^4) \pap{f}{\rho} \pap{\o}{\rho}\Y2)\Y3) +
2 b^2 f^2 \X3(e^m (1 + \rho^2 + b^2) \pap{b}{z} \pap{\o}{z}\\
{} + 12 \rho^2 \pap{b}{\rho} \pap{\o}{\rho}
+ 2 \rho b^3 \X2(1 - 2 \rho \pap{m}{\rho}\Y2) \pap{\o}{\rho} + b \X3(-(-1 + \rho^2) \pap{f}{z} \pap{\o}{z} \\
{}+ \rho \X2(3 \rho \pap{m}{z} \pap{\o}{z} +
\X2(2 - 8 \rho \pap{m}{\rho}\Y2) \pap{\o}{\rho} + 4 \rho \pap{^2\o}{\rho^2}\Y2)\Y3)\Y3)\Y4)\Y4)=0
\e

\goodbreak
$$
\pap{}z\X1(8\pi\Tem_{[14]} - \ov R_\[14] \Y1) + \pap{}\rho \X1(\ov R_\[24] - 8\pi\Tem_{[24]}\Y1) = 0
$$
or
\bml A.13
\pap{}z \,\X4(- \frac{e^{-2m}bf^{3/2}\X1((-1+b^4)(\pap{a_3}z)^2 + (9-11b^2)(\pap{a_3}\rho)^2\Y1)}{(-1+b^2)^2}\\
{}-\frac1{32 \rho^4 f^{5/2}}\, e^{-2m} \X3(-32 e^m \rho^4 b^6 f^2
\pap{b}{z} \X2((e^m + f) \pap{f}{z} - 2 f^2 \pap{m}{z}\Y2) \\
{}-8 \rho^4 b^7 f \X2(e^m \pap{f}{z} - 2 f^2 \pap{m}{z}\Y2)
\X2((e^m + 2 f) \pap{f}{z} - 2 f^2 \pap{m}{z}\Y2) - 8 e^m \rho^4
b^4 f \pap{b}{z} \X2(e^m (1 + f (6 + f)) \pap{f}{z}\\
{} + f (1 + f (4 + f)) \X2(\pap{f}{z} - 2 f \pap{m}{z}\X2)\X2) -
4 e^m \rho^3 b f^2 \X2(e^m \rho (1 + f) (2 + f) \pae{b}{z}2 + 5\rho \pae{f}{z}2 \\
{}- 4 \rho f \pap{f}{z} \pap{m}{z} - 2 \rho f \pap{^2f}{z^2} -
4 f \pap{f}{\rho} + 6 \rho \pae{f}{\rho}2 - 4 \rho f \pap{f}{\rho} \pap{m}{\rho}\Y2) \\
{}- 4 e^m \rho^3 b^2 f \X3(\rho \pap{b}{z} \X2(\X1(e^m (1 + f) (2 + f) + f (2 + f (7 + 2 f))\Y1)
\pap{f}{z} - 2 f^2 (1 + 2 f (3 + f)) \pap{m}{z}\Y2) \\
{}- 4 f^2 \pap{b}{\rho} \X2(-3 \rho \pap{f}{\rho} + 2 f \X2(1 + \rho \pap{m}{\rho}\Y2)\Y2)\Y3) +
16 e^m \rho^3 f^3 \X3(-\rho \pap{b}{z} \X2(\pap{f}{z} - 2 f \pap{m}{z}\Y2) -
3 \rho \pap{b}{\rho} \pap{f}{\rho}\\
{} + f \X2(\rho \pap{^2b}{z^2} + 2 \pap{b}{\rho} \X2(1 + \rho \pap{m}{\rho}\Y2)\Y2)\Y3) - 2 \rho^3 b^5
\X3(e^{2 m} \rho \X2(16 f^3 \pae{b}{z}2 + (1 + f (6 + f)) \pae{f}{z}2\Y2) \\
{}+ 2 e^m \rho f (1 + f (4 + f)) \pap{f}{z} \X2(\pap{f}{z} - 2 f \pap{m}{z}\Y2) +
4 f^3 \X3(\rho \pae{f}{z}2 - \rho (1 + f)^2 \pap{f}{z} \pap{m}{z} +2 \rho f^2 \pae{m}{z}2 \\
{}+ \rho f^3 \pae{m}{z}2 + 3 \rho \pae{f}{\rho}2 +
f \X2(\rho \pae{m}{z}2 - 2 \pap{f}{\rho} \X2(1 + \rho \pap{m}{\rho}\Y2)\Y2)\Y3)\Y3) \\
{}+ b^3 \X3(-e^{2m} \rho^4 \X2(8 f^2 (1 + f (6 + f)) \pae{b}{z}2 + (1 + f) (2 + f) \pae{f}{z}2\Y2)\\
{}+ 2 e^m \rho^3 f \X3(-\rho (2 + f (7 + 2 f)) \pae{f}{z}2 +
2 \rho f (1 + 2 f (3 + f)) \pap{f}{z} \pap{m}{z} +
4 f \pap{f}{\rho} \X2(-3 \rho \pap{f}{\rho} + 2 f \X2(1 + \rho \pap{m}{\rho}\Y2)\Y2)\Y3) \\
{}+ 4 f^3 \X2(\rho^4 (-4 + f) \pae{f}{z}2 + \rho^4 (7 - 2 f) f
\pap{f}{z} \pap{m}{z} - 5 \rho^4 f^2 \pae{m}{z}2 +
\rho^4 f^3 \pae{m}{z}2 - 6 \rho^4 \pae{f}{\rho}2 \\
\hskip30pt {}+ 4 \rho^3 f \pap{f}{\rho} \X2(1 + \rho \pap{m}{\rho}\Y2) + \rho^2 f^4 \X2(\pae{\o}{z}2 + 2 \pae{\o}{\rho}2\Y2) +
f^5 \X2(\pae{\o}{z}2 + 4 \pae{\o}{\rho}2\Y2)\Y2)\Y3)\Y3)\Y4)\\
{}+\pap{}\rho \,\X4(\frac1{16 \rho^4 f^{3/2}}\, e^{-2m}
\X4(e^{2m} \rho^4 b^3 (3 + 4 b^2) \pap{f}{z} \pap{f}{\rho} + e^m \rho^4 b
f \X3((-2 - 15 b^2 - b^4 + 6 b^6) \pap{f}{z} \pap{f}{\rho} \hskip20pt \\
{}+ e^m b \X2(2 (3 + 4 b^2) \pap{b}{z} \pap{f}{\rho} + \pap{f}{z}
\X2(2 (3 + 4 b^2) \pap{b}{\rho} + (b + 2 b^3) \pap{f}{\rho}\Y2)\Y2)\Y3) -
2 \rho^2 b^3 (2 + b^2) f^6 \pap{\o}{z} \pap{\o}{\rho}\\
{} + 4 b^3 f^7 \pap{\o}{z} \pap{\o}{\rho} +
2 \rho^3 f^3 \X3(2 e^{2m} \rho (b + 2 b^3) \pap{b}{z} \pap{b}{\rho} +
b^2 \X3(-6 \rho b^5 \pap{m}{z} \pap{f}{\rho} +
12 \rho \X2(\pap{f}{z} \pap{b}{\rho} - \pap{b}{z} \pap{f}{\rho}\Y2) \\
{}+ b \X2(3 \rho \pap{m}{z} \pap{f}{\rho} + \pap{f}{z} \X2(4 - \rho \pap{f}{\rho}\Y2)\Y2) +
b^3 \X2(5 \rho \pap{m}{z} \pap{f}{\rho} +
\pap{f}{z} \X2(4 - \rho \pap{f}{\rho} + 2 \rho \pap{m}{\rho}\Y2)\Y2)\Y3) \\
{}+ e^m \X2(\pap{b}{z} \X2(4 + b \X2(4 b (2 + b^2) +
8 \rho \pap{b}{\rho} - \rho (b + 3 b^3) \pap{f}{\rho}\Y2) +
2 \rho (4 + 3 b^2 + b^4) \pap{m}{\rho}\Y2) \\
{}+ \rho \X2(2 \X2(b^2 (1 + 2 b^2) \pap{f}{z} - (4 + b^2 + 4 b^4)
\pap{m}{z}\Y2) \pap{b}{\rho} - b^3 (1 + 2 b^2)
\pap{m}{z} \pap{f}{\rho} + 4 (-1 + b^2) \pap{^2b}{\rho\pa z}\Y2)\Y2)\Y3) \\
{}+ \rho^3 f^2 \X3(-4 \rho b^3 (1 + 2 b^2) \pap{f}{z} \pap{f}{\rho} + 2 e^{2m} \rho
b \X2(\X2((6 + 8 b^2) \pap{b}{z} + b (1 + 2 b^2) \pap{f}{z}\Y2)
\pap{b}{\rho} + (b + 2 b^3) \pap{b}{z} \pap{f}{\rho}\Y2) \\
{}+ e^m \X3(-2 \rho b^4 \pap{b}{z} \pap{f}{\rho} + 12 \rho b^6 \pap{b}{z} \pap{f}{\rho} +
2 \rho b^2 \X2(6 \pap{f}{z} \pap{b}{\rho} - 11 \pap{b}{z} \pap{f}{\rho}\Y2) +
4 \rho \X2(3 \pap{f}{z} \pap{b}{\rho} - 7 \pap{b}{z} \pap{f}{\rho}\Y2) \\
{}+ b^5 \X2(-8 \rho \pap{m}{z} \pap{f}{\rho} + \pap{f}{z} \X2(4 + \rho \pap{f}{\rho} +
2 \rho \pap{m}{\rho}\Y2)\Y2) + b^3 \X2(-2 \rho \pap{m}{z} \pap{f}{\rho} +
\pap{f}{z} \X2(8 + \rho \pap{f}{\rho} + 6 \rho \pap{m}{\rho}\Y2) + 4 \rho \pap{^2f}{\rho\pa z}\Y2) \\
{}+ 4 b \X2(\pap{f}{z} \X2(1 + 2 \rho \pap{m}{\rho}\Y2) -
\rho \X2(2 \pap{m}{z} \pap{f}{\rho} + \pap{^2f}{\rho\pa z}\Y2)\Y2)\Y3)\Y3) +
2 \rho^3 b^2 f^4 \X3(-2 (6 + e^m) \rho \pap{m}{z} \pap{b}{\rho} - 4 e^m \rho b^2 \pap{m}{z} \pap{b}{\rho} \\
{}+ b^3 \pap{m}{z} \X2(-4 + \rho \pap{f}{\rho} - 2 \rho \pap{m}{\rho}\Y2) +
b \X2(\pap{m}{z} \X2(-4 + \rho \pap{f}{\rho} - 2 \rho \pap{m}{\rho}\X2) - 4 \rho \pap{^2m}{\rho\pa z}\Y2)\Y3)\Y4)\\
{}-\frac{10e^{-2m}bf^{3/2}\pap{a_3}{z}\pap{a_3}\rho}{-1+b^2}\Y4)=0
\e

Let us consider the second pair of Maxwell \e s in our theory
\beq A.14
\bga
\pa_\mu \fal W^\m=0\\
\fal W^\m = -\fal W^{\nu\mu},
\ega
\e
$\fal W^\m = \sqrt{-g}\X1(H^\m - 2g^\[\m](g^\[\a\b]F_{\a\b})\Y1)$. One gets
\beq A.15
\bga
\pap{}\rho \X3(\frac{2e^m b\sqrt f\,\pap{a_3}z}{b^2-1}\Y3) = 0\\
\pap{}\rho \X3(\frac{e^{-m}(1+b^2)\pap{a_3}z}{b^2-1}\Y3) =0\\
\pap{}\rho \X3(\frac{e^{-m}\pap{a_3}\rho}{b^2-1}\Y3) + \pap{}z \X3(\frac{e^{-m}(1+b^2)\pap{a_3}z}{b^2-1}\Y3)=0
\ega
\e
From \er{A.15} one gets
\beq A.16
\bga
\frac{e^{-m}(1+b^2)\pap{a_3}z}{b^2-1} = g(z)\\
\frac{2e^m b\sqrt f\,\pap{a_3}z}{b^2-1} = h(z)\\
\frac{e^{-m}\pap{a_3}\rho}{b^2-1} = -\frac{dg}{dz}\,\rho +C(z)
\ega
\e
and eventually
\beq A.17
b=\sqrt{\frac{g(z)}{-\frac{dg}{dz}\,\rho +C(z)}}
\e
where $g(z)$, $h(z)$ and $C(z)$ are arbitrary \f s of $z$.

We can introduce \pt s $A_1=A_1(\rho,z)$ and $A_2=A_2(\rho,z)$ for Eqs \er{A.12} and \er{A.13} \st
$$
\ov R_\[23]=\pap{A_1}z\,,\q \ov R_\[13]=\pap{A_1}\rho \qh{and}
\pap{A_2}\rho = 8\pi \Tem_{[14]} - \ov R_\[14], \q \pap{A_2}z = 8\pi\Tem_{[24]} - \ov R_\[24].
$$

In particular one gets
\bml A.18
\pap{A_1}z = \frac1{16 \rho^2}\,e^{-2 m} \sqrt f \X4(16 e^m \rho^2 f \pap{\o}{z} \pap{b}{\rho} + 8 e^m \rho^2 b
\pap{\o}{z} \pap{f}{\rho} - 16 e^m \rho^2 b^6 f \pap{b}{z} \pap{\o}{\rho} \\
{}+ 2 e^m b^4 f (-5 \rho^2 + (-2 + \rho^2) f) \pap{b}{z} \pap{\o}{\rho} -
2 b^2 f \X1(-24 \rho^2 f + e^m (-\rho^2 + (2 + \rho^2) f)\Y1) \pap{b}{z} \pap{\o}{\rho} \\
{}- 8 \rho^2 b^7 \X2((e^m + f) \pap{f}{z} - 2 f^2 \pap{m}{z}\Y2) \pap{\o}{\rho} +
b^5 \X2(\X1(-2 \rho^2 (-1 + f) f + e^m (-5 \rho^2 + (-2 + \rho^2) f)\Y1) \pap{f}{z} \\
{}+ 2 f^2 (-3 \rho^2 + (2 + \rho^2) f) \pap{m}{z}\Y2) \pap{\o}{\rho} +
b^3 \X3(-e^m (-\rho^2 + (2 + \rho^2) f) \pap{f}{z} \pap{\o}{\rho} +
2 f \X3(\rho^2 \X2(3 \pap{\o}{z} \pap{f}{\rho} + 11 \pap{f}{z} \pap{\o}{\rho}\Y2) \\
{}+ f^2 \X2(2 \pap{\o}{z} \pap{m}{\rho} - (2 + \rho^2) \pap{m}{z} \pap{\o}{\rho}\Y2) +
f \X2(2 \pap{\o}{z} \X2(\rho - \pap{f}{\rho}\Y2) + \X2((2 + \rho^2) \pap{f}{z} -
9 \rho^2 \pap{m}{z}\Y2) \pap{\o}{\rho} + 8 \rho^2 \pap{^2\o}{\rho z}\Y2)\Y3)\Y3)\Y4)
\e

\bml A.19
\pap{A_1}\rho =
-\frac1{16 \rho^2}\,e^{-2m} \sqrt f \X4(-2 b^3 (1 - \rho^2 + b^2) f^3 \pap{m}{z} \pap{\o}{z} -
e^m \rho^2 b \X2((4 + b^2) \pap{f}{z} \pap{\o}{z} +
4 (-1 - b^2 + b^4) \pap{f}{\rho} \pap{\o}{\rho}\Y2) \\
{}+ f \X3(e^m \X2(\X2(-2 \rho^2 (4 + b^2) \pap{b}{z} + b^3 (1 + \rho^2 + b^2) \pap{f}{z}\Y2)
\pap{\o}{z} + 8 \rho^2 (1 + b^2 - b^4) \pap{b}{\rho} \pap{\o}{\rho}\Y2) \\
{}+ 2 \rho^2 b^3 \X2(-\pap{f}{z} \pap{\o}{z} + (12 + 5 b^2 - 2 b^4) \pap{f}{\rho} \pap{\o}{\rho}\Y2)\Y3) +
2 b^2 f^2 \X3(e^m (1 + \rho^2 + b^2) \pap{b}{z} \pap{\o}{z}\\
{} + 12 \rho^2 \pap{b}{\rho} \pap{\o}{\rho}
+ 2 \rho b^3 \X2(1 - 2 \rho \pap{m}{\rho}\Y2) \pap{\o}{\rho} + b \X3(-(-1 + \rho^2) \pap{f}{z} \pap{\o}{z} \\
{}+ \rho \X2(3 \rho \pap{m}{z} \pap{\o}{z} +
\X2(2 - 8 \rho \pap{m}{\rho}\Y2) \pap{\o}{\rho} + 4 \rho \pap{^2\o}{\rho^2}\Y2)\Y3)\Y3)\Y4)
\e

\bml A.20
\pap{A_2}\rho = - \frac{e^{-2m}bf^{3/2}\X1((-1+b^4)(\pap{a_3}z)^2 + (9-11b^2)(\pap{a_3}\rho)^2\Y1)}{(-1+b^2)^2}\\
{}-\frac1{32 \rho^4 f^{5/2}}\, e^{-2m} \X3(-32 e^m \rho^4 b^6 f^2
\pap{b}{z} \X2((e^m + f) \pap{f}{z} - 2 f^2 \pap{m}{z}\Y2) \\
{}-8 \rho^4 b^7 f \X2(e^m \pap{f}{z} - 2 f^2 \pap{m}{z}\Y2)
\X2((e^m + 2 f) \pap{f}{z} - 2 f^2 \pap{m}{z}\Y2) - 8 e^m \rho^4
b^4 f \pap{b}{z} \X2(e^m (1 + f (6 + f)) \pap{f}{z}\\
{} + f (1 + f (4 + f)) \X2(\pap{f}{z} - 2 f \pap{m}{z}\X2)\X2) -
4 e^m \rho^3 b f^2 \X2(e^m \rho (1 + f) (2 + f) \pae{b}{z}2 + 5\rho \pae{f}{z}2 \\
{}- 4 \rho f \pap{f}{z} \pap{m}{z} - 2 \rho f \pap{^2f}{z^2} -
4 f \pap{f}{\rho} + 6 \rho \pae{f}{\rho}2 - 4 \rho f \pap{f}{\rho} \pap{m}{\rho}\Y2) \\
{}- 4 e^m \rho^3 b^2 f \X3(\rho \pap{b}{z} \X2(\X1(e^m (1 + f) (2 + f) + f (2 + f (7 + 2 f))\Y1)
\pap{f}{z} - 2 f^2 (1 + 2 f (3 + f)) \pap{m}{z}\Y2) \\
{}- 4 f^2 \pap{b}{\rho} \X2(-3 \rho \pap{f}{\rho} + 2 f \X2(1 + \rho \pap{m}{\rho}\Y2)\Y2)\Y3) +
16 e^m \rho^3 f^3 \X3(-\rho \pap{b}{z} \X2(\pap{f}{z} - 2 f \pap{m}{z}\Y2) -
3 \rho \pap{b}{\rho} \pap{f}{\rho}\\
{} + f \X2(\rho \pap{^2b}{z^2} + 2 \pap{b}{\rho} \X2(1 + \rho \pap{m}{\rho}\Y2)\Y2)\Y3) - 2 \rho^3 b^5
\X3(e^{2 m} \rho \X2(16 f^3 \pae{b}{z}2 + (1 + f (6 + f)) \pae{f}{z}2\Y2) \\
{}+ 2 e^m \rho f (1 + f (4 + f)) \pap{f}{z} \X2(\pap{f}{z} - 2 f \pap{m}{z}\Y2) +
4 f^3 \X3(\rho \pae{f}{z}2 - \rho (1 + f)^2 \pap{f}{z} \pap{m}{z} +2 \rho f^2 \pae{m}{z}2 \\
{}+ \rho f^3 \pae{m}{z}2 + 3 \rho \pae{f}{\rho}2 +
f \X2(\rho \pae{m}{z}2 - 2 \pap{f}{\rho} \X2(1 + \rho \pap{m}{\rho}\Y2)\Y2)\Y3)\Y3) \\
{}+ b^3 \X3(-e^{2m} \rho^4 \X2(8 f^2 (1 + f (6 + f)) \pae{b}{z}2 + (1 + f) (2 + f) \pae{f}{z}2\Y2)\\
{}+ 2 e^m \rho^3 f \X3(-\rho (2 + f (7 + 2 f)) \pae{f}{z}2 +
2 \rho f (1 + 2 f (3 + f)) \pap{f}{z} \pap{m}{z} +
4 f \pap{f}{\rho} \X2(-3 \rho \pap{f}{\rho} + 2 f \X2(1 + \rho \pap{m}{\rho}\Y2)\Y2)\Y3) \\
{}+ 4 f^3 \X2(\rho^4 (-4 + f) \pae{f}{z}2 + \rho^4 (7 - 2 f) f
\pap{f}{z} \pap{m}{z} - 5 \rho^4 f^2 \pae{m}{z}2 +
\rho^4 f^3 \pae{m}{z}2 - 6 \rho^4 \pae{f}{\rho}2 \\
\hskip30pt {}+ 4 \rho^3 f \pap{f}{\rho} \X2(1 + \rho \pap{m}{\rho}\Y2) + \rho^2 f^4 \X2(\pae{\o}{z}2 + 2 \pae{\o}{\rho}2\Y2) +
f^5 \X2(\pae{\o}{z}2 + 4 \pae{\o}{\rho}2\Y2)\Y2)\Y3)\Y3)
\e

\bml A.21
\pap{A_2}z = -\frac1{16 \rho^4 f^{3/2}}\, e^{-2m}
\X4(e^{2m} \rho^4 b^3 (3 + 4 b^2) \pap{f}{z} \pap{f}{\rho} + e^m \rho^4 b
f \X3((-2 - 15 b^2 - b^4 + 6 b^6) \pap{f}{z} \pap{f}{\rho} \hskip20pt \\
{}+ e^m b \X2(2 (3 + 4 b^2) \pap{b}{z} \pap{f}{\rho} + \pap{f}{z}
\X2(2 (3 + 4 b^2) \pap{b}{\rho} + (b + 2 b^3) \pap{f}{\rho}\Y2)\Y2)\Y3) -
2 \rho^2 b^3 (2 + b^2) f^6 \pap{\o}{z} \pap{\o}{\rho}\\
{} + 4 b^3 f^7 \pap{\o}{z} \pap{\o}{\rho} +
2 \rho^3 f^3 \X3(2 e^{2m} \rho (b + 2 b^3) \pap{b}{z} \pap{b}{\rho} +
b^2 \X3(-6 \rho b^5 \pap{m}{z} \pap{f}{\rho} +
12 \rho \X2(\pap{f}{z} \pap{b}{\rho} - \pap{b}{z} \pap{f}{\rho}\Y2) \\
{}+ b \X2(3 \rho \pap{m}{z} \pap{f}{\rho} + \pap{f}{z} \X2(4 - \rho \pap{f}{\rho}\Y2)\Y2) +
b^3 \X2(5 \rho \pap{m}{z} \pap{f}{\rho} +
\pap{f}{z} \X2(4 - \rho \pap{f}{\rho} + 2 \rho \pap{m}{\rho}\Y2)\Y2)\Y3) \\
{}+ e^m \X2(\pap{b}{z} \X2(4 + b \X2(4 b (2 + b^2) +
8 \rho \pap{b}{\rho} - \rho (b + 3 b^3) \pap{f}{\rho}\Y2) +
2 \rho (4 + 3 b^2 + b^4) \pap{m}{\rho}\Y2) \\
{}+ \rho \X2(2 \X2(b^2 (1 + 2 b^2) \pap{f}{z} - (4 + b^2 + 4 b^4)
\pap{m}{z}\Y2) \pap{b}{\rho} - b^3 (1 + 2 b^2)
\pap{m}{z} \pap{f}{\rho} + 4 (-1 + b^2) \pap{^2b}{\rho\pa z}\Y2)\Y2)\Y3) \\
{}+ \rho^3 f^2 \X3(-4 \rho b^3 (1 + 2 b^2) \pap{f}{z} \pap{f}{\rho} + 2 e^{2m} \rho
b \X2(\X2((6 + 8 b^2) \pap{b}{z} + b (1 + 2 b^2) \pap{f}{z}\Y2)
\pap{b}{\rho} + (b + 2 b^3) \pap{b}{z} \pap{f}{\rho}\Y2) \\
{}+ e^m \X3(-2 \rho b^4 \pap{b}{z} \pap{f}{\rho} + 12 \rho b^6 \pap{b}{z} \pap{f}{\rho} +
2 \rho b^2 \X2(6 \pap{f}{z} \pap{b}{\rho} - 11 \pap{b}{z} \pap{f}{\rho}\Y2) +
4 \rho \X2(3 \pap{f}{z} \pap{b}{\rho} - 7 \pap{b}{z} \pap{f}{\rho}\Y2) \\
{}+ b^5 \X2(-8 \rho \pap{m}{z} \pap{f}{\rho} + \pap{f}{z} \X2(4 + \rho \pap{f}{\rho} +
2 \rho \pap{m}{\rho}\Y2)\Y2) + b^3 \X2(-2 \rho \pap{m}{z} \pap{f}{\rho} +
\pap{f}{z} \X2(8 + \rho \pap{f}{\rho} + 6 \rho \pap{m}{\rho}\Y2) + 4 \rho \pap{^2f}{\rho\pa z}\Y2) \\
{}+ 4 b \X2(\pap{f}{z} \X2(1 + 2 \rho \pap{m}{\rho}\Y2) -
\rho \X2(2 \pap{m}{z} \pap{f}{\rho} + \pap{^2f}{\rho\pa z}\Y2)\Y2)\Y3)\Y3) +
2 \rho^3 b^2 f^4 \X3(-2 (6 + e^m) \rho \pap{m}{z} \pap{b}{\rho} - 4 e^m \rho b^2 \pap{m}{z} \pap{b}{\rho} \\
{}+ b^3 \pap{m}{z} \X2(-4 + \rho \pap{f}{\rho} - 2 \rho \pap{m}{\rho}\Y2) +
b \X2(\pap{m}{z} \X2(-4 + \rho \pap{f}{\rho} - 2 \rho \pap{m}{\rho}\X2) - 4 \rho \pap{^2m}{\rho\pa z}\Y2)\Y3)\Y4)\\
{}+\frac{10e^{-2m}bf^{3/2}\pap{a_3}{z}\pap{a_3}\rho}{-1+b^2}
\e

The above \e s can be solved exactly or numerically. In order to solve them exactly we should develop new methods of exact \so\ similar to the
inverse scattering method, B\"acklund \tf\ or Hirota method. First of all we should transform the system of the \e s to some kind Ernst-like \e\
(see Refs~\cite{15}--\cite{27}). This will be a subject of a future development.

\def\hsm{\vrule height14pt depth4pt width0pt }
\def\hsd{\vrule height21pt depth11pt width0pt }
\section{\E\cy\ \gelm c waves\hfil\break in the \NK{}}
Let us consider the \fw\ \nos\ metric tensor (\cy\ \gelm c wave)
\bg4.1
g_{\a\b} = \left(
\begin{array}{c|c|c|c}
\hsm \ -e^{2l-2n}\ & 0 & se^{2n} & 0 \\
\hline
\hsm 0 &\ -r^2e^{-2n}\ & 0 & 0 \\
\hline
\hsm -se^{2n} & 0 & -e^{2n} & se^{2n} \\
\hline
\hsm 0 & 0 &\ -se^{2n}\ &\ e^{2l-2n}\
\end{array}\right) \\
\bga
g = \det g_\m = e^{-4l+4n}r^2 \\
\sqrt{-g} = e^{-2l+2n}r
\ega \lb4.2
\e
in \cy\ \cd s $r,\t,z$.

$l,n,s$ are \f s of $r$ and~$t$. The strength of an \elm c field takes a form
\bg4.3
F_\m = \left(\begin{array}{c|c|c|c}
\hsd 0 & 0 &\ \dsp\pap br\ & 0 \\
\hline
\hsm 0 &\ \ 0\ \ & 0 & 0 \\
\hline
\hsd \ \dsp-\pap br\ & 0 & 0 & \ \dsp-\pap bt\ \\
\hline
\hsd 0 & 0 & \dsp\pap bt & 0
\end{array}\right) \\
\bga
\os B = (-F_{23},-F_{31},-F_{12}) = (B_1,B_2,B_3) \\
\os E = (F_{41},F_{42},F_{43}) = (E_1,E_2,E_3)
\ega \lb4.4 \\
\bga
E_z = -b_1 = \pap bt\\
B_\t = -a_1 = -\pap br\,.
\ega \lb4.5
\e
$b$ is a \f\ of $r$ and $t$.

One gets
\beq4.6
g^\m = e^{2n-2l} \left(\begin{array}{c|c|c|c}
\hsm\ -1+s^2e^{4n-2l}\ & 0 & s & s^2e^{4n-2l} \\
\hline
\hsd 0 &\ \dsp -\frac{e^{2l}}{r^2}\ & 0 & 0 \\
\hline
\hsm -s & 0 &\ -e^{2l-4n}\ & -s \\
\hline
\hsm s^2e^{4n-2l} & 0 & s &\ 1+s^2e^{4n-2l}\
\end{array}\right)
\e
(see Ref. \cite{k}),
\beq4.7
g^\[\m]=\left(\begin{array}{c|c|c|c}
\hsd 0 &\ \ 0\ \  & \ -\dfrac{e^{2n-2l}}r\,s \ & 0 \\
\hline
\hsm 0 & 0 & 0 & 0 \\
\hline
\hsd \ \dfrac{e^{2l-2n}}r\,s\ & 0 & 0 & \ \dfrac{e^{2l-2n}}r\,s\ \\
\hline
\hsd 0 & 0 & \ -\dfrac{e^{2l-2n}}r\,s\ & 0
\end{array}\right)
\e
From the \e
\beq4.8
\gd\falg,[\m],\nu,=0
\e
one gets that
\beq4.9
s(r,t) = \frac{f(r-t)}r
\e
(see Ref.\ \cite k).

One gets
\beq4.10
g^\[\m]F_\m = -2\,\frac{2l-2n}r\,s\Bigl(\pap br+\pap bt\Bigr).
\e
It is easy to see that
\beq4.10a
F_{[\m,\la]} = 0.
\e
The first pair of Maxwell \e s is satisfied.
\bg4.11
H^\m = \left(\begin{array}{c|c|c|c}
\hsm 0 &\ \ 0\ \ & H^{13} & H^{14} \\
\hline \hsm
0 & 0 & 0 & 0 \\
\hline \hsm
\ -H^{13}\ & 0 & 0 & \ H^{34} \ \\
\hline \hsm
-H^{14} & 0 & \ -H^{34}\ & 0
\end{array}\right)\\
\bal
H^{13} &= -H^{31} = e^{-2n}\,\pap br + \frac{e^{4l-4n}}{r^2}\,s^2\X2(\pap bt+ \pap br\Y2)\\
H^{14} &= -H^{41} = -\frac{2e^{4l-4n}}{r^2}\,s^2\X2(\pap bt+\pap br\Y2) \\
H^{34} &= -H^{43} = e^{-2n}\,\pap bt - \frac{e^{4l-4n}}{r^2}\,s^2\X2(\pap bt+\pap br\Y2)
\eal \lb4.12
\e

Let us calculate
\bea4.13
W^\m &= H^\m - 2g^\[\m](g^\[\a\b]F_{\a\b})\\
W^{13} &= -W^{31} = e^{-2n}\,\pap br - 3\,\frac{e^{4l-4n}}{r^2}\,s^2\X2(\pap bt+\pap br\Y2) \lb4.14 \\
W^{14} &= -W^{41} = 2\,\frac{e^{4l-4n}}{r^2}\,s^2\X2(\pap bt+\pap br\Y2) \lb4.15 \\
W^{34} &= -W^{43} = e^{-2n}\,\pap bt + 3\,\frac{e^{4l-4n}}{r^2}\,s^2\X2(\pap bt+\pap br\Y2) \lb4.16
\e

From the second pair of Maxwell \e s in the \NK{} one gets
\beq4.17
\pa_\mu\X1(\sqrt{-g}\,W^\m \Y1) =0.
\e
For $\nu=1$
\beq4.18
\pap{}t\,\X3(\frac{e^{2l-2n}}{r^2}\,s^2\X2(\pap bt+\pap br\Y2)\Y3).
\e

$\nu=3$
\beq4.19
\pap{}r\,\X3(e^{-2l}r\,\pap br- 3e^{2l-2n}\,\frac{s^2}r\,\X2(\pap bt+\pap br\Y2)\Y3) - \pap{}t\,\X3(e^{-2l}r\,\pap bt + 3e^{2l-2n}\,
\frac{s^2}r \,\X2(\pap bt+\pap br\Y2)\Y3)=0.
\e

$\nu=4$
\beq4.20
\pap{}r \X3(\frac{e^{2l-2n}}{r^2}\,s^2\X2(\pap bt+\pap br\Y2)\Y3) =0.
\e

From Eq.~\er{4.18} and Eq.~\er{4.20} one gets
\beq4.21
\pap bt+\pap br = \frac{Cr}{s^2}\,e^{2n-2l}
\e
where $C=\rm const$.

Eq.~\er{3.17a} gives us an existence of a \pt\ $A=A(r,t)$ \st
\beq4.22
\bal
\pap br &= \frac{e^{2l}}r\,\pap At \\
\pap bt &= \frac{e^{2l}}r\,\pap Ar
\eal
\e
The \f\ $A(r,t)$ \sf ies the \e
\beq4.23
\pap At + \pap Ar = \frac{Cr^2}{s^2}\,e^{2n-4l}.
\e
Moreover $b(r,t)$ \sf ies the \e
\beq4.24
\pap{^2b}{r^2} - \pap{^2b}{t^2} + \frac1r\,\pap br - 2\,\pap lr\,\pap br + 2\,\pap lt\,\pap bt = 0.
\e
\E\e s \er{4.21}, \er{4.22}, \er{4.23} and \er{4.24} give us an additional \ia\ between \gr al and \elm c fields which is absent in GR theory.

All \f s are \f s of $r,t$ in the \fw\ way:
\beq4.25
\bal
b&=b(r,r-t)\\
l&=l(r,r-t)\\
n&=n(r,r-t)\\
s&=s(r,r-t)=\frac{f(r-t)}r\,.
\eal
\e

Let us write the remaining field \e s
\bg4.26
\ov R_\(\a\b) = 8\pi\Tem_{\a\b}\\
\ov R_{[\[\a\b],\g]} = 8\pi\Tem_{[\[\a\b],\g]}. \lb4.27
\e

Using the programme from Appendix B one gets field \e s \er{4.26}--\er{4.27}.

There are some zeros:
\bml4.28
\ov R_\(12) = \ov R_\(23) = \ov R_\(24) = \ov R_\[12] = \ov R_\[23] = \ov R_\[24]\\ {}=
\Tem_{(12)} = \Tem_{(23)} = \Tem_{(24)} = \Tem_{[12]} = \Tem_{[13]} = \Tem_{[23]} = \Tem_{[24]} = 0.
\e
In the remaining cases we get
$$
\ov R_{11}=8\pi \Tem_{11}
$$
or
\bml4.29
\frac1{4 r^2}\, e^{-2 n} \X4(-8 e^{8 l} r f^3 \frac{df}{du}
\X3(-2 e^{4 l} + 4 (2 e^{4 l} - e^{2 n}) r \pap lu +
4 e^{2 n} r \pap{n}{u} + 4 e^{4 l} r \pap lr -
e^{2 n} r \pap lr + 2 e^{2 n} r \pap nr\Y3)\\
{}+ 4 e^{4 l+2 n} r^3 f
\X3(2 e^{2 n} r \frac{d^2f}{du^2} + \frac{df}{du} \X2(e^{4 l} - 9 e^{2 n} +
22 e^{2 n} r \pap lu - 4 e^{2 n} r \pap{n}{u} -
8 e^{4 l} r \pap lr + 20 e^{2 n} r \pap lr \\
{}+4 e^{4 l} r \pap nr - 6 e^{2 n} r \pap nr\Y2)\Y3) -
e^{8 l} f^4 \X3(9 e^{4 l} - 3 e^{2 n} +
48 \X2(e^{4 l} - 2 e^{2 n}\Y2) r^2 \X2(\pap lu\Y2)^2 \\
{}-26 e^{4 l} r \pap lr + 12 e^{2 n} r \pap lr +
20 e^{4 l} r^2 \X2(\pap lr\Y2)^2 - 20 e^{2 n} r^2 \X2(\pap lr\Y2)^2 +
16 e^{2 n} r \pap{n}{u} \X2(-1 + 3 r \pap lr\Y2) \\
{}- 8 e^{2 n} r \pap nr + 24 e^{2 n} r^2 \pap lr
\pap nr + 8 r \pap lu \X2(-5 e^{4 l} + 2 e^{2 n} +
12 e^{2 n} r \pap{n}{u} + (8 e^{4 l} - 9 e^{2 n})
r \pap lr + 6 e^{2 n} r \pap nr\Y2)\Y3)\\
{}+ 4 e^{4 n} r^4 \X3(2 e^{4 l} \X2(\frac{df}{du}\Y2)^2 + e^{2 n} r
\X3(2 r \X2(\pap lu\Y2)^2 - \pap{n}{u} - \pap lr +
2 r \X2(\pap lr\Y2)^2 \\
{}+ \pap lu \X2(-1 + 4 r \pap lr\Y2) - \pap nr
- 2 r \pap{^2l}{r\pa u} + 2 r \pap{^2n}{r\pa u} -
r \pap{^2l}{r^2} + r \pap{^2n}{r^2}\Y3)\Y3) \\
{}- 2 e^{4 l} r^2 f^2 \X3(8 e^{8 l} \X2(\frac{df}{du}\Y2)^2 + e^{2 n}
\X3(e^{4 l} - 12 e^{2 n} - 60 e^{2 n} r^2 \X2(\pap lu\Y2)^2 -
12 e^{2 n} r^2 \pap{^2l}{u^2} - 8 e^{4 l} r \pap lr \\
{} + 38 e^{2 n} r \pap lr + 14 e^{4 l} r^2 \X2(\pap lr\Y2)^2 -
38 e^{2 n} r^2 \X2(\pap lr\Y2)^2 +
20 e^{2 n} r \pap{n}{u} \X2(-1 + r \pap lr\Y2) + 2 e^{4 l} r \pap nr \\
{}- 16 e^{2 n} r \pap nr -
6 e^{4 l} r^2 \pap lr \pap nr + 22 e^{2 n}
r^2 \pap lr \pap nr + 2 r \pap lu
\X2(-e^{4 l} + 27 e^{2 n} + 12 e^{2 n} r \pap{n}{u} \\
{} +2 (4 e^{4 l} - 27 e^{2 n}) r \pap lr - 4 e^{4 l}
r \pap nr + 18 e^{2 n} r \pap nr\Y2) +
2 e^{4 l} r^2 \pap{^2l}{r\pa u} - 26 e^{2 n} r^2 \pap{^2l}{r\pa u} \\
\hskip30pt {}+ 4 e^{2 n} r^2 \pap{^2n}{r\pa u} + 2 e^{4 l} r^2 \pap{^2l}{r^2} -
10 e^{2 n} r^2 \pap{^2l}{r^2} + 2 e^{2 n} r^2 \pap{^2n}{r^2}\Y3)\Y3)\Y4)\\
{}=
e^{-2 l}
\X3(-2 e^{2 n} r^2 (e^{2 n} r^2 + 2 e^{4 l} f^2) \X2(\pap bu\Y2)^2 -
2 e^{2 n} r^2 (e^{2 n} r^2 + 2 e^{4 l} f^2)\pap bu \pap br \\
{}- \X1(e^{4 n} r^4 - e^{4 l+2 n} r^2 f^2 - 2 e^{8 l} f^4\Y1)
\X2(\pap br\Y2)^2\Y3)
\e

$$
\ov R_\(13) = 8\pi \Tem_{(13)}
$$
or
\bml4.30
\frac1{8 r^4}\, e^{2 l-4 n} \X4(8 e^{8 l+2 n} r f^4 \frac{df}{du} +
4 e^{6 n} r^5 \frac{df}{du} \X2(-1 + 4 r \pap lr\Y2) +
2 e^{8 l} f^5 \X2(e^{4 l} - 2 e^{2 n} +
12 e^{2 n} r \pap lu + 6 e^{2 n} r \pap lr\Y2) \\
{}-4 e^{4 l+2 n} r^3 f^2 \frac{df}{du}
\X2(2 e^{4 l} - 7 e^{2 n} + 24 e^{2 n} r \pap lu +
(-4 e^{4 l} r + 22 e^{2 n} r) \pap lr\Y2) \\
{}+ e^{2 n} r^2 f^3 \X3(3 e^{8 l} - 15 e^{4 l+2 n} -
144 e^{4 l+2 n} r^2 \X2(\pap lu\Y2)^2 -
6 e^{8 l} r \pap lr - 8 e^{4 n} r \pap lr +
76 e^{4 l+2 n} r \pap lr \\
{} + 8 e^{8 l} r^2 \X2(\pap lr\Y2)^2 + 16 e^{4 n} r^2 \X2(\pap lr\Y2)^2 -
96 e^{4 l+2 n} r^2 \X2(\pap lr\Y2)^2 -
16 e^{4 n} r \pap{n}{u} \X2(-1 + 2 r \pap lr\Y2) \\
{}+ 4 r \pap lu \X2(-2 e^{8 l} - 4 e^{4 n} + 21 e^{4 l+2 n} +
2 (2 e^{8 l} + 4 e^{4 n} - 33 e^{4 l+2 n}) r \pap lr\Y2) \\
{}+ 8 e^{4 n} r \pap nr -
16 e^{4 n} r^2 \pap lr \pap nr\Y3) +
8 e^{4 n} r^4 f \X3(-2 e^{4 l} \X2(\frac{df}{du}\Y2)^2 +
e^{2 n} \X2(1 - 3 r \pap lr + 2 r^2 \X2(\pap lr\Y2)^2 \\
{}+ r \pap lu \X2(-1 + 2 r \pap lr\Y2) + \pap{n}{u}
\X2(r - 2 r^2 \pap lr\Y2) + r \pap nr - 2 r^2 \pap lr \pap nr + r^2 \pap{^2l}{r\pa u} + r^2 \pap{^2l}{r^2}\Y2)\Y3)\Y4)\\
{}=
2 f
\pap br \X2(e^{2 n} r^2 \pap bu +
- e^{4 l} f^2 \pap br\Y2)
\e

$$
R_{22}=8\pi \Tem_{22}
$$
or
\bml4.31
\frac1{2 r^2}\, e^{-2 n} \X3(-4 e^{4 l+2 n} r f
\frac{df}{du} \X2(-1 + 2 r \pap lu + r \pap lr\Y2) +
e^{4 l} f^2 \X2(-24 e^{2 n} r^2 \X2(\pap lu\Y2)^2
+ 2 r \pap lu \X2(e^{4 l} + 8 e^{2 n} \\
{} - 12 e^{2 n} r \pap lr\Y2) + \X2(-1 + r \pap lr\Y2)
\X2(e^{4 l} + e^{2 n} + 2 e^{2 n} r \pap lr\Y2)\Y2) -
2 e^{4 n} r^3 \X2(\pap lu + \pap lr +
2 r \pap{^2l}{r\pa u} + r \pap{^2l}{r^2}\Y2)\Y3)\\
{}=
e^{-2 l} \X2(2 \pap bu + \pap br\Y2)
\X2(e^{2 n} r^2 \pap br -
e^{4 l} f^2 \X2(2 \pap bu + \pap br\Y2)\Y2)
\e

$$
\ov R_\(14) = 8\pi\Tem_{(14)}
$$
or
\bml4.32
\frac1{4 r^6}\, e^{-4 n} \X4(-4 e^{6 n} r^5 \X2(2 r \X2(\pap lu\Y2)^2 -
\pap{n}{u} + 2 r \pap lu \pap lr\Y2) -
16 e^{8 l+2 n} r^2 f^3 \frac{df}{du} \X2(2 \pap lu -
2 \pap{n}{u} + \pap lr - \pap nr\Y2) \\
{}+ 4 e^{4 (l+n)} r^3 f \frac{df}{du}
\X2(1 + 5 r \pap lr - 4 r \pap nr\Y2) +
e^{8 l} f^4 \X3(-3 e^{2 n} - 96 e^{2 n} r^2 \X2(\pap lu\Y2)^2 -
4 e^{4 l} r \pap lr \\
{}+ 16 e^{2 n} r \pap lr +
8 e^{4 l} r^2 \X2(\pap lr\Y2)^2 - 32 e^{2 n} r^2 \X2(\pap lr\Y2)^2
+16 e^{2 n} r \pap{n}{u} \X2(-1 + 3 r \pap lr\Y2) - 8 e^{2 n} r \pap nr \\
{}+ 24 e^{2 n} r^2 \pap lr
\pap nr + 8 r \pap lu \X2(-e^{4 l} + 2 e^{2 n} +
12 e^{2 n} r \pap{n}{u} + 2 (e^{4 l} - 6 e^{2 n})
r \pap lr + 6 e^{2 n} r \pap nr\Y2)\Y3) \\
{} + e^{4 l+2 n} r^2 f^2 \X3(5 e^{4 l} - e^{2 n} -
18 e^{4 l} r \pap lr - 6 e^{2 n} r \pap lr +
12 e^{4 l} r^2 \X2(\pap lr\Y2)^2 + 36 e^{2 n} r^2 \X2(\pap lr\Y2)^2 \\
{}+ 16 e^{2 n} r \pap{n}{u} \X2(-2 + r \pap lr\Y2) +
8 e^{4 l} r \pap nr - 8 e^{2 n} r \pap nr -
4 e^{4 l} r^2 \pap lr \pap nr -
16 e^{2 n} r^2 \pap lr \pap nr \\
{} + 8 e^{2 n} r \pap lu \X2(2 + 9 r \pap lr - 6 r \pap nr\Y2) -
4 e^{2 n} r^2 \pap{^2l}{r\pa u} + 8 e^{2 n} r^2 \pap{^2n}{r\pa u} +
2 e^{4 l} r^2 \pap{^2l}{r^2} \\
{}- 2 e^{2 n} r^2 \pap{^2l}{r^2} + 4 e^{2 n} r^2 \pap{^2n}{r^2}\Y3)\Y4)
=
2 \X3(\X2(\pap bu\Y2)^2 + \pap bu \pap br -
\frac1{r^4}\, e^{8 l-4 n} f^4 \X2(\pap br\Y2)^2\Y3)
\e

\goodbreak
$$
\ov R_{33} = 8\pi\Tem_{33}
$$
or
\bml4.33
\frac1{2 r^4}\, e^{2 l-4 n}
\X4(-4 e^{6 n} r^3 f \frac{df}{du} \X2(-1 + 2 r \pap lr\Y2) -
e^{4 l} f^4 \X2(-e^{8 l} + e^{4 n} - 8 e^{4 l+2 n} r \pap lu \\
{}+ 8 e^{4 l+2 n} r \pap{n}{u} +
2 e^{8 l} r \pap lr - 4 e^{4 n} r \pap lr -
4 e^{4 l+2 n} r \pap lr + 4 e^{4 n} r^2
\X2(\pap lr\Y2)^2 + 4 e^{4 l+2 n} r \pap nr\Y2)\\
{} + 2 e^{6 n} r^5 \X2(\pap lu + \pap lr +
2 r \pap{^2l}{r\pa u} + r \pap{^2l}{r^2}\Y2) -
e^{2 n} r^2 f^2 \X3(2 e^{4 n} - e^{4 l+2 n} +
2 (e^{8 l} - 4 e^{4 n}) r \pap lu \\
{}+ 8 e^{4 n} r \pap{n}{u} + e^{8 l} r \pap lr - 6 e^{4 n}
r \pap lr + e^{4 l+2 n} r \pap lr -
2 e^{4 l+2 n} r^2 \X2(\pap lr\Y2)^2 + 4 e^{4 n} r
\pap nr - 2 e^{4 l+2 n} r \pap nr \\
{}+ 4 e^{4 l+2 n} r^2 \pap lr \pap nr +
12 e^{4 n} r^2 \pap{^2l}{r\pa u} - 8 e^{4 n} r^2 \pap{^2n}{r\pa u} +
6 e^{4 n} r^2 \pap{^2l}{r^2} - 2 e^{4 l+2 n}
r^2 \pap{^2l}{r^2} - 4 e^{4 n} r^2 \pap{^2n}{r^2}\Y3)\Y4)\\
{}=
\X3(e^{2 n} r^2 \pap br \X2(2 \pap bu + \pap br\Y2) +
e^{4 l} f^2 \X2(4 \X2(\pap bu\Y2)^2 +
4 \pap bu \pap br - \X2(\pap br\Y2)^2\Y2)\Y3)
\e

$$
\ov R_\(34) = 8\pi\Tem_{(34)}
$$
or
\bml4.34
\frac1{8 r^4}\, e^{2 l-4 n} \X4(8 e^{8 l+2 n} r f^4 \frac{df}{du} +
4 e^{6 n} r^5 \frac{df}{du} \X2(-1 + 4 r \pap lr\Y2) + 2 e^{8 l}
f^5 \X2(e^{4 l}  - 2 e^{2 n} \\
{}+ 12 e^{2 n} r \pap lu + 6 e^{2 n} r \pap lr\Y2) - 4 e^{4 l+2 n}
r^3 f^2 \frac{df}{du} \X2(24 e^{2 n} r \pap lu -
(2 e^{4 l} - e^{2 n}) \X2(-1 + 2 r \pap lr\Y2)\Y2) \\
{}+ e^{2 n} r^2 f^3 \X3(5 e^{8 l} - 3 e^{4 l+2 n} -
144 e^{4 l+2 n} r^2 \X2(\pap lu\Y2)^2 -
18 e^{8 l} r \pap lr - 8 e^{4 n} r \pap lr +
8 e^{8 l} r^2 \X2(\pap lr\Y2)^2 \\
{}+ 16 e^{4 n} r^2 \X2(\pap lr\Y2)^2 +
24 e^{4 l+2 n} r^2 \X2(\pap lr\Y2)^2 +
4 (2 e^{8 l} + 4 e^{4 n} - 3 e^{4 l+2 n}) r \pap lu
\X2(-1 + 2 r \pap lr\Y2) \\
{}- 16 e^{4 n} r \pap{n}{u}
\X2(-1 + 2 r \pap lr\Y2) + 8 e^{4 n} r \pap nr -
16 e^{4 n} r^2 \pap lr \pap nr\Y3)\\
{} + 8 e^{4 n} r^4 f \X3(-2 e^{4 l} \X2(\frac{df}{du}\Y2)^2 +
e^{2 n} r \X2(\pap{n}{u} \X2(1 - 2 r \pap lr\Y2) +
\pap lu \X2(-1 + 2 r \pap lr\Y2) + r \pap{^2l}{r\pa u}\Y2)\Y3)\Y4)\\
{}=  2 f
\pap br \X2(e^{2 n} r^2 \pap bu +
\X2(e^{2 n} r^2 - e^{4 l} f^2\Y2) \pap br\Y2)
\e

$$
\ov R_{44} = 8\pi\Tem_{44}
$$
or
\bml4.35
\frac1{4 r^2}\, e^{-2 n} \X4(8 e^{8 l} r f^3 \frac{df}{du}
\X2(-2 e^{4 l} + 4 \X2(2 e^{4 l} + e^{2 n}\Y2) r \pap lu -
4 e^{2 n} r \pap{n}{u} + 4 e^{4 l} r \pap lr +
3 e^{2 n} r \pap lr - 2 e^{2 n} r \pap{n}{r}\Y2) \\
{}+ e^{8 l} f^4 \X3(9 e^{4 l} + 3 e^{2 n} +
48 (e^{4 l} + 2 e^{2 n}) r^2 \X2(\pap lu\Y2)^2 -
18 e^{4 l} r \pap lr - 20 e^{2 n} r \pap lr +
4 e^{4 l} r^2 \pap lr^2 + 44 e^{2 n} r^2 \X2(\pap lr\Y2)^2 \\
{}- 16 e^{2 n} r \pap{n}{u} \X2(-1 + 3 r \pap lr\Y2) +
8 e^{2 n} r \pap{n}{r} - 24 e^{2 n} r^2 \pap lr
\pap{n}{r} + 8 r \pap lu \X2(-3 e^{4 l} - 2 e^{2 n} -
12 e^{2 n} r \pap{n}{u} \\
{}+ (4 e^{4 l} + 15 e^{2 n})
r \pap lr - 6 e^{2 n} r \pap{n}{r}\Y2)\Y3) -
4 e^{4 l+2 n} r^3 f \X2(2 e^{2 n} r \frac{d^2f}{du^2} +
\frac{df}{du} \X2(e^{4 l} + e^{2 n} + 22 e^{2 n} r \pap lu \\
{}- 4 e^{2 n} r \pap{n}{u} - 8 e^{4 l} r \pap lr +
2 e^{2 n} r \pap lr + 4 e^{4 l} r \pap{n}{r} +
2 e^{2 n} r \pap{n}{r}\Y2)\Y2) +
4 e^{4 n} r^4 \X3(-2 e^{4 l} \X2(\frac{df}{du}\Y2) \\
{}+ e^{2 n} r \X2(\pap lu + 2 r \X2(\pap lu\Y2)^2 - \pap{n}{u} +
\pap lr - \pap{n}{r} + 2 r \pap{^2l}{r \pa u} -
2 r \pap{^2n}{r \pa u} + r \pap{^2l}{r^2} - r \pap{^2n}{r^2}\Y2)\Y3) \\
{}+ 2 e^{4 l} r^2 f^2 \X3(8 e^{8 l} \X2(\frac{df}{du}\Y2)^2 +
e^{2 n} \X3(2 e^{4 l} + e^{2 n} -
60 e^{2 n} r^2 \X2(\pap lu\Y2)^2 - 12 e^{2 n} r^2 \pap{^2l}{u^2} - 10 e^{4 l} r \pap lr \\
{} - 8 e^{2 n} r \pap lr +
2 e^{4 l} r^2 \X2(\pap lr\Y2)^2 + 10 e^{2 n} r^2 \X2(\pap lr\Y2)^2 +
4 e^{2 n} r \pap{n}{u} \X2(3 + r \pap lr\Y2) +
6 e^{4 l} r \pap{n}{r} + 8 e^{2 n} r \pap{n}{r} \\
{} -2 e^{4 l} r^2 \pap lr \pap{n}{r} - 10 e^{2 n} r^2 \pap lr \pap{n}{r} + 2 r \pap lu
\X2(-e^{4 l} - 5 e^{2 n} + 12 e^{2 n} r \pap{n}{u} +
2 (4 e^{4 l} - 3 e^{2 n}) r \pap lr - 4 e^{4 l} r \pap{n}{r} \\
{}- 6 e^{2 n} r \pap{n}{r}\Y2) +
2 e^{4 l} r^2 \pap{^2l}{r \pa u} + 2 e^{2 n} r^2 \pap{^2l}{r \pa u} -
4 e^{2 n} r^2 \pap{^2n}{r \pa u} + 4 e^{2 n} r^2 \pap{^2l}{r^2} -
2 e^{2 n} r^2 \pap{^2n}{r^2}\Y3)\Y3)\Y4)\\
{}= e^{-2 l}
\X3(\X2(-2 e^{4 n} r^4 + 4 e^{4 l+2 n} r^2 f^2\Y2) \X2(\pap bu\Y2)^2
- 2 e^{2 n} r^2 \X2(e^{2 n} r^2 - 2 e^{4 l} f^2\Y2) \pap bu \pap br \\
{} - \X2(e^{4 n} r^4 + e^{4 l+2 n} r^2 f^2 - 2 e^{8 l} f^4\Y2)
\X2(\pap br\Y2)^2\Y3)
\e

$$
\pap{}t \X1(\ov R_{[13]} - 8\pi\Tem_{[13]}\Y1) + \pap{}r\X1(\ov R_\[34] - 8\pi\Tem_{[34]}\Y1) =0
$$
or
\bml4.36
\pap{}t \X4(
\frac1{8 r^7}\, e^{4 l-8 n}
\X4(4 e^{6 n} r^5 \frac{df}{du} - 8 e^{8 l+2 n} r f^4 \frac{df}{du} +
2 e^{8 l} f^5 \X2(e^{4 l}  + 2 e^{2 n} -
12 e^{2 n} r \pap lu - 6 e^{2 n} r \pap lr\Y2) \\
{}+ 4 e^{4 l+2 n} r^3 f^2 \frac{df}{du} \X2(-3 e^{2 n} + 24 e^{2 n}
r \pap lu - 2 (2 e^{4 l} - 5 e^{2 n}) r \pap lr\Y2) \\
{} +e^{2 n} r^2 f^3 \X3(e^{8 l} + 2 e^{4 n} + 21 e^{4 l+2 n} +
144 e^{4 l+2 n} r^2 \X2(\pap lu\Y2)^2 + 6 e^{8 l} r
\pap lr - 60 e^{4 l+2 n} r \pap lr - 8 e^{8 l} r^2 \X2(\pap lr\Y2)^2 \\
{}- 8 e^{4 n} r^2 \X2(\pap lr\Y2)^2 +
48 e^{4 l+2 n} r^2 \X2(\pap lr\Y2)^2 +
16 e^{4 n} r \pap{n}{u} \X2(-1 + 2 r \pap lr\Y2) \\
{}- 4 r \pap lu \X2(-4 e^{4 n} + 9 e^{4 l+2 n} + 2 (2 e^{8 l} +
4 e^{4 n} - 15 e^{4 l+2 n}) r \pap lr\Y2) -
8 e^{4 n} r \pap nr + 8 e^{4 l+2 n} r
\pap nr + 16 e^{4 n} r^2 \pap lr \pap nr\Y3) \\
{}- 8 e^{4 n} r^4 f \X3(-2 e^{4 l} \X2(\frac{df}{du}\Y2)^2 +
e^{2 n} \X2(1 - 4 r \pap lr + 4 r^2 \X2(\pap lr\Y2)^2 +
2 r \pap lu \X2(-1 + r \pap lr\Y2) \\
{} + \pap{n}{u} \X2(r - 2 r^2 \pap lr\Y2) + r \pap nr -
2 r^2 \pap lr \pap nr - r^2 \pap{^2l}{r\pa u}\Y2)\Y3)\Y4)\\
{}-\frac1{r^3} e^{2 l-4 n} f
\X3(e^{4 l}  f^2 \X2(2 \pap bu + \pap br\Y2)^2 +
e^{2 n} r^2 \pap br \X2(8 \pap bu + 9 \pap br\Y2)\Y3)\Y4)\\
{}+\pap{}r\X4(
\frac1{8 r^7} e^{4 l-8 n}
\X3(4 e^{6 n} r^5 \frac{df}{du} - 8 e^{8 l+2 n} r f^4 \frac{df}{du} +
2 e^{8 l} f^5 \X2(e^{4 l}  + 2 e^{2 n} -
12 e^{2 n} r \pap lu - 6 e^{2 n} r \pap lr\Y2) \\
{}+ 4 e^{4 l+2 n} r^3 f^2 \frac{df}{du} \X2(-5 e^{2 n} + 24 e^{2 n} r \pap lu - 2 (2 e^{4 l}  - 7 e^{2 n}) r \pap lr\Y2) \\
{}- e^{2 n} r^2 f^3 \X3(e^{8 l} + 2 e^{4 n} - 25 e^{4 l+2 n} -
144 e^{4 l+2 n} r^2 \X2(\pap lu\Y2)^2 -
10 e^{8 l} r \pap lr - 16 e^{4 n} r \pap lr + 80 e^{4 l+2 n} r \pap lr \\
{}+ 8 e^{8 l} r^2 \X2(\pap lr\Y2)^2 + 24 e^{4 n} r^2 \X2(\pap lr\Y2)^2 -
72 e^{4 l+2 n} r^2 \X2(\pap lr\Y2)^2 - 16 e^{4 n} r
\pap{n}{u} \X2(-1 + 2 r \pap lr\Y2) \\
{} + 4 r \pap lu \X2(-4 e^{4 n} + 15 e^{4 l+2 n} + 2 \X2(2 e^{8 l} +
4 e^{4 n} - 21 e^{4 l+2 n}\Y2) r \pap lr\Y2) +
8 e^{4 n} r \pap nr - 8 e^{4 l+2 n} r
\pap nr - 16 e^{4 n} r^2 \pap lr \pap nr\Y3) \\
{}- 4 e^{4 n} r^4 f \X3(-4 e^{4 l}  \X2(\frac{df}{du}\Y2)^2 +
e^{2 n} \X2(-1 + 2 r \pap lr - 4 r^2 \X2(\pap lr\Y2)^2 +
2 r \pap{n}{u} \X2(1 - 2 r \pap lr\Y2) +
4 r \pap lu \X2(-1 + r \pap lr\Y2) \\
{}-2 r^2 \pap{^2l}{r\pa u} - 2 r^2 \pap{^2l}{r^2}\Y2)\Y3)\Y3)\\
{}-\frac1{r^3}\, e^{2 l-4 n} f
\X3(e^{2 n} r^2 \X2(8 \pap bu - \pap br\X2) \pap br +
e^{4 l}  f^2 \X2(2 \pap bu + \pap br\Y2)^2\Y3)\Y4) = 0
\e

We can introduce a \pt\ $B=B(r,t)$ \st
$$
\pap Br = \ov R_\[13] - 8\pi\Tem_{[13]} \qh{and} \pap Bt = 8\pi \Tem_{[34]}-\ov R_\[34].
$$
In particular one gets
\bml 4.38
\pap Br = \frac1{8 r^7}\, e^{4 l-8 n}
\X4(4 e^{6 n} r^5 \frac{df}{du} - 8 e^{8 l+2 n} r f^4 \frac{df}{du} +
2 e^{8 l} f^5 \X2(e^{4 l}  + 2 e^{2 n} -
12 e^{2 n} r \pap lu - 6 e^{2 n} r \pap lr\Y2) \\
{}+ 4 e^{4 l+2 n} r^3 f^2 \frac{df}{du} \X2(-3 e^{2 n} + 24 e^{2 n}
r \pap lu - 2 (2 e^{4 l} - 5 e^{2 n}) r \pap lr\Y2) \\
{} +e^{2 n} r^2 f^3 \X3(e^{8 l} + 2 e^{4 n} + 21 e^{4 l+2 n} +
144 e^{4 l+2 n} r^2 \X2(\pap lu\Y2)^2 + 6 e^{8 l} r
\pap lr - 60 e^{4 l+2 n} r \pap lr - 8 e^{8 l} r^2 \X2(\pap lr\Y2)^2 \\
{}- 8 e^{4 n} r^2 \X2(\pap lr\Y2)^2 +
48 e^{4 l+2 n} r^2 \X2(\pap lr\Y2)^2 +
16 e^{4 n} r \pap{n}{u} \X2(-1 + 2 r \pap lr\Y2) \\
{}- 4 r \pap lu \X2(-4 e^{4 n} + 9 e^{4 l+2 n} + 2 (2 e^{8 l} +
4 e^{4 n} - 15 e^{4 l+2 n}) r \pap lr\Y2) -
8 e^{4 n} r \pap nr + 8 e^{4 l+2 n} r
\pap nr + 16 e^{4 n} r^2 \pap lr \pap nr\Y3) \\
{}- 8 e^{4 n} r^4 f \X3(-2 e^{4 l} \X2(\frac{df}{du}\Y2)^2 +
e^{2 n} \X2(1 - 4 r \pap lr + 4 r^2 \X2(\pap lr\Y2)^2 +
2 r \pap lu \X2(-1 + r \pap lr\Y2) \\
{} + \pap{n}{u} \X2(r - 2 r^2 \pap lr\Y2) + r \pap nr -
2 r^2 \pap lr \pap nr - r^2 \pap{^2l}{r\pa u}\Y2)\Y3)\Y4)\\
{}-\frac1{r^3} e^{2 l-4 n} f
\X3(e^{4 l}  f^2 \X2(2 \pap bu + \pap br\Y2)^2 +
e^{2 n} r^2 \pap br \X2(8 \pap bu + 9 \pap br\Y2)\Y3)
\e

\bml4.39
\pap Bt = -\frac1{8 r^7} e^{4 l-8 n}
\X3(4 e^{6 n} r^5 \frac{df}{du} - 8 e^{8 l+2 n} r f^4 \frac{df}{du} +
2 e^{8 l} f^5 \X2(e^{4 l}  + 2 e^{2 n} -
12 e^{2 n} r \pap lu - 6 e^{2 n} r \pap lr\Y2) \\
{}+ 4 e^{4 l+2 n} r^3 f^2 \frac{df}{du} \X2(-5 e^{2 n} + 24 e^{2 n} r \pap lu - 2 (2 e^{4 l}  - 7 e^{2 n}) r \pap lr\Y2) \\
{}- e^{2 n} r^2 f^3 \X3(e^{8 l} + 2 e^{4 n} - 25 e^{4 l+2 n} -
144 e^{4 l+2 n} r^2 \X2(\pap lu\Y2)^2 -
10 e^{8 l} r \pap lr - 16 e^{4 n} r \pap lr + 80 e^{4 l+2 n} r \pap lr \\
{}+ 8 e^{8 l} r^2 \X2(\pap lr\Y2)^2 + 24 e^{4 n} r^2 \X2(\pap lr\Y2)^2 -
72 e^{4 l+2 n} r^2 \X2(\pap lr\Y2)^2 - 16 e^{4 n} r
\pap{n}{u} \X2(-1 + 2 r \pap lr\Y2) \\
{} + 4 r \pap lu \X2(-4 e^{4 n} + 15 e^{4 l+2 n} + 2 \X2(2 e^{8 l} +
4 e^{4 n} - 21 e^{4 l+2 n}\Y2) r \pap lr\Y2) +
8 e^{4 n} r \pap nr - 8 e^{4 l+2 n} r
\pap nr - 16 e^{4 n} r^2 \pap lr \pap nr\Y3) \\
{}- 4 e^{4 n} r^4 f \X3(-4 e^{4 l}  \X2(\frac{df}{du}\Y2)^2 +
e^{2 n} \X2(-1 + 2 r \pap lr - 4 r^2 \X2(\pap lr\Y2)^2 +
2 r \pap{n}{u} \X2(1 - 2 r \pap lr\Y2) +
4 r \pap lu \X2(-1 + r \pap lr\Y2) \\
{}-2 r^2 \pap{^2l}{r\pa u} - 2 r^2 \pap{^2l}{r^2}\Y2)\Y3)\Y3)\\
{}+\frac1{r^3}\, e^{2 l-4 n} f
\X3(e^{2 n} r^2 \X2(8 \pap bu - \pap br\X2) \pap br +
e^{4 l}  f^2 \X2(2 \pap bu + \pap br\Y2)^2\Y3)
\e

In future papers we will try to solve these \e s to get \cy\ \gelm c waves, possibly exactly or numerically.

Let us consider material sources for a \cy ly \s ic \gelm c waves in the \NK{}. In \cy\ coordinates we write an electric current
\beq4.40
j^\nu=\left(\begin{array}{c} j_r \\ j_\t \\ j_z \\ j_t \end{array}\right).
\e
$j_r, j_\t, j_z$ and $j_t$ are \f s of $r$ and $t$ only.

Let us consider a conservation law for an electric charge
\beq4.41
\pa_\mu\bigl(\sqrt{-g}\,j^\mu(r,t)\bigr) = 0.
\e
Let us suppose that $j_\t=0$. Eventually one gets
\beq4.42
\pap{}r J_r(r,t) + \pap{}t J_t(r,t) =0.
\e
where
\bea4.43
J_r(r,t)&= e^{-2l+2n} r j_r(r,t) \\
J_t(r,t)&= e^{-2l+2n} r j_t(r,t). \lb4.44
\e
We introduce a \pt\ $d(r,t)$ \st
$$
J_r=\pap dt\,, \q J_t=-\pap dr\,.
$$
One gets
\beq4.45
j^\mu = \left(\begin{array}{c}
\frac1r\,e^{2l-2n}\pap dt \\ 0 \\ j_z(r,t) \\ -\frac1r\,e^{2l-2n}\,\pap dr
\end{array}\right)
\e
where $d=d(r,t)$ is a \pt. From Eq.\ \er{2.69} one gets
\bg4.46
\pap{}t\biggl(\frac{e^{2l-2n}}{r^2}\,s^2\Bigl(\pap bt+\pap br\Bigr) -d \biggr) = 0 \\
\pap{}r\biggl(\frac{e^{2l-2n}}{r^2}\,s^2\Bigl(\pap bt+\pap br\Bigr) +d \biggr) = 0 \lb4.47 \\
\pap{}r \X3(e^{-2l}r\,\pap br - 3e^{2l-2n}\,\frac {f^2}{r^3}\X2(\pap bt+\pap br\Y2)\Y3)-
\pap{}t \X3(e^{-2l}r\,\pap bt + 3e^{2l-2n}\,\frac {f^2}{r^3}\X2(\pap bt+\pap br\Y2)\Y3) \hskip40pt \nonumber\\
\hskip240pt {}=e^{-2l+2n}rj_z(r,t).\lb4.48
\e
One also gets
\beq4.49
d(r,t)= \frac{\wt A(r)-\wt B(t)}2
\e
where $\wt A(r)$, $\wt B(t)$ are arbitrary \f s of one variable and eventually
\bg4.50
\falj{}_r(r,t) = -\frac12\,\frac{d\wt B}{dt} \\
j_t(r,t) = -\frac12\,\frac{d\wt A}{dr} \lb4.51 \\
\pap{}t\X3(\frac{e^{2l-2n}}{r^4}\,f^2 \,\X2(\pap bt+\pap br\Y2)+\frac12\,\wt B(t)\Y3)=0 \lb4.52 \\
\pap{}r\X3(\frac{e^{2l-2n}}{r^4}\,f^2 \,\X2(\pap bt+\pap br\Y2)+\frac12\,\wt A(r)\Y3)=0. \lb4.53
\e
Thus an electric current has a shape
\beq4.54
j^\mu = \left(\begin{array}{c}
-\frac1{2r}\,e^{2l-2n}\frac{d\wt B}{dt} \\ 0 \\ j_z(r,t) \\ \frac1{2r}\,e^{2l-2n}\,\frac{d\wt A}{dr}
\end{array}\right)
\e
An interacting \Lg\ in this case has the \fw\ shape
\beq4.55
\cL_{\rm int} = j_zb\sqrt{-g}
\e
and
\beq4.56
\nad{\rm int}T_{\!\m} = g_\m (j_zb).
\e
If we suppose that $j_z=0$
\bg4.57
\nad{\rm eff}T_{\!\m}=\Tem_\m \\
j^\mu = \left(\begin{array}{c}
-\frac1{2r}\,e^{2l-2n}\frac{d\wt B}{dt} \\ 0 \\ 0 \\ \frac1{2r}\,e^{2l-2n}\,\frac{d\wt A}{dr}
\end{array}\right). \lb4.58
\e

The simplest example of an \ia\ of \elm c and \gr al (skewon-\gr al) waves in the \NK{} gives us a possibility to generate \gelm c wave by an
external source. The \fw\ \e s give us in some sense a theory of such a generator.
In this case our \e s (the second pair of Maxwell \e s) look as follows:
\bg4.59
\pap{}t\X3(\frac{e^{2l-2n}}{r^4}\,f^2 \,\X2(\pap bt+\pap br\Y2)+\frac12\,\wt B(t)\Y3)=0  \\
\pap{}r\X3(\frac{e^{2l-2n}}{r^4}\,f^2 \,\X2(\pap bt+\pap br\Y2)+\frac12\,\wt A(r)\Y3)=0 \lb4.60 \\
\pap{}r \X3(e^{-2l}r\,\pap br - 3e^{2l-2n}\,\frac {f^2}{r^3}\X2(\pap bt+\pap br\Y2)\Y3)-
\pap{}t \X3(e^{-2l}r\,\pap bt + 3e^{2l-2n}\,\frac {f^2}{r^3}\X2(\pap bt+\pap br\Y2)\Y3) =0 \lb4.61
\e
where as usual
$$
f=f(r-t), \q b=b(r,r-t), \q l=l(r,r-t), \q n=n(r,r-t).
$$

Moreover a more interesting case can be derived if we suppose that $\wt A(r)=\wt B(t)=0$ and $j_z(r,t)\ne0$ in such a way that
\bg4.62
j_z(r,t) = \wt c(r,r-t) \\
j^\mu = \left( \begin{array}{c} 0\\ 0 \\ \wt c(r,r-t) \\ 0 \end{array}\right).
\e
In this case one gets
\bg4.64
\pap{}t\X3(\frac{e^{2l-2n}}{r^4}\,f^2 \,\X2(\pap bt+\pap br\Y2)\Y3)=0  \\
\pap{}r\X3(\frac{e^{2l-2n}}{r^4}\,f^2 \,\X2(\pap bt+\pap br\Y2)\Y3)=0 \lb4.65
\e
and
\bml4.66
\pap{}r \X3(e^{-2l}r\,\pap br - 3e^{2l-2n}\,\frac {f^2}{r^3}\X2(\pap bt+\pap br\Y2)\Y3)-
\pap{}t \X3(e^{-2l}r\,\pap bt + 3e^{2l-2n}\,\frac {f^2}{r^3}\X2(\pap bt+\pap br\Y2)\Y3) \\
{}=e^{-2l+2n}r\wt c(r,r-t).
\e
One can even design an experiment to check whether we are right or not in \gelm c waves generation problem. Moreover we should pass in all \e s
\er{4.30}--\er{4.39} from $\Tem_\m$ to
\beq4.67
\nad{\rm eff}T_{\!\m} = \Tem_m + g_\m \X1(\wt c(r,r-t)b(r,r-t)\Y1)
\e
where $g_\m$ is given by Eq.\ \er{4.1}.

One gets:
\beq4.68
T_\m = \left(\begin{array}{c|c|c|c}
\hsm\ -e^{2l-2n}\wt cb\ & 0 &\ se^{2n}\wt c b\ & 0 \\\hline
\hsm 0 &\ -r^2e^{-2n}\wt c b\ & 0 & 0 \\\hline
\hsm -se^{2n}\wt c b & 0 & -e^{-2n}\wt c b & se^{2n}\wt c b \\\hline
\hsm 0 & 0 &\ -se^{2n}\wt c b\ &\ e^{2l-2n}\wt c b \
\end{array}\right).
\e

Let us consider an external \elm c (it can be in reality e.g.\ a laser field) field \st
\beq4.69
\pa_\mu F^\m_{\rm ex}\ne0
\e
and let $j_z(r,t)$ be a source of this field
\beq4.70
\pa_\mu F^\m_{\rm ex}=j_z(r,t).
\e
Let \setbox0=\hbox{$\dsp\hphantom{-\pap ht}$}
\def\pud#1{\hbox to\wd0{\hfil#1\hfil}}
\beq4.71
F^\m_{\rm ex} = \left(\begin{array}{c|c|c|c}
\hsd 0 & \pud 0 & \dsp\pap hr & 0 \\\hline
\hsd 0 & 0 & \pud 0 & \pud 0 \\\hline
\hsd \dsp-\pap hr & 0 & 0 & \dsp-\pap ht \\\hline
\hsd 0 & \pud 0 & \dsp\pap ht & 0
\end{array}\right), \qquad h=h(r,r-t).
\e
One gets in the place of Eq.\ \er{4.66}
\bml4.72
\pap{}r\X3(e^{-2l}r\,\pap br - 3e^{2l-2n}\,\frac{f^2}{r^3}\X2(\pap bt+\pap br\Y2)\Y3)
-\pap{}t\X3(e^{-2l}r\,\pap bt + 3e^{2l-2n}\,\frac{f^2}{r^3}\X2(\pap bt+\pap br\Y2)\Y3)\\
{}=e^{-2l+2n}r\X2(\pap{^2h}{r^2} + \pap{^2h}{t^2}\Y2).
\e
Eqs \er{4.64} and \er{4.65} do not change.

Let us consider an initial value problem for Eqs \er{4.64}--\er{4.65} and Eq.~\er{4.72}.
In this way we should consider an initial value problem for a metric tensor $g_\m$ given by Eq.~\er{4.1}.
The best way to do this is to take $g_{tt}$ in \cy\ \cd s from spherically \s ic \so\ in NGT
\beq4.73
\X3(1+\frac{\wt l{}^4}{(r^2+z^2)^2}\Y3)\X3(1-\frac{2m}{(z^2+r^2)^{1/2}}\Y3)
\e
($m$ is a mass and $\wt l$ is an additional parameter) and to equal it to $g_{tt}|_{t=0}$ of the metric \er{4.1}, i.e.
\beq4.74
\X3(1+\frac{\wt l{}^4}{(r^2+z^2)^2}\Y3)\X3(1-\frac{2m}{(z^2+r^2)^{1/2}}\Y3) = e^{2l-2n}\big|_{t=0}\,.
\e
Supposing $|z|\ll r$ one gets
\beq4.75
\X2(1+\frac{\wt l{}^4}{r^4}\Y2)\X2(1-\frac{2m}r\Y2) = e^{2(l-n)} \big|_{t=0}\,.
\e
In this case we do not add any energy-momentum tensor to $\Tem_{\a\b}$ and $\nad{\rm eff}T_{\!\a\b} = \Tem_{\a\b}$.

We suppose that external \elm c field does not change a geometry of \spt. In this way Eqs \er{4.29}--\er{4.36} are considered as consistency
conditions for \e s of a generator of gravito-\elm c (\elm c-\gr al) waves.

In some applications one can suppose
\beq4.76
\X2(1+\frac{\wt l{}^4}{r^4}\Y2) = e^{2l} \big|_{t=0}\,, \q \X2(1-\frac{2m}r\Y2) e^{-2n} \big|_{t=0}\,.
\e

\section{A \gz\ \KW \e\ as a possibility to travel in higher \di s}
Let us consider \gz\ \KW \e\ in the \NK{} for GSW (Glashow--Salam--Weinberg) model, i.e.\ Eqs (C.49)--(C.51) and Eqs (C.53)--(C.58) from Ref.~\cite8.
Let us quote those \e s here.

\begin{equation}\label{imp}
\gathered
\frac{\wt{\ov D}u^\mu}{d\tau} - \frac{\wt Q{}^{iu}}{m_0}\wt g{}^\(\mu\d) \d_{ij}u^\b W_{\b\d}^{iu} - \frac{Q^{0u}}{m_0}\,\wt
g{}^\(\mu\d)u^\b Z_{\b\d}^{0u} -\frac q{m_0}\,\wt g{}^\(\mu\d)u^\b F_{\b\d}\hskip100pt \\
{}+\frac1{\sqrt2\,m_0}\,\wt
g{}^\(\mu\d)\biggl(u^5\Bigl(\frac1{\a\cdot\a} (q_{-\a}W_\d^{-u} {-} q_\a W_\d^{+u}) + \frac1{\b\cdot\b}\bigl( h(x_\a,x_\a)q_\a W_\d^{-u}
+h(x_{-\a},x_{-\a})q_{-\a}W_\d^{+u}\bigr) \Bigr) \\
\kern-19pt{}+u^6 \sin\psi\Bigl(\frac{v{+}H(x)}{\a\cdot\a}(q_\a W_\d^{+u}{-}q_{-\a}W_\d^{-u}
)-\frac1{2(\b\cdot\b)}\bigl(h(x_\a,x_\a)q_{-\a}(v{+}H(x))\bigr)W_\d^{+u}\Bigr) \biggr)=0\kern-3pt  \\
\frac{\wt Du^5}{d\tau} -
\frac1{r^2}\,\frac{u^\b}{\sqrt2\,m_0} \Bigl( \frac1{\a\cdot\a}(q_{-\a}W_\b^{-u} - q_\a W_\b^{+u})\hskip80pt \\
\hskip80pt
{}+\frac1{\b\cdot\b}\bigl(h(x_\a,x_\a)q_\a W_\b^{-u} + h(x_{-\a},x_{-\a}) q_{-\a}W_\b^{+u}\bigr)\Bigr)=0  \\
\frac{\wt Du^6}{d\tau} - \frac
1{r^2}\,\frac{u^\b}{\sqrt2\,m_0\sin\psi}\Bigl( \frac{v+H(x)}{\a\cdot\a}(q_\a W_\b^{+u} - q_{-\a}W_\b^{-u})\hskip80pt \\
\hskip80pt
{}+\frac1{2(\b\cdot\b)}\bigl(-h(x_\a,x_\a)q_\a W_\b^{-u} + h(x_{-\a},x_{-\a}) (v+H(x))W_\b^{+u}\bigr)\Bigr)=0
\endgathered
\end{equation}

\begin{equation}
\gathered
\frac{dQ_\g^u}{d\tau} - i\biggl(\frac12(Z_\mu^{0u}\sqrt3 - A_\mu)Q_\g^u -W_\mu^{-u}\wt q\biggr)u^\mu \hskip100pt \\
\hskip100pt {}-
\frac{3(v+H(x))}{\sqrt2}\,q_\a \Bigl(u^5+i\,\frac{\sin\psi}2\,u^6\Bigr) - i\cos\psi Q_\g^u=0\\
\frac{dQ_{-\g}^u}{d\tau} - iu^\mu
\Bigl(W_\mu^{+u}\wt Q{}^u -\frac12(\sqrt3\,Z_\mu^{0u}-A_\mu)Q_{-\g}^u\Bigr)\hskip100pt \\
\hskip100pt
{}+\frac{3u^5}{\sqrt2}(v+H(x))q_{-\a}\Bigl(u^5-i\,\frac{\sin\psi} 2\, u^6\Bigr) + iu^6\cos\psi Q_{-\g}^u=0 \\
\frac{dq}{d\tau}=0 \\
\frac{d\wt Q{}^u}{d\tau} - \frac12 (W_\mu^{+u}Q_\g^u - W_\mu^{-u}Q_{-\g}^u)u^\mu=0  \\
\frac{dq_\a}{d\tau} + \frac12\Bigl(\frac32(\sqrt3\,A_\mu -
Z_\mu^{0u}) -\frac{\sqrt3}6 (\sqrt3\,Z_\mu^{0u}-A_\mu)\Bigr)q_\a u^\mu \hskip100pt \\
\hskip100pt {}+\frac{u^5(v+H(x))}{\sqrt2}\,Q_\g^u \Bigl( u^5
+ \frac{i\sin\psi}{2\sqrt2}\,u^6\Bigr) -\frac 12\,u^6\cos\psi q_\a=0  \\
\frac{dq_{-\a}}{d\tau} + \frac 12\Bigl(\frac{\sqrt3}6(\sqrt3\,A_\mu -
Z_\mu^{0u}) + \frac32(\sqrt3\,Z_\mu^{0u} - A_\mu)\Bigr) u^\mu q_{-\a} \hskip100pt \\
\hskip100pt {}+\frac{v+H(x)}{\sqrt2}\,Q_{-\g}^u \Bigl(u^5 -
\frac{i\sin\psi}2 \,u^6\Bigr)+\frac 12 \cos\psi q_{-\a}=0.
\endgathered \tag{$5.1_{\rm cont}$}
\end{equation}

All the details concerning the \NK{} (in general) with and without \ssb\ can be found in Ref.~\cite8. Some applications for GSW model of the theory
can be found also here. Let us make some simplification of the \KW \e\ in a particular application in GSW model.

Let $F_{\b\g}=0$ ($A_\mu=0$), and $Z^{0u}_{\b\g}=0$ ($Z^{0u}_\mu=0$). Simultaneously let some gauge charges be equal to zero,
$q=\wt Q{}^{iu}=\wt Q{}^{0u}=0$ (i.e.\ an electric charge and weak \ia\ charges) and also $\gd Q,u,\g,=\gd Q,u,-\g,=q_{-\a}=0$ (i.e.\ some charges
connected to Higgs' field). Let us remind to the reader that Higgs' field in our approach is also a gauge field, but over additional \di s,
in particular for GSW model over fifth and sixth \di s. In this case additional \di s are \di s of~$S^2$ sphere ($\psi$~and~$\vf$, $\psi\in
\langle-\frac\pi2,\frac\pi2\rangle$, $\vf\in\langle0,2\pi)$).

One gets the following set of \e s:
\bg5.1
{}\ddf{^2x^\mu}{\tau^2} + \biggl\{{\mu\atop\b\g}\biggr\} \ddf{x^\b}\tau\,\ddf{x^\g}\tau + \frac{q_\a}{2\sqrt2\,m_0}\,\wt g^\(\mu\d)
\biggl(\ddf\vf\tau\, h(x_\a,x_\a)\gd W,-u,\d, - \gd W,+u,\d,\biggr)\hskip70pt \nonumber\\
\hskip130pt {}+ \ddf\vf\tau \sin\psi(v+H(x))\gd W,+\mu,\d,\biggr)=0 \lb5.1a \\
\ddf{^2\psi}{\tau^2}+\frac12\sin2\psi\X3(\ddf\vf\tau\Y3)^2 - \frac{q_\a}{2\sqrt2\,r^2m_0}\X3(\ddf{x^\b}{\tau}\Y3)
\bigl(\gd W,+u,\b,+h(x_\a,x_\a)\gd W,-u,\b,\bigr)=0  \lb5.2 \\
\ddf{^2\vf}{\tau^2}+2\cot\vf\X3(\ddf \psi\tau\Y3)\X3(\ddf\vf\tau\Y3) \hskip170pt \nonumber \\
\hskip70pt {}- \frac{q_\a}{2\sqrt2\,m_0\sin\psi}\X3(\ddf{x^\b}t\Y3)
\bigl((v+H(x))\gd W,+u,\b, - \frac12h(x_\a,x_\a)\gd W,-u,\b,\bigr)=0 \lb5.3 \\
\ddf{q_\a}\tau + \frac12\X3(\ddf\vf\tau\Y3)\cos \psi q_\a=0. \lb5.4
\e
In the \e s from \cite8 we have
\bg5.5
u^5 = \ddf\psi\tau, \q u^6=\ddf\vf\tau\\
\frac{\wt Du^5}{d\tau}= \ddf{^2x^5}{\tau^2} + \wt\G{}^5_{\td a\td b}\,\ddf{x^{\td a}}\tau \, \ddf{x^{\td b}}\tau =
\ddf{^2\psi}{\tau^2} + \wt\G{}^5_{66}\X3(\ddf\vf\tau\Y3)^2 = \ddf{^2\psi}{\tau^2}+\frac12\sin2\psi\X3(\ddf\vf\tau\Y3)^2 \lb5.6 \\
\frac{\wt Du^6}{d\tau}= \ddf{^2x^6}{\tau^2}+\wt\G{}^6_{\td a\td b}\,\ddf{x^{\td a}}\tau\,\ddf{x^{\td b}}\tau = \ddf{^2\vf}{\tau^2}
+2\wt\G{}^6_{56}\X3(\ddf\psi\tau\Y3)\X3(\ddf\vf\tau\Y3) = \ddf{^2\vf}{\tau^2}+2\cot\psi\X3(\ddf\psi\tau\Y3)\X3(\ddf\vf\tau\Y3) \lb5.7 \\
\frac{\wt{\ov D}u^\a}{d\tau} = \ddf{^2x^\a}\tau + \X3\{{\a\atop\b\g}\Y3\}\ddf{x^\b}\tau\, \ddf{x^\g}\tau \lb5.8 \\
u^\a=\ddf{x^\a}\tau \nonumber
\e
where $\{{\a\atop\b\g}\}$ are Christoffel's symbols formed for $g_\(\a\b)$, $\gd W,\pm u,\g,$ are charged weak \ia s gauge fields in the gauge~$U$,
$H(x)$ is a Higgs' field after a \ssb, $v$~is a vacuum expectation value of a Higgs' field, $r$~is the radius of the sphere $S^2$. Let us notice
the following fact: a~charge $q_\a$ is a reason to connect the motion on \spt\ $E$ to higher \di\ on~$S^2$, $m_0$~is a mass of a~test \pc.
Let us remind to the reader the following fact: Eqs \er{imp} are \e s of a test \pc\ equipped with many
charges in our theory.

After some simplifications given here we get \e s in six \di s on $E\times S^2$ (not four as on~$E$). Additional \di s, fifth and sixth, are \di s
on~$S^2$. Moreover, the radius $r$ of~$S^2$ is very small ($r\approx2.39\cdot10^{-18}$~m).

The set of those \e s gives us a possibility of travelling in higher \di s due to the existence of Higgs' field and a charge $q_\a$ connected to it.
This is in some sense a~Lorentz force term involving Higgs' field. For a~Higgs' field is a gauge field over fifth and sixth \di s, the force
involves higher \di s. Such a situation can give a possibility to go to distant points in space crosspassing in short (shortcut),
travelling in time, or to go out from compact,
3-\di al objects without crossing boundaries. Such \so s of Eqs \er{5.1a}--\er{5.4} are under considerations. One easily gets from Eq.~\er{5.4}
\bg5.9
q_\a = C_\a\exp\X4(-\frac12 \int_{\tau_0}^\tau \X3(\ddf\vf\tau\Y3)\cos\psi\,ds \Y4) \\
C_\a = q_\a(\tau_0) = {\rm const.}, \q C_\a>0 \lb5.10
\e
If $C_\a=0$, an effect of higher \di s disappears.

Let us consider a more general situation where not only $q_\a\ne0$, but also $q_{-\a}\ne0$. In this case from Eqs~\er{imp} one gets supposing
as before $Z_\mu^{0u}=Z_{\b\g}^{0u}=F_{\b\g}=A_\mu=q=\gd Q,u,\g, =\gd Q,u,-\g,=0$:
\bml5.11
\X3(\ddf{^2x^\mu}{\tau^2} + \X3\{{\mu\atop\b\g}\Y3\}\ddf{x^\b}\tau \,\ddf{x^\g}\tau \Y3)
+\frac1{\sqrt2\,m_0} \,\wt g{}^\(\mu\d) \X3(\ddf\psi\tau \Y3)\frac12(q_{-\a}W^{-u}_\d - q_\a W^{+u}_\d)\\
{}+\frac12\X1(h(x_\a,x_\a)q_\a W^{-u}_\d - h(x_\a,x_{-\a})q_{-\a} W^{+u}_\d\Y1) \\ {}+ \ddf\vf\tau \sin\psi
\X3(\frac{v+H(x)}2 (q_aW^{+u}_\d - q_{-\a}W^{-u}_\d ) - \frac14\X1(h(x_\a,x_\a)q_{-\a}(v+H(x))W^{+u}_\d\Y1)\Y3)=0
\e
\bml5.12
\ddf{^2\psi}{\tau^2} + \frac12\sin 2\psi \X2(\ddf\psi\tau\Y2)^2 - \frac1{r^2}\X2(\ddf{x^\b}\tau\Y2)^2\frac1{\sqrt2\,m_0}
\X3(\frac12(q_{-\a}W^{-u}_\b + q_\a W^{+u}_\b)\\
{} + \frac12\X1(h(x_\a,x_\a)q_\a W^{-u}_\b + h(x_{-\a},x_{-\a})q_{-\a}W^{+u}_\b\Y1)\Y3)=0
\e
\bml5.13
\ddf{^2\vf}{\tau^2}+2 \cot\vf\X2(\ddf\psi\tau\Y2)\X2(\ddf\vf\tau\Y2) - \frac1{r^2}\X2(\ddf{x^\b}\tau\Y2)
\frac1{\sqrt2\,m_0\sin\psi} \X3(\frac{v+H(x)}2 (q_\a W^{+u}_\b - q_{-\a}W^{-u}_\b)\\
{}+\frac14\X1(-h(x_\a,x_\a)q_\a W^{-u}_\b + h(x_{-\a},x_{-\a})q_\a(v+H(x))W^{+u}_\b\Y1)\Y3)=0
\e
\beq5.14
\ddf{q_{-\a}}\tau - \frac12\cos\psi q_{-\a}=0
\e
and Eqs \er{5.9}--\er{5.10} for $q_\a$. From Eq.~\er{5.14} one easily gets
\beq5.15
q_{-\a} = C_{-\a}\exp\biggl(\frac12\int_{\tau_0}^\tau \cos\psi\,ds\biggr)
\e
where $C_{-\a}=q_{-\a}(\tau_0)={\rm const.}$, $C_{-\a}>0$. If $C_{-\a}=C_\a=0$, an effect of higher \di s disappears.

Let us notice that such a physics was considered in~\cite{x,x1,x2}. Here we give mathematical and physical reasons to realize such situations.

Let us consider a very special situation where $C_\a=C_{-\a}=0$, i.e.\ $q_\a=q_{-\a}=0$. In this case one gets the following set of \e s
from Eqs \er{5.11}--\er{5.13}:
\bg5.17
\ddf{^2x^\a}{\tau^2}+\X2\{{\a\atop\b\g}\Y2\}\ddf{x^\b}\tau \, \ddf{x^\g}\tau =0\\
\ddf{^2\psi}{\tau^2}+\frac12\sin\psi \X2(\ddf\psi\tau\Y2)^2=0 \lb5.18 \\
\ddf{^2\vf}{\tau^2} + 2\cot\psi \X2(\ddf\psi\tau\Y2)\X2(\ddf\vf\tau\Y2)=0. \lb5.19
\e
Eq.\ \er{5.17} is a geodetic \e\ on~$E$ with a metric tensor $g_\(\a\b)$. Eqs \er{5.18}--\er{5.19} are geodetic \e s on~$S^2$. The last can easily
be solved and we get
\bea5.20
\psi(t)&=\psi_0={\rm const.} \\
\vf(t)&=a(\tau-\tau_0) \lb5.21
\e
where $a$ and $\tau_0$ are constants.

In the case where all the charges are zero the movements on~$E$ and on~$S^2$ decouple. Those movements also decouple in the case of non-zero
electric charge~$q$. Moreover, if the charges $\wt Q{}^{iu}, \wt Q{}^{0u}, Q^u_\g, Q^u_{-\g}$ are non-zero the coupling between movement on~$E$
and on~$S^2$ is evident. Let us notice that a value of a radius of a sphere enters the \e s.

Let $g_\(\a\b)=\eta_{\a\b}={\rm diag}(-1,-1,-1,+1)$ (a Minkowski space) and let $\psi$ and~$\vf$ be described by Eqs \er{5.20}--\er{5.21}. Let us
consider Eq.~\er{5.11}. One gets
\beq5.22
\ddf{^2x^\mu}{\tau^2}+a \sin\psi_0\, \eta^{\mu\d}\X3(\frac{v+H(x)}2(q_\a W^{+u}_\d -q_{-\a}W^{-u}_\d)
- \frac14\X1(h(x_\a,x_\a)q_{-\a}(v+H(x))\Y1)W^{+u}_\d\Y3).
\e
Moreover,
\bea5.23
q_\a &= C_\a \exp\X2(-\frac12\,a(\tau-\tau_0)\cos\psi_0\Y2) \\
q_{-\a} &= C_{-\a} \exp\X2(\frac12(\tau-\tau_0)\cos\psi_0\Y2). \lb5.24
\e
In this way one gets
\bml5.25
\ddf{^2x^\mu}{\tau^2} + a\sin\psi_0 \eta^{\mu\d} \X3(\frac{v+H(x)}2 \X2(C_\a\exp\X2(-\frac12\,a(\tau-\tau_0)\cos\psi_0\Y2)W^{+u}_\d\\
{}-C_{-\a}\exp\X2(\frac12(\tau-\tau_0)\cos\psi_0\Y2)W_\d^{-u}\Y2)\\
{}- \frac14\X2(h(x_\a,x_\a)C_{-\a}\exp\X2(\frac12(\tau-\tau_0)\Y2)(v+H(x))\Y2)W^{+u}_\d\Y3).
\e

Eqs \er{5.25} and \er{5.20}--\er{5.21} govern a motion in higher \di s under some assumptions. We can simplify Eq.~\er{5.25} even more
taking $H(x)=0$ and $W^{-u}_\d=0$. One gets
\beq5.26
\ddf{^2x^\mu}{\tau^2} +\frac{av}2 \sin\psi_0 \eta^{\mu\d}\X2(C_\a\exp\X2(-\frac12\,a(\tau-\tau_0)\cos\psi_0\Y2)W^{+u}_\d
-\frac12\,C_{-\a}\exp\X2(\frac12(\tau-\tau_0)\Y2)W^{+u}_\d\Y2)=0.
\e
If $C_{-\a}=0$ ($q_{-\a}=0$) one gets even more simple
\beq5.27
\ddf{^2x^\mu}{\tau^2}+\frac{avC_\a}2 \exp\X2(-\frac12\,a(\tau-\tau_0)\cos\psi_0\Y2) \eta^{\mu\d}W^{+u}_\d=0
\e
and Eqs \er{5.20}--\er{5.21}.

Let $W^{+u}_\d={\rm const}$. One gets
\bg5.28
\ddf{^2x^i}{\tau_2} - \frac{avC_\a}2 \exp\X2(-\frac a2(\tau-\tau_0)\cos\psi_0\Y2)W^{+u}_i =0, \q i=1,2,3,\\
\ddf{^2x^4}{\tau^2} + \frac{avC_\a}2 \exp\X2(-\frac a2(\tau-\tau_0)\cos\psi_0\Y2)W^{+u}_4 =0. \lb5.29
\e
These \e s are easy to be solved exactly.
\bg5.30
x^i(\tau)=\a^i (\tau-\tau_0) + \frac{2vC_\a W^{+u}_i}{a\cos^2\psi_0}\exp\X2(-\frac a2\cos\psi_0(\tau-\tau_0)\Y2), \q i=1,2,3, \\
x^4(\tau)=\a^4 (\tau-\tau_0) - \frac{2vC_\a W^{+u}_4}{a\cos^2\psi_0}\exp\X2(-\frac a2\cos\psi_0(\tau-\tau_0)\Y2), \lb5.31
\e
where $\a^i$, $i=1,2,3$, and $\a^4$ are constants.

$\tau$ is here of course a proper time. A \cd\ time is $t_1=x^4$.

Thus we should solve a transcendent \e
\beq5.32
t_1=\a^4(\tau-\tau_0) - \frac{2vC_\a W^{+u}_4}{a\cos^2\psi_0} \exp\X2(-\frac a2 \cos\psi_0(\tau-\tau_0)\Y2)
\e
getting
\bea5.33
\tau &= f(t_1)+\tau_0, \\
\wt x{}^i(t_1)&= x^i(f(t_1)+\tau_0), \q i=1,2,3, \lb5.34 \\
\wt\psi(t_1) &= \psi_0, \lb5.35 \\
\wt\vf(t_1) &=a f(t_1), \lb5.36
\e
where $f(t)$ is a \so\ of the transcendental \e:
\beq5.37
t = \a^4 f(t_1) - \frac{2vC_\a W^{+u}_4 }{a\cos^2\psi_0} \exp\X2(-\frac a2\cos\psi_0 f(t_1)\Y2).
\e
Taking for simplicity $\a^4=0$ one finds
\beq5.38
f(t_1)= - \frac{2}{a\cos\psi_0} \log\X3(-\frac{a\cos\psi_0 t_1}{2vC_\a W^{+u}_4}\Y3)
\e
under the condition
\beq5.39
\frac{a\cos\psi_0 t_1}{2vC_\a W^{+u}_4}<0.
\e
In this case one gets
\bg5.40
\wt x{}^i(t_1) = -\frac{2\a^i}{a\cos\psi_0} \log\X3(-\frac{a\cos\psi_0t_1}{2vC_\a W^{+u}_4}\Y3) - \frac{W^{+u}_i}{W^{+u}_4}\,t_1,
\q=1,2,3, \\
\wt\psi(t_1) = \psi_0 \lb5.41 \\
\wt \vf(t_1) = -\frac2{\cos\psi_0}\log \X3(-\frac{a\cos\psi_0t_1}{2v C_\a W^{+u}_4}\Y3) \lb5.42
\e
under the condition \er{5.39}.

Eqs \er{5.40}--\er{5.42} give us a trajectory of a travel in higher \di s on $E\times S^2$ in a \f\ of a \cd\ time on~$E$ (a~Minkowski space).

Let us identify $\wt x{}^1=x$, $\wt x{}^2=y$, $\wt x{}^3=z$; $(x,y,z)$---a Cartesian \cd\ system in~$E^3$. Let us introduce a Cartesian \cd\ system in the 3-\di al
$\mathbb R^3$ space where our $S^2$ sphere is located.
\beq5.43
\bal
\wt \xi&= r\cos\psi \sin\vf\\
\wt\eta &= r\cos\psi \cos\vf\\
\wt\z &= r\sin\psi.
\eal
\e
In this way our trajectory looks like
\beq5.44
\bal
x(t_1) &=  -\frac{2\a_x}{a\cos\psi_0} \log\X3(-\frac{a\cos\psi_0t_1}{2vC_\a W^{+u}_4}\Y3) - \frac{W^{+u}_x}{W^{+u}_4}\,t_1\\
y(t_1) &=  -\frac{2\a_y}{a\cos\psi_0} \log\X3(-\frac{a\cos\psi_0t_1}{2vC_\a W^{+u}_4}\Y3) - \frac{W^{+u}_y}{W^{+u}_4}\,t_1\\
z(t_1) &=  -\frac{2\a_z}{a\cos\psi_0} \log\X3(-\frac{a\cos\psi_0t_1}{2vC_\a W^{+u}_4}\Y3) - \frac{W^{+u}_z}{W^{+u}_4}\,t_1\\
\wt\xi(t_1) &= -r\cos\psi_0 \sin\X3(\frac2{\cos\psi_0}\log\X3(-\frac{a\cos\psi_0t_1}{2vC_\a W^{+u}_4}\Y3)\Y3)\\
\wt\eta(t_1) &= r\cos\psi_0 \cos\X3(\frac2{\cos\psi_0}\log\X3(-\frac{a\cos\psi_0t_1}{2vC_\a W^{+u}_4}\Y3)\Y3) \\
\wt\z(t_1) &= r\sin\psi_0,
\eal
\e
where $r\approx 2.39\cdot 10^{-18}$ m is a scale of GSW-model,
$$
\a^1=\a_x,\ \a^2=\a_y,\ \a^3=\a_z,\ W^{+u}_1=W^{+u}_x,\ W^{+u}_2=W^{+u}_y,\ W^{+u}_3=W^{+u}_z,
$$
under the condition \er{5.39}.

Let us take the \fw\ transformation for a time. Let
\bea5.40a
t&= \frac2{\cos\psi_0}\log\X3(-\frac{a\cos\psi_0t_1}{2vC_\a W^{+u}_4}\Y3), \\
t_1&= -\frac{2vC_\a W^{+u}_4}{a\cos\psi_0} \exp\X2(\frac{\cos\psi_0}2\,t\Y2). \lb5.41a
\e
In this way one gets
\bea5.42a
x(t)&=-\frac{\a_x}a\,t + \frac{2vC_\a W^{+u}_x}{a\cos\psi_0}\exp\X2(\frac{\cos\psi_0}2\,t\Y2)\\
y(t)&=-\frac{\a_y}a\,t + \frac{2vC_\a W^{+u}_y}{a\cos\psi_0}\exp\X2(\frac{\cos\psi_0}2\,t\Y2) \lb5.43a \\
z(t)&=-\frac{\a_z}a\,t + \frac{2vC_\a W^{+u}_z}{a\cos\psi_0}\exp\X2(\frac{\cos\psi_0}2\,t\Y2) \lb5.44a \\
\wt\xi(t) &= -r\cos\psi_0 \sin t \lb5.45 \\
\wt\eta(t) &= r\cos\psi_0 \cos t \lb5.46 \\
\wt \z(t) &= r\sin\psi_0. \lb5.47
\e

Let us consider the condition \er{5.39} in more details
\beq5.48
\ve \frac{|a|\cdot|{\cos\psi_0}|\cdot t_1}{2|v|\cdot|C_\a|\cdot|W^{+u}_4}| < 0
\e
where
\beq5.49
\ve=\sgn(a)\sgn(\cos\psi_0)\sgn(v)\sgn(C_\a)\sgn(W^{+u}_4).
\e
One gets
\beq5.50
\ve t_1 <0 .
\e
If $\ve>0$, $t_1<0$; if $\ve<0$, $t_1>0$.

The trajectory \er{5.40a}--\er{5.47} is a trajectory in six \di s. Moreover, an additional movement in higher \di s is on a sphere~$S^2$
embedded in~$E^3$. We can get more complicated trajectory if we disturb the \so\ \er{5.20}--\er{5.21}, allowing in Eqs \er{5.12}--\er{5.13}
some additional terms, i.e.
\bea5.51
&\ddf{^2\psi}{\tau^2} + \frac12\sin\psi \X2(\ddf \psi\tau\Y2)^2 - \frac{q_\a}{2\sqrt2\, m_0r^2}\,W^{+u}_\b \X2(\ddf{x^\b}\tau\Y2)=0\\
&\ddf{^2\vf}{\tau^2}+2\cot\psi \X2(\ddf\psi\tau\Y2)\X2(\ddf\vf\tau\Y2) - \frac{q_\a v(2+h(x_{-\a},h_{-\a}))}
{4\sqrt2\,m_0r^2\sin\psi}\,W^{+u}_\b \X2(\ddf{x^\b}\tau\Y2)=0 \lb5.52
\e
and Eqs \er{5.9}--\er{5.10}.
In this case we expect some additional effects of impossible physics. We will consider such possibility in further papers.

\def\labelenumi{{\rm\arabic{enumi}$^\circ$}}
\section*{Conclusions and prospects for further research}
In the paper we consider \as\ and \sy\ fields in the \NK{}, i.e.\  a \nos\ metric (\s ic metric and skew\s ic part---skewon part) and an \elm c
field. We derive \e s of the theory using symbolic manipulation programme written in \ti{Mathematica}. There are some prospects for future research:
\ben
\item Solve these \e s exactly or numerically.
\item Extend this formalism for \nA\ \NK{} also with \ssb\ and \Hm. In this case we should consider also spherically \s ic and \sy\ \so s,
cylindrical and spherical \YM' waves.
\een

We consider also in the paper a \cy\ \gelm c wave in the \NK{}. Why it is called \gelm c wave is explained in Ref.~\cite7 and we do not repeat our
explanation in details here. There are some prospects for future research.

\ben
\item Find exact (or numerical) \so s for field \e s.
\item Extend a formalism to spherical \gelm c wave with derivations of exact (or numerical \so s).
\een

For \gr al waves have been detected (three events: GW150914, LVT151012 (a~lower significance), GW151226)
by LIGO-Virgo interferometers (see Refs \cite{alfa}, \cite{beta}, \cite{a}, \cite{b}), the existence of \gelm c waves is highly
probable. We concentrate in future investigations on a possible detection of these waves.

Let us notice the \fw\ fact. One considers a notion of \elm c-\gr al waves earlier as a~conclusion of ``interference effects'' in the \NK{\JT}.

In Ref.~\cite1, p.~289, point 3) one writes in the \fw\ way on a further prospect of research:
{\sl To find wave-like \so s of the field \e s in the Abelian and \nA\ cases. This could, in the \elm c case offer a \so\ which could be treated as
a~kind of electromagneto\gr al wave (nonlinear wave) with nontrivial \ia s between fields. The objective of this hope is related to points 4) and
5) (see p.~6 of Ref.~\cite1) in the list of ``interference effects'' (we recall that the displacement current in the classical Maxwell \e s leads
us to the nontrivial \ia\ between the electric and magnetic field---the {\it raison d'\^etre} of the wave \so s for Maxwell \e s. However, this
is only a historical remark.  By a nontrivial \ia, we mean that the flow of energy is possible from one field to another in a quasi-periodic way.
One can try to use the \fw\ Ansatz for the \dots}.

Points 4) and 5) from page~6 of Ref.~\cite1 are as follows:

{\sl 4) The source in the second pair of Maxwell (\YM) \e s, i.e.\ a current $j_\mu$ $(j^a_\mu)$.

5) The polarization of vacuum $M_\m=-\frac1{4\pi}(H_\m-F_\m)$ $(M^a_\m=-\frac1{4\pi}(L^a_\m-H^a_\m))$ with an interpretation as the torsion in the
5th \di\ (in higher \di s in \YM\ case).}

Let us remind to the reader that in Ref.~\cite7 we consider gravito-\elm c waves in the \NK{}. In some sense they are similar to Ansatz from
Ref.~\cite1, p.~289. They are generalized plane waves for $g_\m$ (\nos) and generalized \elm c plane waves. We use some achievements from Einstein
Unified Field Theory to find \so s for \gelm c waves (generalized plane \gelm c waves). Let us notice that in the limit of zero
skewon and \elm c fields we get generalized plane waves from General Relativity (see Ref.~\cite{xb}). The \elm c field of our \so\ has remarkable
properties to have both invariants of the \elm c field $S=F_\m F^\m$ and $P=F_\m F^{\ast\m}$ equal to zero. The \elm c field depends on arbitrary
\f\ of $(z-t)$. Skewon part of the \so\ describes a wave in this sense that it depends on $(z-t)$ (a~Riemann invariant, see Ref.~\cite{xa}). It
depends also on an arbitrary \f\ of $(z-t)$. Travelling in higher \di s can also be considered.

Recently the LIGO and Virgo collaborations achieved a three-detector observation of \gr al waves from a binary black hole coalescence (see
Ref.~\cite{47}). This is the first case of \gr al waves detection with three-detector observation.

\section*{Appendix A}
In this appendix we give a programme written in \ti{Mathematica} ({\tt kalinowski\_1044pr1w2.nb})
to calculate $\ov R_\(\a\b)$, $\ov R_\[\a\b]$, $\Tem_{(\a\b)}$, $\Tem_{[\a\b]}\frac{}{}$, $W^\m$ for \as, \sy\ case.

{\tt \small
\begin{verbatim}
x[1] = x
x[2] = y
x[3] = z
x[4] = t1
r2[x,y]=f2[x,y]^(1/2)
e2[x,y]=D[w1[x,y],x[1]]*Exp[-n1[x,y]]*f2[x,y]^(1/2)
e3[x,y]=D[w1[x,y],x[2]]*Exp[-n1[x,y]]*f2[x,y]^(1/2)
r1[x, y]=x*f2[x,y]^(-1/2)
tt = Table[0, {i, 1, 4}, {j, 1, 4}]
tt[[1, 1]] = -1
tt[[2, 2]] = -1
tt[[3, 3]] =-r1[x, y]^2
tt[[4, 4]] = r2[x, y]^2
tt[[1, 4]] =b2[x,y]*f2[x,y]^(1/2)
tt[[4, 1]] =-b2[x,y]*f2[x,y]^(1/2)
n1[x,y]=m1[x,y]-1/2*Log[f2[x,y]]
ff = Table[0, {i, 1, 4}, {j, 1, 4}]
ff[[2, 4]] = D[a3[x, y], x[2]]*Exp[-n1[x, y]]*f2[x,y]^(1/2)
ff[[4, 1]] = - D[a3[x, y], x[1]]*Exp[-n1[x, y]]*f2[x,y]^(1/2)
ff[[4, 2]] = - D[a3[x, y], x[2]]*Exp[-n1[x, y]]*f2[x,y]^(1/2)
ff[[1, 4]] = D[a3[x, y], x[1]]*Exp[-n1[x, y]]*f2[x,y]^(1/2)
MatrixForm[tt]
MatrixForm[ff]
h1h = Table[0, {i, 1, 4}, {j, 1, 4}, {k, 1, 4}]
h1h[[1, 2, 1]] = (D[n1[x, y], x[2]]
               - 1/2*D[f2[x, y], x[2]]*f2[x, y]^(-1))*Exp[-n1[x, y]]*f2[x,y]^(1/2)
h1h[[1, 1, 2]] = -(D[n1[x, y], x[2]]
               - 1/2*D[f2[x, y], x[2]]*f2[x, y]^(-1))*Exp[-n1[x, y]]*f2[x,y]^(1/2)
h1h[[2, 1, 2]] = (D[n1[x, y], x[1]]
               - 1/2*D[f2[x, y], x[1]]*f2[x, y]^(-1))*Exp[-n1[x, y]]*f2[x,y]^(1/2)
h1h[[2, 2, 1]] = -(D[n1[x, y], x[1]]
               - 1/2*D[f2[x, y], x[1]]*f2[x, y]^(-1))*Exp[-n1[x, y]]*f2[x,y]^(1/2)
h1h[[4,1,3]]=-e2[x,y]
h1h[[4,3,1]]=e2[x,y]
h1h[[4,2,3]]=-e3[x,y]
h1h[[4,3,2]]=e3[x,y]
gam = Table[0, {i, 1, 4}, {j, 1, 4}, {k, 1, 4}]
gam[[1, 2, 2]] = -(D[n1[x, y], x[1]]
               - 1/2*D[f2[x, y], x[1]]*f2[x, y]^(-1))*Exp[-n1[x, y]]*f2[x,y]^(1/2)
gam[[2, 1, 1]] = -(D[n1[x, y], x[2]]
               - 1/2*D[f2[x, y], x[2]]*f2[x, y]^(-1))*Exp[-n1[x, y]]*f2[x,y]^(1/2)
gam[[1, 2, 1]] =(D[n1[x, y], x[2]]
               - 1/2*D[f2[x, y], x[2]]*f2[x, y]^(-1))*Exp[-n1[x, y]]*f2[x,y]^(1/2)
gam[[2, 1, 2]] =(D[n1[x, y], x[1]]
               - 1/2*D[f2[x, y], x[1]]*f2[x, y]^(-1))*Exp[-n1[x, y]]*f2[x,y]^(1/2)
gam[[1, 3, 3]] =- r1[x, y]*D[r1[x, y], x[1]]*Exp[-n1[x, y]]*f2[x,y]^(1/2)
gam[[2, 3, 3]] = -r1[x, y]*D[r1[x, y], x[2]]*Exp[-n1[x, y]]*f2[x,y]^(1/2)
gam[[1, 4, 4]] =r2[x, y]*D[r2[x, y], x[1]]*Exp[-n1[x, y]]*f2[x,y]^(1/2)
gam[[2, 4, 4]] = r2[x, y]*D[r2[x, y], x[2]]*Exp[-n1[x, y]]*f2[x,y]^(1/2)
gam[[3, 1, 3]] = 1/r1[x, y]*D[r1[x, y], x[1]]*Exp[-n1[x, y]]*f2[x,y]^(1/2)
gam[[3, 3, 1]] = 1/r1[x, y]*D[r1[x, y], x[1]]*Exp[-n1[x, y]]*f2[x,y]^(1/2)
gam[[4, 1, 4]] = 1/r2[x, y]*D[r2[x, y], x[1]]*Exp[-n1[x, y]]*f2[x,y]^(1/2)
gam[[4, 4, 1]] = 1/r2[x, y]*D[r2[x, y], x[1]]*Exp[-n1[x, y]]*f2[x,y]^(1/2)
gam[[3, 2, 3]] = 1/r1[x, y]*D[r1[x, y], x[2]]*Exp[-n1[x, y]]*f2[x,y]^(1/2)
gam[[3, 3, 2]] = 1/r1[x, y]*D[r1[x, y], x[2]]*Exp[-n1[x, y]]*f2[x,y]^(1/2)
gam[[4, 4, 2]] = 1/r2[x, y]*D[r2[x, y], x[2]]*Exp[-n1[x, y]]*f2[x,y]^(1/2)
gam[[4, 2, 4]] = 1/r2[x, y]*D[r2[x, y], x[2]]*Exp[-n1[x, y]]*f2[x,y]^(1/2)
gam[[4, 1, 3]] =-1/2* e2[x, y]
gam[[4, 3, 1]] =1/2*e2[x, y]
gam[[4, 3, 2]] =1/2*e3[x, y]
gam[[4, 2, 3]] =-1/2*e3[x, y]
gam[[3, 4, 1]] =1/2*r2[x, y]^2*r1[x, y]^(-2)* e2[x, y]
gam[[3, 1, 4]] =1/2*r2[x, y]^2*r1[x, y]^(-2)* e2[x, y]
gam[[3, 4, 2]] =1/2*r2[x, y]^2*r1[x, y]^(-2)* e3[x, y]
gam[[3, 2, 4]] =1/2*r2[x, y]^2*r1[x, y]^(-2)* e3[x, y]
gam[[1, 3, 4]] =-1/2*r2[x, y]^2* e2[x, y]
gam[[1, 4, 3]] =-1/2*r2[x, y]^2* e2[x, y]
gam[[2, 3, 4]] =-1/2*r2[x, y]^2* e3[x, y]
gam[[2, 4, 3]] =-1/2*r2[x, y]^2* e3[x, y]
tts = FullSimplify[1/2*(tt + Transpose[tt])]
tti = FullSimplify[Inverse[tt]]
ttsi = FullSimplify[Inverse[tts]]
sdet = FullSimplify[Det[tt]]
sdets = FullSimplify[Det[tts]]
tta = FullSimplify[1/2*(tt - Transpose[tt])]
ttia = FullSimplify[1/2*(tti - Transpose[tti])]
ccudd = h1h;
gsuu = ttsi;
gdd = tt;
guu = tti;
ffdd = ff;
MatrixForm[tts]
MatrixForm[ttsi]
gsdd = 1/2*(gdd + Transpose[gdd]);
kkdd = 1/2*(gdd - Transpose[gdd]);
tkkuu[a_, b_] :=
Sum[Sum[gsuu[[l, a]]*gsuu[[m, b]]*kkdd[[l, m]], {m, 1, 4}], {l, 1, 4}];
kkuu = Table[tkkuu[a, b], {a, 1, 4}, {b, 1, 4}];
aguu= 1/2*(-Transpose[guu] + guu);
tkkddd[a_, b_, c_] := D[kkdd[[b, c]], x[a]] -
    Sum[(gam[[d, b, a]]*kkdd[[d, c]] - gam[[d, c, a]]*kkdd[[b, d]]), {d, 1, 4}];
kkddd = Table[tkkddd[a, b, c], {a, 1, 4}, {b, 1, 4}, {c, 1, 4}];
takkddd[a_, b_, c_] := -kkddd[[a, b, c]] + kkddd[[c, a, b]] - kkddd[[b, c, a]];
akkddd = Table[takkddd[a, b, c], {a, 1, 4}, {b, 1, 4}, {c, 1, 4}];
takkddu[n_, w_, m_] := Sum[gsuu[[m, p]]*akkddd[[n, w, p]], {p, 1, 4}];
akkddu = Table[takkddu[n, w, m], {n, 1, 4}, {w, 1, 4}, {m, 1, 4}];
takkdud[l_, a_, w_] := Sum[gsuu[[a, p]]*akkddd[[l, p, w]], {p, 1, 4}];
akkdud = Table[takkdud[l, a, w], {l, 1, 4}, {a, 1, 4}, {w, 1, 4}];
tkkdu[m_, a_] := Sum[gsuu[[a, p]]*kkdd[[m, p]], {p, 1, 4}];
kkdu = Table[tkkdu[m, a], {m, 1, 4}, {a, 1, 4}];
tgamudd[n_, w_, m_] := gam[[n, w, m]] + 1/2*(akkddu[[w, m,n]] -
    Sum[Sum[kkdu[[m, a]]*akkddd[[w, a, b]]*kkuu[[n, b]] -
    kkdu[[w, a]]*akkddd[[m, a, b]]*kkuu[[n, b]], {a, 1, 4}], {b, 1, 4}]) +
    Sum[Sum[gsuu[[n,l]]*(1/2*(akkddu[[l, a, w]]*kkdd[[m, a]] +
    akkddu[[l, m, a]]*kkdd[[ w,a]])), {a, 1, 4}], {l, 1, 4}] +
    Sum[Sum[Sum[Sum[1/2*(gsuu[[n, l]]*(kkdu[[c,b]]*
    (kkdu[[m, c]]*akkddd[[w, a, b]]*kkdu[[l, a]] +
    kkdu[[w, c]]*akkddd[[m, a, b]]*kkdu[[l, a]] -
    akkddd[[l, a, b]]*kkdu[[w, a]]*kkdu[[m, c]] -
    akkddd[[l,a,b]]*kkdu[[m, a]]*kkdu[[w, c]]))),
    {c, 1, 4}], {l, 1, 4}], {a, 1, 4}], {b, 1, 4}];
gamudd = Table[tgamudd[n, w, m], {n, 1, 4}, {w, 1, 4}, {m, 1, 4}];
ruddd[a_, b_, r_, s_] := (-D[gamudd[[a, b, r]], x[s]] + D[gamudd[[a, b, s]], x[r]])*
    Exp[-n1[x, y]]*f2[x,y]^(1/2) + Sum[(gamudd[[a, c, r]]*gamudd[[c, b, s]] -
    gamudd[[a, c, s]]*gamudd[[c, b, r]]), {c, 1, 4}] +
    Sum[gamudd[[a, b, c]]*ccudd[[c, r, s]], {c, 1, 4}];
tabruddd = Table[ruddd[a, b, r, s], {a, 1, 4}, {b, 1, 4}, {r, 1, 4}, {s, 1, 4}];
ricdd[b_, r_] := Sum[tabruddd[[a, b, r, a]] + 1/2*tabruddd[[a, a, b, r]], {a, 1, 4}];
tabricdd = Table[ricdd[b, r], {b, 1, 4}, {r, 1, 4}] ;
sricdd =Simplify[ 1/2*(tabricdd + Transpose[tabricdd])];
aricdd =Simplify[ 1/2*(tabricdd - Transpose[tabricdd])];
hhdd[w_, m_] := ffdd[[w, m]] +
    Sum[Sum[-gsuu[[t, a]]*ffdd[[a, w]]*kkdd[[m, t]] +
    gsuu[[t, a]]*ffdd[[a, m]]*kkdd[[w, t]], {a, 1, 4}], {t, 1, 4}];
tabhhdd = Table[hhdd[w, m], {w, 1, 4}, {m, 1, 4}];
hhguu[m_, a_] :=
    Sum[Sum[guu[[b, m]]*guu[[c, a]]*tabhhdd[[b, c]], {b, 1, 4}], {c, 1, 4}];
tabhhguu = Table[hhguu[m, a], {m, 1, 4}, {a, 1, 4}]
aadd[a_, b_] :=
    Sum[Sum[gdd[[c, b]]*tabhhdd[[m, a]]*tabhhguu[[m, c]], {m, 1, 4}], {c, 1, 4}];
tabaadd = Table[aadd[a, b], {a, 1, 4}, {b, 1, 4}];
bbdd[a_, b_] :=
    -2*Sum[Sum[aguu[[m, c]]*ffdd[[m, c]]*ffdd[[a, b]], {m, 1, 4}], {c, 1, 4}];
tabbbdd = Table[bbdd[a, b], {a, 1, 4}, {b, 1, 4}];
ccdd[a_, b_] := -1/4*gdd[[a, b]]*Sum[Sum[tabhhguu[[m, n]]*ffdd[[m, n]] -
    2*(aguu[[m, n]]*ffdd[[m, n]])^2, {n, 1, 4}], {m, 1, 4}];
tabccdd = Table[ccdd[a, b], {a, 1, 4}, {b, 1, 4}];
tedd = tabaadd + tabbbdd + tabccdd;
tesdd =Simplify[ 1/2*(tedd + Transpose[tedd])];
teadd =Simplify[1/2*(tedd - Transpose[tedd])];
wwuu[a_, m_] := (-sdet)^(1/2)*(tabhhguu[[a, m]] -
    2*aguu[[a, m]]*Sum[Sum[aguu[[n, b]]*ffdd[[n, b]], {n, 1, 4}], {b, 1, 4}]);
wwu[a_] := Sum[D[wwuu[a, m], x[m]], {m, 1, 4}];
ee=Table[0,{a,1,4},{b,1,4}]
ee = sricdd ;
ee1=2*tesdd;
eea=Table[0,{a,1,4},{b,1,4}]
eea = aricdd ;
eea1=2*teadd;
vv[a_] := Sum[D[(-sdet)^(1/2)*aguu[[a, b]], x[b]], {b, 1, 4}];
ww = FullSimplify[Table[wwu[a], {a, 1, 4}]]
vs = FullSimplify[Table[vv[a], {a, 1, 4}]]
\end{verbatim}

}
\def\tto#1;#2;{{\tt#1}\to#2}
Let us notice the \fw\ fact. In our programme to calculate symbolically we use several notions. Moreover, in the text we change a notation
for convenience. In this way one gets
\bgg
\tto x;\rho;, \qquad \tto y;z;\\
\tto b2[x,y];b;,\qquad \tto m1[x,y];m;\\
\tto b2^{(1,0)}[x,y];\pap b\rho;,\qquad \tto m1^{(1,0)}[x,y];\pap m\rho;\\
\tto b2^{(0,1)}[x,y];\pap bz;,\qquad \tto m1^{(0,1)}[x,y];\pap mz;\\
\tto b2^{(2,0)}[x,y];\pap{^2b}{\rho^2};,\qquad \tto m1^{(2,0)}[x,y];\pap{^2m}{\rho^2};\\
\tto b2^{(1,1)}[x,y];\pap{^2b}{\rho\pa z};,\qquad \tto m1^{(1,1)}[x,y];\pap{^2m}{\rho\pa z};\\
\tto b2^{(0,2)}[x,y];\pap{^2b}{z^2};,\qquad \tto m1^{(0,2)}[x,y];\pap{^2m}{z^2};\\
\tto f2[x,y];f;,\qquad \tto w1[x,y];\o;\\
\tto f2^{(1,0)}[x,y];\pap f\rho;,\qquad \tto w1^{(1,0)}[x,y];\pap \o \rho;\\
\tto f2^{(0,1)}[x,y];\pap fz;,\qquad \tto w1^{(0,1)}[x,y];\pap \o z;\\
\tto f2^{(2,0)}[x,y];\pap{^2f}{\rho^2};,\qquad \tto w1^{(2,0)}[x,y];\pap{^2\o }{\rho^2};\\
\tto f2^{(1,1)}[x,y];\pap{^2f}{\rho\pa z};,\qquad \tto w1^{(1,1)}[x,y];\pap{^2\o }{\rho\pa z};\\
\tto f2^{(0,2)}[x,y];\pap{^2f}{z^2};,\qquad \tto w1^{(0,2)}[x,y];\pap{^2\o }{z^2};\\
\tto a3^{(1,0)}[x,y];\pap {a_3}\rho;,\qquad \tto a3^{(0,1)}[x,y];\pap {a_3}z;
\e

\section*{Appendix B}
In this appendix we give a programme ({\tt kalinowski\_24mmkw.nb}) written in \ti{Mathematica} to calculate symbolically all
quantities from Section~2 for \cy\ gravito-\elm c wave.

{\small
\begin{verbatim}
x[1] = r1
x[2] = th
x[3] = z1
x[4] = t1
n1 = n2[x[1], (x[1] - x[4])]
a1 = D[b2[x[1], (x[1] - x[4])], x[1]]
s1 = s2[x[1] - x[4]]*(x[1])^(-1)
l1 = l2[x[1], (x[1] - x[4])]
b1 = -D[b2[x[1], (x[1] - x[4])], x[4]]
tt = Table[0, {i, 1, 4}, {j, 1, 4}]
tt[[1, 1]] = -Exp[2*(n1 - l1)]
tt[[1, 3]] = s1*Exp[2*l1]
tt[[2, 2]] = -(x[1])^2*Exp[-2*l1]
tt[[3, 1]] = -s1*Exp[2*l1]
tt[[3, 3]] = -Exp[2*l1]
tt[[3, 4]] = s1*Exp[2*l1]
tt[[4, 4]] = Exp[2*(n1 - l1)]
tt[[4, 3]] = -s1*Exp[2*l1]
ff = Table[0, {i, 1, 4}, {j, 1, 4}]
ff[[1, 3]] = a1
ff[[3, 1]] = -a1
ff[[3, 4]] = b1
ff[[4, 3]] = -b1
MatrixForm[tt]
MatrixForm[ff]
sdet = FullSimplify[Det[tt]]
gdd = tt;
ffdd = ff;
tti = FullSimplify[Inverse[tt]]
guu = tti;
Table[0, {i, 1, 4}, {j, 1, 4}]
tts = FullSimplify[1/2*(tt + Transpose[tt])]
MatrixForm[tts]
sdets = FullSimplify[Det[tts]]
ttsi = FullSimplify[Inverse[tts]]
gsuu = ttsi;
MatrixForm[ttsi]
gsdd =1/2*(gdd + Transpose[gdd]);
kkdd =1/2*(gdd - Transpose[gdd]);
tkkuu[a_, b_] :=
    Sum[Sum[gsuu[[l, a]]*gsuu[[m, b]]*kkdd[[l, m]], {m, 1, 4}], {l, 1, 4}];
kkuu = Table[tkkuu[a, b], {a, 1, 4}, {b, 1, 4}];
aguu = 1/2*(-Transpose[guu] + guu);
tchrudd[c_, b_, m_] :=
    Sum[1/2*gsuu[[a, c]]*(D[gsdd[[a, b]], x[m]] +
    D[gsdd[[a, m]], x[b]] - D[gsdd[[b, m]], x[a]]), {a, 1, 4}];
chrudd = Table[tchrudd[c, b, m], {c, 1, 4}, {b, 1, 4}, {m, 1, 4}];
tkkddd[a_, b_, c_] :=
    D[kkdd[[b, c]], x[a]] - Sum[(chrudd[[d, b, a]]*kkdd[[d, c]] +
    chrudd[[d, c, a]]*kkdd[[b, d]]), {d, 1, 4}];
kkddd = Table[tkkddd[a, b, c], {a, 1, 4}, {b, 1, 4}, {c, 1, 4}];
takkddd[a_, b_, c_] := -kkddd[[a, b, c]] + kkddd[[c, a, b]] - kkddd[[b, c, a]];
akkddd = Table[takkddd[a, b, c], {a, 1, 4}, {b, 1, 4}, {c, 1, 4}];
takkddu[ w_, m_,n_] := Sum[gsuu[[n, p]]*akkddd[[ w, m,p]], {p, 1, 4}];
akkddu = Table[takkddu[n, w, m], {n, 1, 4}, {w, 1, 4}, {m, 1, 4}];
chrudd = Table[tchrudd[c, b, m], {c, 1, 4}, {b, 1, 4}, {m, 1, 4}];
tkkddd[a_, b_, c_] :=
    D[kkdd[[b, c]], x[a]] - Sum[(chrudd[[d, b, a]]*kkdd[[d, c]] +
    chrudd[[d, c, a]]*kkdd[[b, d]]), {d, 1, 4}];
kkddd = Table[tkkddd[a, b, c], {a, 1, 4}, {b, 1, 4}, {c, 1, 4}];
takkddd[a_, b_, c_] := -kkddd[[a, b, c]] + kkddd[[c, a, b]] - kkddd[[b, c, a]];
akkddd = Table[takkddd[a, b, c], {a, 1, 4}, {b, 1, 4}, {c, 1, 4}];
takkddu[ w_, m_,n_] := Sum[gsuu[[n, p]]*akkddd[[ w, m,p]], {p, 1, 4}];
akkddu = Table[takkddu[n, w, m], {n, 1, 4}, {w, 1, 4}, {m, 1, 4}];
takkdud[l_, a_, w_] := Sum[gsuu[[a, p]]*akkddd[[l, p, w]], {p, 1, 4}];
akkdud = Table[takkdud[l, a, w], {l, 1, 4}, {a, 1, 4}, {w, 1, 4}];
tkkdu[m_, a_] := FullSimplify[Sum[gsuu[[a, p]]*kkdd[[m, p]], {p, 1, 4}]];
kkdu = Table[tkkdu[m, a], {m, 1, 4}, {a, 1, 4}];
tgamudd[n_, w_, m_] := chrudd[[n, w, m]] + 1/2*(akkddu[[w, m,n]] -
    Sum[Sum[kkdu[[m, a]]*akkddd[[w, a, b]]*kkuu[[n, b]] -
    kkdu[[w, a]]*akkddd[[m, a, b]]*kkuu[[n, b]], {a, 1, 4}], {b, 1, 4}]) +
    Sum[Sum[gsuu[[n, l]]*(1/2*(akkddu[[l, a, w]]*kkdd[[m, a]] +
    akkddu[[l, m, a]]*kkdd[[ w,a]])), {a, 1, 4}], {l, 1, 4}] +
    Sum[Sum[Sum[Sum[1/2*(gsuu[[n, l]]*(kkdu[[c,
    b]]*(kkdu[[m, c]]*akkddd[[w, a, b]]*kkdu[[l, a]] +
    kkdu[[w, c]]*akkddd[[m, a, b]]*kkdu[[l, a]] -
    akkddd[[l, a, b]]*kkdu[[w, a]]*kkdu[[m, c]] - akkddd[[l,a,b]]*
    kkdu[[m, a]]*kkdu[[w, c]]))), {c, 1, 4}], {l, 1, 4}], {a, 1, 4}], {b, 1, 4}];
gamudd = Table[tgamudd[n, w, m], {n, 1, 4}, {w, 1, 4}, {m, 1, 4}];
truddd[a_, b_, r_, s_] := FullSimplify[-D[gamudd[[a, b, r]], x[s]] +
    D[gamudd[[a, b, s]], x[r]] + Sum[(gamudd[[a, c, r]]*gamudd[[c, b, s]] -
    gamudd[[a, c, s]]*gamudd[[c, b, r]]), {c, 1, 4}]];
tabruddd= Table[truddd[a, b, r, s], {a, 1, 4}, {b, 1, 4}, {r, 1, 4}, {s, 1, 4}];
ricdd[b_, r_] := Sum[tabruddd[[a, b, r, a]] + 1/2*tabruddd[[a, a, b, r]], {a, 1, 4}];
tabricdd = Table[ricdd[b, r], {b, 1, 4}, {r, 1, 4}] ;
sricdd =Simplify[ 1/2*(tabricdd + Transpose[tabricdd])];
aricdd =Simplify[ 1/2*(tabricdd - Transpose[tabricdd])];
hhdd[w_, m_] := ffdd[[w, m]] +
    Sum[Sum[-gsuu[[t, a]]*ffdd[[a, w]]*kkdd[[m, t]] +
    gsuu[[t, a]]*ffdd[[a, m]]*kkdd[[w, t]], {a, 1, 4}], {t, 1, 4}];
tabhhdd = Table[hhdd[w, m], {w, 1, 4}, {m, 1, 4}];
hhguu[m_, a_] :=
    Sum[Sum[guu[[b, m]]*guu[[c, a]]*tabhhdd[[b, c]], {b, 1, 4}], {c, 1, 4}];
tabhhguu = Table[hhguu[m, a], {m, 1, 4}, {a, 1, 4}];
aricdd =Simplify[ 1/2*(tabricdd - Transpose[tabricdd])];
hhdd[w_, m_] := ffdd[[w, m]] + Sum[Sum[-gsuu[[t, a]]*ffdd[[a, w]]*kkdd[[m, t]] +
    gsuu[[t, a]]*ffdd[[a, m]]*kkdd[[w, t]], {a, 1, 4}], {t, 1, 4}];
tabhhdd = Table[hhdd[w, m], {w, 1, 4}, {m, 1, 4}];
hhguu[m_, a_] :=
    Sum[Sum[guu[[b, m]]*guu[[c, a]]*tabhhdd[[b, c]], {b, 1, 4}], {c, 1, 4}];
tabhhguu = Table[hhguu[m, a], {m, 1, 4}, {a, 1, 4}]
aadd[a_, b_] :=
    Sum[Sum[gdd[[c, b]]*tabhhdd[[m, a]]*tabhhguu[[m, c]], {m, 1, 4}], {c, 1, 4}];
tabaadd = Table[aadd[a, b], {a, 1, 4}, {b, 1, 4}];
bbdd[a_, b_] := -2* Sum[Sum[aguu[[m, c]]*ffdd[[m, c]]*ffdd[[a, b]],
    {m, 1, 4}], {c, 1, 4}];
tabbbdd = Table[bbdd[a, b], {a, 1, 4}, {b, 1, 4}];
ccdd[a_, b_] := -1/4*gdd[[a, b]]*Sum[Sum[tabhhguu[[m, n]]*ffdd[[m, n]] -
    2*(aguu[[m, n]]*ffdd[[m, n]])^2, {n, 1, 4}], {m, 1, 4}];
tabccdd = Table[ccdd[a, b], {a, 1, 4}, {b, 1, 4}];
tedd = tabaadd + tabbbdd + tabccdd;
tesdd = Simplify[ 1/2*(tedd + Transpose[tedd])];
teadd =Simplify[ 1/2*(tedd - Transpose[tedd])];
wwuu[a_, m_] :=FullSimplify[(-sdet)^(1/2)*(tabhhguu[[a, m]] - 2*aguu[[a, m]]*
    Sum[Sum[aguu[[n, b]]*ffdd[[n, b]], {n, 1, 4}], {b, 1, 4}])];
wwu[a_] := Sum[D[wwuu[a, m], x[m]], {m, 1, 4}];
ee=Table[0,{a,1,4},{b,1,4}]
ee = sricdd ;
ee1=2*tesdd;
eea=Table[0,{a,1,4},{b,1,4}]
eea = aricdd ;
eea1=2*teadd;
vv[a_] := FullSimplify[
Sum[D[(-sdet)^(1/2)*aguu[[a, b]], x[b]], {b, 1, 4}]];
ww = Table[wwu[a], {a, 1, 4}]
vs = Table[vv[a], {a, 1, 4}]
\end{verbatim}

}
Here also we change a notation in the text for convenience:
\bgg
\tto r1;r;,\qquad \tto l2[r1,r1-t1];l;\\
\tto n2[r1,r1-t1];n;,\qquad \tto s2[r1,r1-t1];f;\\
\tto s2'[r1,r1-t1];\frac{df}{du};,\qquad \tto s2''[r1,r1-t1];\frac{d^2f}{du^2};\\
\tto l2^{(1,0)}[r1,r1-t1];\pap lr;,\qquad \tto n2^{(1,0)}[r1,r1-t1];\pap nr;\\
\tto l2^{(0,1)}[r1,r1-t1];\pap lu;,\qquad \tto n2^{(0,1)}[r1,r1-t1];\pap nu;\\
\tto l2^{(2,0)}[r1,r1-t1];\pap{^2l}{r^2};,\qquad \tto n2^{(2,0)}[r1,r1-t1];\pap{^2n}{r^2};\\
\tto l2^{(1,1)}[r1,r1-t1];\pap{^2l}{r\pa u};,\qquad \tto n2^{(1,1)}[r1,r1-t1];\pap{^2n}{r\pa u};\\
\tto l2^{(0,2)}[r1,r1-t1];\pap{^2l}{u^2};,\qquad \tto n2^{(0,2)}[r1,r1-t1];\pap{^2n}{u^2};\\
\tto b2^{(1,0)}[r1,r1-t1];\pap br;,\qquad \tto b2^{(0,1)}[r1,r1-t1];\pap bu;
\e

\section*{Acknowledgement}
I would like to thank Professor B. Lesyng for the opportunity to carry out
computations using Mathematica\TM~9\footnote{Mathematica\TM\ is the
registered mark of Wolfram Co.} in the Centre of Excellence
BioExploratorium, Faculty of Physics, University of Warsaw, Poland.


\begin{thebibliography}{11}

\bibitem1
M. W. {Kalinowski},
{\it Nonsymmetric Fields Theory and its Applications\/},
World Scientific, Singapore 1990.

\bibitem{2}
M. W. {Kalinowski},
{\it Nonsymmetric Kaluza--Klein $($Jordan--Thiry$)$ Theory in a general nonabelian case}\/,
Int. J. Theor. Phys. {\bf30}, p.~281 (1991).

\bibitem{3}
M. W. {Kalinowski},
{\it Nonsymmetric Kaluza--Klein $($Jordan--Thiry$)$ Theory in the electromagnetic case}\/,
Int. Journal of Theor. Phys. {\bf31}, p.~611 (1992).

\bibitem{3a}
M. W. Kalinowski,
{\it Can we get confinement from extra dimensions}\/,
in: Physics of Elementary Interactions (ed. Z.~Ajduk, S.~Pokorski,
A.~K.~Wr\'oblewski), World Scientific, Singapore, New
Jersey, London, Hong Kong 1991, p.~294.

\bibitem4
M. W. {Kalinowski},
{\it Scalar fields
in the Nonsymmetric Kaluza--Klein $($Jordan--Thiry$)$ Theory}\/,
arXiv: hep-th/0307242v10, 7~Jul 2015.

\bibitem{4a}
M. {Kalinowski},
{\it Scalars Fields
in the Nonsymmetric Kaluza--Klein $($Jordan--Thiry$)$ Theory}\/,
Scholars' Press, Saarbr\"ucken 2016.

\bibitem{m}
M. W. Kalinowski,
\ti{Gauge fields with torsion} (in Polish),
PhD thesis submitted to the Faculty of Physics of the University of Warsaw (June 1978); published in full in Englis:
Int. J. Theor. Phys. {\bf20}, p.~563 (1981).

\bibitem{5}
J. W. Moffat,
{\it Generalized theory of gravitation and its physical consequences}\/,
in: Proceeding of the VII International School of Gravitation and Cosmology.
Erice, Sicilly, ed.  by V.~de~Sabbata, World  Scientific, Singapore, p.~127, 1982.

\bibitem{6}
M. W. {Kalinowski},
{\it Pioneer 10 and 11 spacecraft anomalous acceleration in the light of the Nonsymmetric
Kaluza--Klein $($Jordan--Thiry$)$ Theory},
Fortschrifte der Physik {\bf63}, p.~711 (2015).

\bibitem7
M. W. {Kalinowski},
{\it On some developments in the Nonsymmetric Kaluza--Klein Theory}\/,
The European Physical Journal {\bf C74}, id.~2742 (2014).

\bibitem{8}
M. W. Kalinowski,
\ti{The Nonsymmetric Kaluza--Klein Theory and Modern Physics. A novel approach},
Fortschritte der Physik {\bf 64}, p.~190 (2016).

\bibitem{9}
M. W. Kalinowski,
\ti{Hierarchy of symmetry breaking in the Nonsymmetric Kaluza--Klein (Jordan--Thiry) Theory},
J.~Phys. Math. {\bf7} (issue~1):152 (2016); arXiv: hep-th/0307135v4.

\bibitem{10}
M. W. Kalinowski,
\ti{Cosmological models in the Nonsymmetric Kaluza--Klein Theory},
Journal of Astrophysics and Aerospace Technology {\bf4} (issue~2), p.~133 (2016);
arXiv: astro-ph/0310745.

\bibitem{11}
M. W. Kalinowski,
{\it The programme of geometrization of physics. Some philosophical remarks},
Synthese {\bf77}, p.~129 (1988).

\bibitem{12}
M. Kalinowski,
{\it Holism and geometrization and unification of physical interactions},
Scholars' Press, Saarbr\"ucken 2016.

\bibitem{l2}
H. Prasad, K. B. Lal,
\ti{On cylindrical wave solutions of Einstein's unified field theory of gravitation and electromagnetism},
Tensor (N.S.) {\bf16}, p.~209 (1965).

\bibitem{l3}
H. Prasad, K. B. Lal,
\ti{On cylindrical wave solutions of Einstein's unified field theory of gravitation and electromagnetism} II,
Tensor (N.S.) {\bf17}, p.~179 (1966).

\bibitem{l1}
K. B. Lal, T. Singh,
\ti{On cylindrical wave solutions of Bonner's and Schr\"odinger's unified field equations},
Tensor (N.S.) {\bf23}, p.~151 (1972).

\bibitem{l4}
N. J. Cornish, J. W. Moffat, D. C. Tatarski,
\ti{Gravitational waves in the nonsymmetric gravitational theory},
arXiv: gr-qc/9211023v3.

\bibitem{l5}
N. J. Cornish, J. W. Moffat, D. C. Tatarski,
\ti{Gravitational waves from an axi-symmetric source in the nonsymmetric gravitational theory},
Gen. Rel. and Gravitation {\bf27}, p.~933 (1995).

\bibitem{l6}
N. J. Cornish, J. W. Moffat, D. C. Tatarski,
\ti{Gravitational waves in the nonsymmetric gravitational theory},
Phys. Lett. A {\bf173}, p.~109 (1993).

\bibitem{13}
V. Hlavat\'y,
{\it Geometry of Einstein Unified Field Theory},
P.~Noordhoff Ltd., Groningen 1957.

\bibitem{14}
R. C. Wrede,
\ti{``$n$'' Dimensional Considerations of Basic Principles A and~B of the
Unified Theory of Relativity}, Ph.D. Thesis submitted to the Faculty of the
Graduate School of Indiana University, August 1956; published partially in Tensor
(N.S.) {\bf8}, p.~95 (1958).

\bibitem{aa}
M. W. Kalinowski,
\ti{On generalization of Einstein--Cartan theory and Kaluza--Klein Theory},
Lett. in Math. Phys. {\bf5}, p.~489 (1981).

\bibitem{ab}
M. W. Kalinowski,
\ti{The Klein--Kaluza Theory with torsion},
Acta Physica Austriaca Suppl. {\bf XXVIII}, p.~641 (1981).

\bibitem{ac}
M. W. Kalinowski,
\ti{Torsion and the Klein--Kaluza Theory},
Acta Physica Austriaca {\bf57}, p.~45 (1985).

\bibitem{15}
F. J. Ernst,
\ti{New formulation of the axially symmetric gravitational field problem},
Phys. Rev. {\bf 167}, p.~1175 (1968).

\bibitem{16}
F. J. Ernst,
\ti{New formulation of the axially symmetric gravitational field problem}~II,
Phys. Rev. {\bf 168}, p.~168 (1968).

\bibitem{17}
J. N. Islam,
\ti{On the stationary axisymmetric Einstein--Maxwell equations},
General Relativity and Gravitation {\bf9}, p.~687 (1978).

\bibitem{18}
K. C. Das, S. Baneri,
\ti{Axially symmetric stationary solutions of Einstein--Maxwell equations},
General Relativity and Gravitation {\bf9}, p.~845 (1978).

\bibitem{19}
E. Kyriakopoulos,
\ti{New series of asymptotically flat axisymmetric and stationary solutions},
General Relativity and Gravitation {\bf20}, p.~1067 (1988).

\bibitem{20}
F. J. Ernst,
\ti{Complex potential formulation of the axially symmetric gravitational field problem},
Journal of Math. Physics {\bf15}, p.~1409 (1974).

\bibitem{21}
C. Reina, A. Treves,
\ti{Axisymmetric gravitational fields},
General Relativity and Gravitation {\bf7}, p.~817 (1976).

\bibitem{22}
T. Adamo, E. T. Newman,
\ti{The Kerr--Newman metric: A Review},
arxiv: 1410.6626v1 [gr-gc], Scolarpedia 9(10): 3791.

\bibitem{23}
H. Erbin,
\ti{Janis--Newman algorithm: simplifications and gauge field transformation},
arxiv: 1410.2602;
General Relativity and Gravitation {\bf47}, p.~19 (2015).

\bibitem{27}
G. Neugebauer, R. Meinel,
\ti{Progress in relativistic gravitational theory using the inverse scattering method},
J. Math. Phys. {\bf44}, p.~3407 (2003).

\bibitem{k}
R. L. Joshi, S. I. Husain,
\ti{Total radiation in Einstein Unified Field Theory},
Tensor (N.S.) {\bf15}, p.~66 (1964).

\bibitem{x}
M. Kaku,
\ti{Physics of the Impossible},
Penguin Books, New York, 2009.

\bibitem{x1}
M. Kaku,
\ti{Physics of the Future},
Penguin Books, New York, 2012.

\bibitem{x2}
M. Kaku,
\ti{Hyperspace},
Oxford Univ. Press, Oxford, New York, 2016.

\bibitem{alfa}
B. P. Abbott et al.,
\ti{Observation of gravitational waves from a binary black hole merger},
Phys. Rev. Lett. {\bf116}, id.~061102 (2016).

\bibitem{beta}
B. P. Abbott et al.,
\ti{Test of General Relativity with GW150914},
Phys. Rev. Lett. {\bf116}, id.~221101 (2016).

\bibitem{a}
B. P. Abbott et al.,
\ti{GW150914: First results from the search for binary black hole coalescence with Advanced LIGO},
Phys. Rev. D {\bf93}, id.~12203 (2016).

\bibitem{b}
B. P. Abbott et al.,
\ti{Binary black hole mergers in the first Advanced LIGO observing run},
Phys. Rev.~X {\bf 6}, id.~041015 (2016).

\bibitem{xb}
V. D. Zakharov,
\ti{Gravitational waves in Einstein's Theory} (in Russian),
Nauka, Moscow;
English transl.: Israel Program for Scientific Translations,
Halsted Press, Jerusalem--London 1972.

\bibitem{xa}
M. Kalinowski,
\ti{Exact solutions for nonlinear wave equations},
Scholars' Press, Saarbr\"ucken 2016.

\bibitem{47}
The LIGO Scientific Collaboration and the Virgo Collaboration,
\ti{GW170814: A three-detector observation of gravitational waves from a binary
black hole coalescence},
arXiv: 1709.09660 [gr-qc].

\end{thebibliography}
\end{document}